\documentclass[aps,prd,twocolumn,showpacs,superscriptaddress,showpacs]{revtex4}  

\pdfoutput=1
\usepackage{hyperref}
\usepackage{verbatim}
\usepackage{graphicx}
\usepackage{color}

\usepackage{epsfig}
\usepackage{colordvi}
\usepackage{amsmath}
\usepackage{dcolumn}
\usepackage{bm}     
\usepackage{url}    
\usepackage{amssymb}

\def\urltilde{\kern -.15em\lower .7ex\hbox{\~{}}\kern .04em}
\def\urldot{\kern -.10em.\kern -.10em}
\def\urlhttp{http\kern -.10em\lower -.1ex\hbox{:}\kern -.12em\lower 0ex\hbox{/}\kern -.18em\lower 0ex\hbox{/}}

\def\urltilde{\kern -.15em\lower .7ex\hbox{\~{}}\kern .04em}
\def\urldot{\kern -.10em.\kern -.10em}
\def\urlhttp{http\kern -.10em\lower -.1ex\hbox{:}\kern -.12em\lower 0ex\hbox{/}\kern -.18em\lower 0ex\hbox{/}}

\def\mevc {\ifmmode {\rm MeV}/c \else MeV$/c$\fi}
\def\mevcc {\ifmmode {\rm MeV}/c^2 \else MeV$/c^2$\fi}
\def\gevc {\ifmmode {\rm GeV}/c \else GeV$/c$\fi}
\def\gevcc {\ifmmode {\rm GeV}/c^2 \else GeV$/c^2$\fi}
\def\tevcc {\ifmmode {\rm TeV}/c^2 \else TeV$/c^2$\fi}

\def\ol   {\overline}

\def\vtd  {\ifmmode |V_{td}| \else $|V_{td}|$\fi}
\def\vtb  {\ifmmode |V_{tb}| \else $|V_{tb}|$\fi}
\def\vts  {\ifmmode |V_{ts}| \else $|V_{ts}|$\fi}
\def\vcb  {\ifmmode |V_{cb}| \else $|V_{cb}|$\fi}

\newcommand{\Ds} {\ifmmode D_{\mbox{\sl s}}^{-}
                       \else $D_{\mbox{\sl s}}^{-}$\fi}
\newcommand{\Bs} {\ifmmode B_{\mbox{\sl s}}^{0}
                       \else $B_{\mbox{\sl s}}^{0}$\fi}
\newcommand{\Bsb} {\ifmmode \ol B_{\mbox{\sl s}}^{0}
                       \else $\ol B_{\mbox{\sl s}}^{0}$\fi}
\newcommand{\Bsh} {\ifmmode B_{\mbox{\sl s}}^H
                       \else $B_{\mbox{\sl s}}^H$\fi}
\newcommand{\Bsl} {\ifmmode B_{\mbox{\sl s}}^L
                       \else $B_{\mbox{\sl s}}^L$\fi}
\newcommand{\Dsl} {\ifmmode D_{\mbox{\sl s}}^{-} \ell^+
                       \else $D_{\mbox{\sl s}}^{-} \ell^+$\fi}
\newcommand{\xs} {\ifmmode x_{\mbox{\sl s}}
                       \else $x_{\mbox{\sl s}}$\fi}
\newcommand{\xd} {\ifmmode x_d \else $x_d$\fi}
\newcommand{\lxy} {\ifmmode L_{\rm xy} \else $L_{\rm xy}$\fi}
\newcommand{\dgam} {\ifmmode \Delta\Gamma \else $\Delta\Gamma$\fi}
\newcommand{\dm} {\ifmmode \Delta m \else $\Delta m$\fi}
\newcommand{\ctau} {\ifmmode c\tau \else $c\tau$\fi}
 
\newcommand{\et}{E_T}

\newcommand{\abseta}{\mid\eta \mid\leq}

\newcommand{\ptran}{\mbox{${p_T}$}}
\newcommand{\etran}{\mbox{${E_T}$}}
\newcommand{\met}{\mbox{$\protect \raisebox{.3ex}{$\not$}\et$}}

\newcommand{\ppbar}{p\bar{p}}
\newcommand{\qqbar}{q\bar{q}} 
\newcommand{\ttbar}{t\bar{t}}
\newcommand{\tbar}{\bar{t}}
\newcommand{\mttb}{M_{t\bar{t}}}
\newcommand{\mttbi}{M_{t\bar{t},i}}

\newcommand{\mtop}{M_{t}}

\newcommand{\yh}{y_{h}}
\newcommand{\yl}{y_{l}}
\newcommand{\yt}{y_{t}}
\newcommand{\ytlab}{y_t^{\rm \ppbar}}
\newcommand{\ytrest}{y_{t}^{\rm \ttbar}}
\newcommand{\ytbar}{y_{\tbar}}
\newcommand{\ytbarlab}{y_{\tbar}^{\rm \ppbar}}
\newcommand{\dylh}{\Delta y_{lh}}
\newcommand{\dy}{\Delta y}
\newcommand{\al}{A^{\rm \ppbar}}
\newcommand{\ad}{A^{\rm{\ttbar}}}
\newcommand{\daddm}{A^{{\rm \ttbar}}(M_{\ttbar,i})}
\newcommand{\daddy}{A^{{\rm \ttbar}}(\dy_i)}

\newcommand{\ifb}{ {\rm fb}^{-1} }

\begin{document}
\hspace{5.2in} \mbox{Fermilab-Pub-10-525-E}

\bibliographystyle{apsrev}


\title{Evidence for a Mass Dependent Forward-Backward Asymmetry \\
in Top Quark Pair Production}


\affiliation{Institute of Physics, Academia Sinica, Taipei, Taiwan 11529, Republic of China} 
\affiliation{Argonne National Laboratory, Argonne, Illinois 60439, USA} 
\affiliation{University of Athens, 157 71 Athens, Greece} 
\affiliation{Institut de Fisica d'Altes Energies, Universitat Autonoma de Barcelona, E-08193, Bellaterra (Barcelona), Spain} 
\affiliation{Baylor University, Waco, Texas 76798, USA} 
\affiliation{Istituto Nazionale di Fisica Nucleare Bologna, $^z$University of Bologna, I-40127 Bologna, Italy} 
\affiliation{University of California, Davis, Davis, California 95616, USA} 
\affiliation{University of California, Los Angeles, Los Angeles, California 90024, USA} 
\affiliation{Instituto de Fisica de Cantabria, CSIC-University of Cantabria, 39005 Santander, Spain} 
\affiliation{Carnegie Mellon University, Pittsburgh, Pennsylvania 15213, USA} 
\affiliation{Enrico Fermi Institute, University of Chicago, Chicago, Illinois 60637, USA}
\affiliation{Comenius University, 842 48 Bratislava, Slovakia; Institute of Experimental Physics, 040 01 Kosice, Slovakia} 
\affiliation{Joint Institute for Nuclear Research, RU-141980 Dubna, Russia} 
\affiliation{Duke University, Durham, North Carolina 27708, USA} 
\affiliation{Fermi National Accelerator Laboratory, Batavia, Illinois 60510, USA} 
\affiliation{University of Florida, Gainesville, Florida 32611, USA} 
\affiliation{Laboratori Nazionali di Frascati, Istituto Nazionale di Fisica Nucleare, I-00044 Frascati, Italy} 
\affiliation{University of Geneva, CH-1211 Geneva 4, Switzerland} 
\affiliation{Glasgow University, Glasgow G12 8QQ, United Kingdom} 
\affiliation{Harvard University, Cambridge, Massachusetts 02138, USA} 
\affiliation{Division of High Energy Physics, Department of Physics, University of Helsinki and Helsinki Institute of Physics, FIN-00014, Helsinki, Finland} 
\affiliation{University of Illinois, Urbana, Illinois 61801, USA} 
\affiliation{The Johns Hopkins University, Baltimore, Maryland 21218, USA} 
\affiliation{Institut f\"{u}r Experimentelle Kernphysik, Karlsruhe Institute of Technology, D-76131 Karlsruhe, Germany} 
\affiliation{Center for High Energy Physics: Kyungpook National University, Daegu 702-701, Korea; Seoul National University, Seoul 151-742, Korea; Sungkyunkwan University, Suwon 440-746, Korea; Korea Institute of Science and Technology Information, Daejeon 305-806, Korea; Chonnam National University, Gwangju 500-757, Korea; Chonbuk National University, Jeonju 561-756, Korea} 
\affiliation{Ernest Orlando Lawrence Berkeley National Laboratory, Berkeley, California 94720, USA} 
\affiliation{University of Liverpool, Liverpool L69 7ZE, United Kingdom} 
\affiliation{University College London, London WC1E 6BT, United Kingdom} 
\affiliation{Centro de Investigaciones Energeticas Medioambientales y Tecnologicas, E-28040 Madrid, Spain} 
\affiliation{Massachusetts Institute of Technology, Cambridge, Massachusetts 02139, USA} 
\affiliation{Institute of Particle Physics: McGill University, Montr\'{e}al, Qu\'{e}bec, Canada H3A~2T8; Simon Fraser University, Burnaby, British Columbia, Canada V5A~1S6; University of Toronto, Toronto, Ontario, Canada M5S~1A7; and TRIUMF, Vancouver, British Columbia, Canada V6T~2A3} 
\affiliation{University of Michigan, Ann Arbor, Michigan 48109, USA} 
\affiliation{Michigan State University, East Lansing, Michigan 48824, USA}
\affiliation{Institution for Theoretical and Experimental Physics, ITEP, Moscow 117259, Russia}
\affiliation{University of New Mexico, Albuquerque, New Mexico 87131, USA} 
\affiliation{Northwestern University, Evanston, Illinois 60208, USA} 
\affiliation{The Ohio State University, Columbus, Ohio 43210, USA} 
\affiliation{Okayama University, Okayama 700-8530, Japan} 
\affiliation{Osaka City University, Osaka 588, Japan} 
\affiliation{University of Oxford, Oxford OX1 3RH, United Kingdom} 
\affiliation{Istituto Nazionale di Fisica Nucleare, Sezione di Padova-Trento, $^{aa}$University of Padova, I-35131 Padova, Italy} 
\affiliation{LPNHE, Universite Pierre et Marie Curie/IN2P3-CNRS, UMR7585, Paris, F-75252 France} 
\affiliation{University of Pennsylvania, Philadelphia, Pennsylvania 19104, USA}
\affiliation{Istituto Nazionale di Fisica Nucleare Pisa, $^{bb}$University of Pisa, $^{cc}$University of Siena and $^{dd}$Scuola Normale Superiore, I-56127 Pisa, Italy} 
\affiliation{University of Pittsburgh, Pittsburgh, Pennsylvania 15260, USA} 
\affiliation{Purdue University, West Lafayette, Indiana 47907, USA} 
\affiliation{University of Rochester, Rochester, New York 14627, USA} 
\affiliation{The Rockefeller University, New York, New York 10065, USA} 
\affiliation{Istituto Nazionale di Fisica Nucleare, Sezione di Roma 1, $^{ee}$Sapienza Universit\`{a} di Roma, I-00185 Roma, Italy} 

\affiliation{Rutgers University, Piscataway, New Jersey 08855, USA} 
\affiliation{Texas A\&M University, College Station, Texas 77843, USA} 
\affiliation{Istituto Nazionale di Fisica Nucleare Trieste/Udine, I-34100 Trieste, $^{ff}$University of Trieste/Udine, I-33100 Udine, Italy} 
\affiliation{University of Tsukuba, Tsukuba, Ibaraki 305, Japan} 
\affiliation{Tufts University, Medford, Massachusetts 02155, USA} 
\affiliation{University of Virginia, Charlottesville, VA  22906, USA}
\affiliation{Waseda University, Tokyo 169, Japan} 
\affiliation{Wayne State University, Detroit, Michigan 48201, USA} 
\affiliation{University of Wisconsin, Madison, Wisconsin 53706, USA} 
\affiliation{Yale University, New Haven, Connecticut 06520, USA} 
\author{T.~Aaltonen}
\affiliation{Division of High Energy Physics, Department of Physics, University of Helsinki and Helsinki Institute of Physics, FIN-00014, Helsinki, Finland}
\author{B.~\'{A}lvarez~Gonz\'{a}lez$^v$}
\affiliation{Instituto de Fisica de Cantabria, CSIC-University of Cantabria, 39005 Santander, Spain}
\author{S.~Amerio}
\affiliation{Istituto Nazionale di Fisica Nucleare, Sezione di Padova-Trento, $^{aa}$University of Padova, I-35131 Padova, Italy} 

\author{D.~Amidei}
\affiliation{University of Michigan, Ann Arbor, Michigan 48109, USA}
\author{A.~Anastassov}
\affiliation{Northwestern University, Evanston, Illinois 60208, USA}
\author{A.~Annovi}
\affiliation{Laboratori Nazionali di Frascati, Istituto Nazionale di Fisica Nucleare, I-00044 Frascati, Italy}
\author{J.~Antos}
\affiliation{Comenius University, 842 48 Bratislava, Slovakia; Institute of Experimental Physics, 040 01 Kosice, Slovakia}
\author{G.~Apollinari}
\affiliation{Fermi National Accelerator Laboratory, Batavia, Illinois 60510, USA}
\author{J.A.~Appel}
\affiliation{Fermi National Accelerator Laboratory, Batavia, Illinois 60510, USA}
\author{A.~Apresyan}
\affiliation{Purdue University, West Lafayette, Indiana 47907, USA}
% ...
\collaboration{CDF Collaboration\footnote{With visitors from $^a$University of Massachusetts Amherst, Amherst, Massachusetts 01003,
$^b$Istituto Nazionale di Fisica Nucleare, Sezione di Cagliari, 09042 Monserrato (Cagliari), Italy,
$^c$University of California Irvine, Irvine, CA  92697, 
$^d$University of California Santa Barbara, Santa Barbara, CA 93106
$^e$University of California Santa Cruz, Santa Cruz, CA  95064,
$^f$CERN,CH-1211 Geneva, Switzerland,
$^g$Cornell University, Ithaca, NY  14853, 
$^h$University of Cyprus, Nicosia CY-1678, Cyprus, 
$^i$University College Dublin, Dublin 4, Ireland,
$^j$University of Fukui, Fukui City, Fukui Prefecture, Japan 910-0017,
$^k$Universidad Iberoamericana, Mexico D.F., Mexico,
$^l$Iowa State University, Ames, IA  50011,
$^m$University of Iowa, Iowa City, IA  52242,
$^n$Kinki University, Higashi-Osaka City, Japan 577-8502,
$^o$Kansas State University, Manhattan, KS 66506,
$^p$University of Manchester, Manchester M13 9PL, England,
$^q$Queen Mary, University of London, London, E1 4NS, England,
$^r$Muons, Inc., Batavia, IL 60510,
$^s$Nagasaki Institute of Applied Science, Nagasaki, Japan, 
$^t$National Research Nuclear University, Moscow, Russia,
$^u$University of Notre Dame, Notre Dame, IN 46556,
$^v$Universidad de Oviedo, E-33007 Oviedo, Spain, 
$^w$Texas Tech University, Lubbock, TX  79609, 
$^x$Universidad Tecnica Federico Santa Maria, 110v Valparaiso, Chile,
$^y$Yarmouk University, Irbid 211-63, Jordan,
$^{gg}$On leave from J.~Stefan Institute, Ljubljana, Slovenia. 
}}
\noaffiliation



\begin{abstract}
We present a new measurement of the inclusive forward-backward $\ttbar$ production asymmetry and its rapidity and mass dependence. The measurements are performed with data corresponding to an integrated luminosity of $5.3~\ifb$ of $\ppbar$ collisions at $\sqrt{s} = 1.96$ TeV, recorded with the CDF II Detector at the Fermilab Tevatron. Significant inclusive asymmetries are observed in both the laboratory frame and the $\ttbar$ rest frame, and in both cases are found to be consistent with CP conservation under interchange of $t$ and $\bar{t}$. In the $\ttbar$ rest frame, the asymmetry is observed to increase with the $\ttbar$ rapidity difference, $\dy$, and with the invariant mass $\mttb$ of the $\ttbar$ system. Fully corrected parton-level asymmetries are derived in two regions of each variable, and the asymmetry is found to be most significant at large $\dy$ and $\mttb$. For $\mttb \geq 450~\gevcc$, the parton-level asymmetry in the $\ttbar$ rest frame is $A^{{\rm \ttbar}} = 0.475\pm 0.114$ compared to a next-to-leading order QCD prediction of $0.088\pm 0.013$.

\end{abstract}  


\maketitle

\date{\today}


\pagebreak

\section{Introduction}\label{intro}

Top quark pair production in $\ppbar$ collisions is a sensitive probe of quantum chromodynamics at high energy.
At lowest order in the standard model (SM), quark pair production is symmetric under charge conjugation. At next-to-leading order (NLO) the interference of processes that differ under charge conjugation leads to a small forward-backward asymmetry of order $0.06\pm 0.01$ in the $\ttbar$ rest frame ~\cite{almeida,kuhn,nlotheory}. An analogous effect is predicted at order $\alpha^3$ in QED and is confirmed in measurements of $e^+e^-\rightarrow \mu^+\mu^-$ \cite{qed}. Study of the NLO QCD asymmetry in inclusive jet events is
 hampered by the difficulty of measuring the jet charge. In pair produced top quarks with one semi-leptonic decay, the top can be tagged according to the well-measured lepton charge, enabling a probe of the NLO QCD effect and a test of charge conjugation symmetry in strong interactions at high energy.

The CDF and D0 experiments have made initial measurements of the $\ttbar$ asymmetry in $\ppbar$ collisions at $\sqrt{s} = 1.96$ TeV at the Fermilab Tevatron \cite{afbprlcdf,afbprld0}. In the $CP$-invariant $\ppbar$ system, the NLO QCD effect appears as a charge dependent forward-backward asymmetry of the top quark direction with respect to the proton direction. Using data samples corresponding to $1.9~\ifb$ and $0.9~\ifb$ respectively, CDF and D0 report positive asymmetries that are consistent with the QCD prediction within large experimental uncertainties. A number of theoretical papers suggest interesting new physics mechanisms including axigluons, diquarks, new weak bosons, and extra-dimensions that can all produce forward-backward $\ttbar$ asymmetries \cite{afbtheory,rodrigo}. The model building must accommodate the observed consistency of the $\ttbar$ cross-section and total invariant mass distribution with the SM QCD prediction \cite{ alice, tZxsec}.

We report here on a new study of the forward-backward asymmetry in $\ppbar$ collisions at $\sqrt{s} = 1.96$ TeV, using data corresponding to an integrated luminosity 
of $5.3 ~\ifb$ recorded with the CDF II Detector. We study events with the lepton+jets topology, where either the $t$ or $\bar{t}$ has decayed semileptonically. The asymmetries are measured in two variables: $\yh$, the rapidity of the hadronically decaying top quark, corresponding to the top rapidity in the laboratory (lab) frame, and $\dy$, the difference of the rapidities of the leptonic and hadronic top systems, which measures the top quark rapidity in the $\ttbar$ rest frame. We show that $t$ and $\bar{t}$ asymmetries are consistent with $CP$ conservation, and combine them to measure the total asymmetry in the sample. We measure the inclusive asymmetries, and the functional dependence of the $\ttbar$ frame asymmetry on $\dy$ and on the total invariant mass of the $\ttbar$ system, $\mttb$. We apply corrections for backgrounds, acceptance, and resolution to calculate parton level measures of the inclusive asymmetry in both the lab and $\ttbar$ rest frames, and in two regions of $\dy$ and $\mttb$ in the $\ttbar$ frame.

\section{Detection, Event Selection and Reconstruction}\label{sec:det}

CDF II is a general purpose, azimuthally and forward-backward symmetric magnetic spectrometer with calorimetry and muon detectors \cite{cdf}. Charged particle trajectories are measured with a silicon-microstrip detector backed by a large open-cell drift chamber in a 1.4 T solenoidal magnetic field. Electromagnetic and hadronic calorimeters located outside the solenoid provide jet and missing energy reconstruction. Outside the calorimeter are multilayer proportional chambers and plastic scintillator hodoscopes that provide muon identification in the pseudorapidity region $\abseta 1.0$. We use a cylindrical coordinate system with origin at the detector center and z-axis along the proton 
direction \cite{coords}. 
 
This measurement uses $t\bar{t}$ candidate events in the ``lepton+jets'' topology, where one top quark decays semileptonically ($t \rightarrow l \nu b$) and the other hadronically ($t \rightarrow q \bar{q}^{\prime} b$) \cite{cc}. We detect the lepton and four jets from top quark decays and quark hadronization, and an inferred neutrino based on the presence of missing energy. The detector is triggered by a high transverse momentum electron(muon) in the central portion of the detector  with $\etran(\ptran) >20$ GeV(GeV/c) and $|\eta| < 1.0$.  We require four or more hadronic jets with $E_T >20$ GeV and $|\eta| < 2.0$, and a large amount of missing transverse energy, $\met \geq 20$ GeV, consistent with the presence of an undetected neutrino. The jets are reconstructed using a cone algorithm with $\delta R = \sqrt{\delta\phi^2+\delta\eta^2} < 0.4$, and calorimeter signals are corrected for detector inefficiencies and for the energy scale factor. The SECVTX algorithm \cite{secvtx} is used to find displaced $b$-decay vertices using the tracks within the jet cones, and at least one jet must contain such a ``$b$-tag''. Jets with $b$-tags are restricted to $|\eta| < 1.0$. 

A total of 1260 events pass this selection. The size of the non-$\ttbar$ background processes in the lepton+jets+$b$-tag selection is derived in precision measurements of the $\ttbar$ production cross-section \cite{tZxsec}. The estimated background in our sample $283.3 \pm 91.2$ events. The predominant backgrounds are from QCD-induced W+multi-parton events containing either $b$-tagged heavy-flavor jets or errantly tagged light-flavor jets. These are modeled using a simulation sample derived from the {\sc alpgen} generator \cite{alpgen} and a data driven technique that derives tagging efficiencies, mis-tagging rates and sample normalizations from direct measurement. A background component from QCD multi-jet events with fake leptons and mis-measured $\met$ is modeled using multi-jet events with lepton candidates that are rejected by our cuts. Other small backgrounds from electroweak processes ($WW, WZ$, single-top) are reliably estimated using Monte Carlo generators. Further details on the sample selection and background modeling can be found in Ref.~\cite{tZxsec}.

The reconstruction (reco) of the $\ttbar$ kinematics uses the measured momenta of the lepton, the $\met$, and the four leading jets in the event. The jet-parton assignment and calculation of the $\ttbar$ four-vectors uses a simple $\chi^2$-based fit of the lepton and jet kinematics to the $\ttbar$ hypothesis, allowing the jet energies to float within their expected uncertainties, and applying the constraints that $M_W = 80.4 ~\gevcc$, $M_t = ~172.5 ~\gevcc$, and $b$-tagged 
jets are associated with b-partons. This algorithm is well understood in the context of precision top mass measurements, where the fit is performed without the top mass constraint~\cite{reco}, and other top physics studies that use the top mass constraint~\cite{alice}. We study the reconstructed top quark rapidity and the difference in the reconstructed top and anti-top rapidities, from which we derive the forward-backward asymmetries in the $\ppbar$ (laboratory) rest frame and in the $\ttbar$ rest frame.

The validity of the analysis is checked at all steps by comparison to a standard prediction made using the {\sc pythia}~\cite{pythia} $\ttbar$ model, the CDF lepton+jets+$b$-tag background model, and a full simulation of the CDF-II detector. We use {\sc pythia} 6.2.16 with CTEQ5L parton distribution functions~\cite{cteq} and $\mtop = 172.5 ~\gevcc$. The background model developed in concert with the precision cross-section studies provides good measures of both the normalizations and shapes of the non-$\ttbar$ processes \cite{tZxsec}. The veracity of the combined {\sc pythia} plus background model, and in particular, its reliability for the estimation of systematic uncertainties, is well verified in many other top-physics studies at CDF~\cite{afbprlcdf,tZxsec,reco,alice,dlm}.

Note that because {\sc pythia} does not include the NLO QCD charge asymmetry, the standard {\sc pythia} prediction is not the SM prediction for the forward-backward asymmetry. Studies with the {\sc mc@nlo} generator \cite{mcnlo} (see Sec.~\ref{sec:mcnlo}) predict that the magnitude of reconstructed QCD asymmetry in our sample is smaller than the current experimental resolution. Symmetric {\sc pythia} is thus a good approximation for SM $\ttbar$ and provides an unbiased control sample for many of our studies. We will compare our measurements to the SM predictions of {\sc mc@nlo} when appropriate.
  
\section{Rapidity Variables and Asymmetry Definitions}
In the lepton+jets decay topology of the $\ttbar$ pair, there is a leptonic decay, 
$t\rightarrow Wb \rightarrow l\nu b$, and a hadronic decay $t\rightarrow Wb \rightarrow q\bar{q}^{\prime} b$. The complications of the central lepton acceptance and the reconstruction of the neutrino from the $\met$ create a difference in the reconstruction resolution for the two different kinds of decay. In order to control effects of this kind, our treatment of top rapidity variables maintains the distinction between the leptonic and hadronic decay systems, with the $t$ and $\bar{t}$ assignments following in accordance with the lepton charge.

The most direct measurement of the top direction with respect to the beamline is the rapidity of the hadronic top 
system in the lab frame, $\yh$, which has acceptance out to $|\eta| = 2.0$ and good directional precision. In events with a negative (positive) lepton, $\yh$ is the lab rapidity of the $t$ quark, $\ytlab$ ($\tbar$ quark, $\ytbarlab$). If $CP$ is a good symmetry, the distributions of 
$\ytbarlab$ and $\ytlab$ are reflections of each other, and we can combine both samples, weighting with the lepton charge, to use $-q\yh$ as the rapidity of the $t$ quark in the lab frame, $\ytlab$.

A frame independent measurement is available in the rapidity difference of the leptonic and hadronic systems $\dy_{lh} = \yl -\yh$. After multiplication by the lepton charge $q$, this variable measures the difference between the top and antitop rapidities: $q\dy_{lh} = q(\yl - \yh) =  \yt - \ytbar = \dy$. The rapidity difference $\dy$ is independent of the $\ttbar$ system longitudinal motion and is simply related to the top quark rapidity in the $\ttbar$ rest frame: $\ytrest = \frac{1}{2}\dy$. Since the rapidity preserves the sign of the production angle, an asymmetry in $\dy$ is identical to the asymmetry in the top quark production angle in the $\ttbar$ rest frame.   

With N as the number of events with a given rapidity, we define the total $\ttbar$ frame asymmetry:

\begin{eqnarray}
   \ad   & =  & \frac{N(\dy > 0) - N(\dy< 0)}{N(\dy > 0) + N(\dy < 0)} \\
         & = & \frac{N(\ytrest> 0) - N(\ytrest< 0)}{N(\ytrest> 0) + N(\ytrest< 0)}  \notag \\
         &    &   \notag
\end{eqnarray}
   
\noindent and the total laboratory frame asymmetry, assuming $CP$ invariance:

\begin{eqnarray}
   \al    & = & \frac{N(-q\yh > 0) - N(-q\yh< 0)}{N(-q\yh > 0) + N(-q\yh < 0)}  \\ 
          & = & \frac{N(\ytlab> 0) - N(\ytlab< 0)}{N(\ytlab > 0) + N(\ytlab< 0)}   \notag \\
          &   &  \notag
\end{eqnarray}  

\noindent Since $\yh$ and $\dy_{lh}$ are identified with either a $t$ or an $\bar{t}$ by the sign of the lepton in the event, they are the primary variables for defining the charge dependence of the asymmetries and testing for $CP$ invariance. We define the forward-backward charge asymmetry in the $\ttbar$ rest frame to be: 

\begin{eqnarray}\label{deltayafb}
   A_{lh}^{\pm} &  = & \frac{N^{\pm}(\dy_{lh} > 0) - N^{\pm}(\dy_{lh}< 0)}{N^{\pm}(\dy_{lh} > 0) + N^{\pm}(\dy_{lh} < 0)}\\
                &    &  \notag
\end{eqnarray}

\noindent and in the laboratory frame to be:

\begin{eqnarray}\label{Yhadafb}
   A_{h}^{\pm} & = & \frac{N^{\pm}(\yh > 0) - N^{\pm}(\yh< 0)}{N^{\pm}(\yh > 0) + N^{\pm}(\yh < 0)}\\
               &   & \notag
\end{eqnarray}

\noindent where the superscript refers to the sign of the lepton charge $q$.

\begin{center} 
\begin{table*}[!th]
\caption{Summary of rapidity variables and asymmetries.} \label{tab:defs}
\begin{tabular}{l l}
\hline
\hline
                &  definition\\
\hline
 $\yh$          &  rapidity of hadronic top system in lab\\
 $\yl$          &  rapidity of leptonic top system in lab\\
 $\dylh$        &  rapidity difference $\yl-\yh$\\
 $\dy$          &  $\ttbar$ rapidity difference: $y_{t}-y_{\bar{t}}= q(\yl-\yh)$\\
 $\ytlab$       &  top quark rapidity in laboratory frame: $-q\yh$\\
 $\ytrest$      &  top quark rapidity in $\ttbar$ rest frame: $\frac{1}{2}\dy$\\
\hline
 $A_{h}$        &  asymmetry in $\yh$\\
 $A_{lh}$       &  asymmetry in $\dylh$\\
 $\al$          &  laboratory frame asymmetry in $\ytlab$ (both charges)\\
 $\ad$          &  $\ttbar$ frame asymmetry in $\ytrest$ (both charges)\\         
\hline
\hline                
\end{tabular}
\end{table*}
\end{center}

The laboratory and $\ttbar$ frame present trade-offs for the asymmetry measurement. The laboratory frame is experimentally simple: the direction of the three-jet hadronic top decay in the detector is well-resolved, with uncertainty dominated by a Gaussian width $\delta \yh \sim 0.034$, and free from the complications of the neutrino reconstruction~\cite{towers}. The $\yh$ distribution is thus the simplest way to test for the presence of an asymmetry. However, as the laboratory frame includes an uncontrolled longitudinal boost from the rest frame of the primary $\qqbar$ interaction, the information on the fundamental production asymmetry is diluted.

Because the momentum scale of initial state radiation is small compared to $\mttb$, the $\qqbar$ frame is well approximated by the $\ttbar$ rest frame. We measure the $\ttbar$ frame rapidity in an experimentally robust way using the difference of two rapidities in the detector frame, $\dy = q(y_l-y_h)$. But the inclusion of $\yl$ and the poorly resolved neutrino reconstruction degrades the precision: the Gaussian part of the $\ttbar$ frame resolution has width $\delta \dy \sim 0.100$ and significant non-Gaussian tails. The $\ttbar$ frame has an advantage in interpretation, but a disadvantage in resolution. 

The frame dependent resolution has to be considered against a possible frame dependence in the size of the asymmetry. In the case of the QCD charge asymmetry, our NLO models (see Table~\ref{tab:nloframes}) suggest that the reconstructed asymmetry is reduced by a factor of $0.6-0.7$ in the transition from the $\ttbar$ to laboratory frame. This roughly balances the resolution difference to give comparable sensitivities to the inclusive QCD asymmetries in the two frames. Asymmetries generated by other processes may produce a different ratio between the two frames, possibly with a $\dy$ or $\mttb$ dependence, and a more precise measurement of the ratio could help to illuminate the underlying physics. We will return to this issue in Sec.~\ref{sec:frames}. 

A summary of the rapidity variable and asymmetry definitions used in this paper is given in Table~\ref{tab:defs}.

\section{Physics Models and Expectations}\label{sec:models}

\begin{figure*}[!ht]
\begin{center}
\includegraphics[height=2.3in, clip]{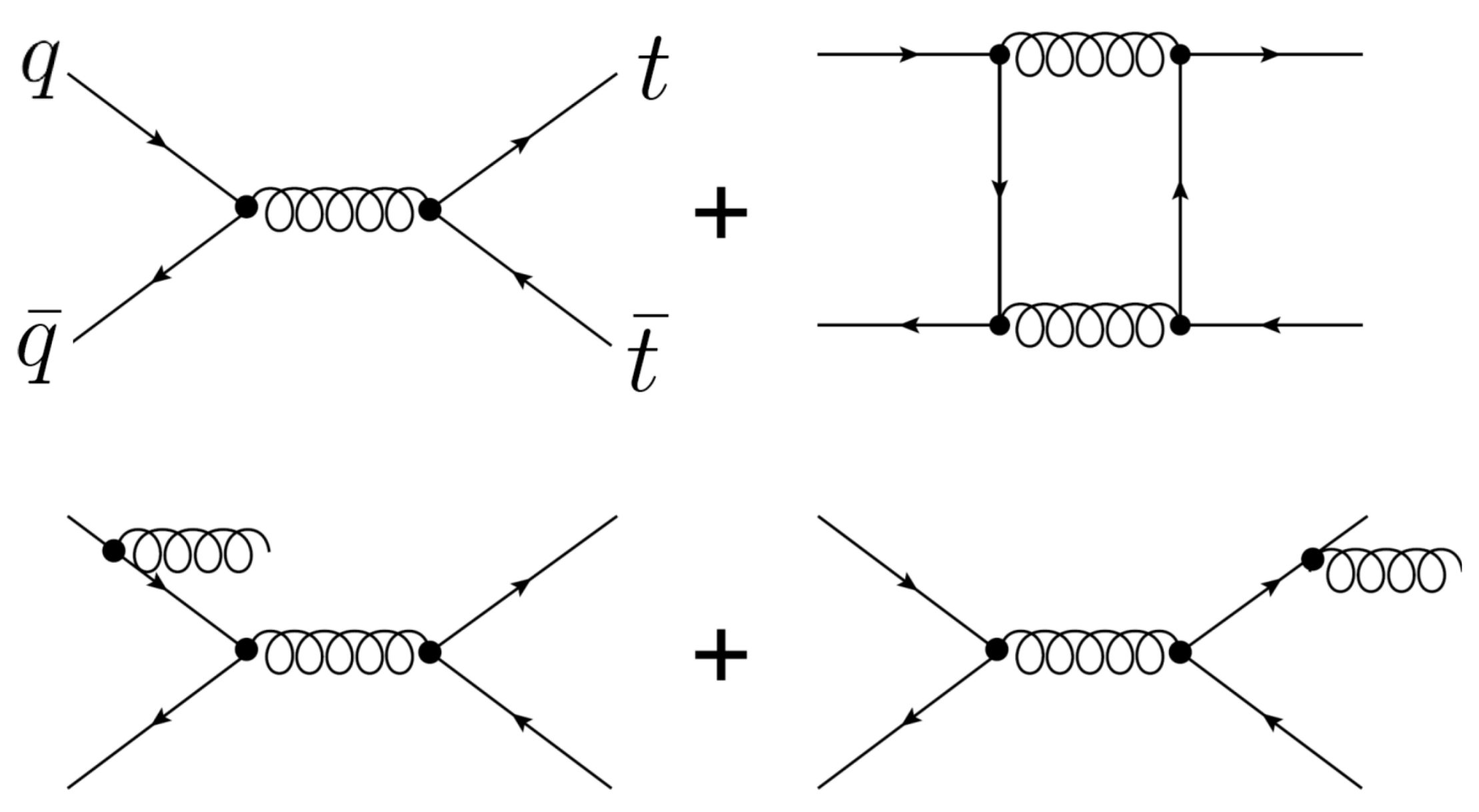}
\caption{{\small Interfering $\qqbar\rightarrow\ttbar$ (above) and $\qqbar\rightarrow\ttbar j$ (below) amplitudes.} \label{fig:nlo}}
\end{center}
\end{figure*}

We briefly describe the theoretical basis for the QCD asymmetry at NLO, the calculation of the theoretical asymmetry using the {\sc MCFM} program~\cite{mcfm}, and use of the {\sc mc@nlo} event generator in creating a simulated NLO sample for input to our analysis. We also describe a simple chiral color-octet model, executed in {\sc madgraph}~\cite{madgraph}, that we use to understand the response of our analysis to a large $\ttbar$ asymmetry.

\subsection{NLO QCD Theory and MCFM}\label{sec:mcfm}

The NLO QCD asymmetry arises in the interference of $\qqbar$ processes that behave differently under charge conjugation. The $gg$ initial state does not contribute to the asymmetry, but does dilute the average value. 

Early, pre-top, treatments of the QCD charge asymmetry discussed measurement of generic heavy quarks in hadron collisions~\cite{prenlotheory}. More recent treatments have focused on the particular case of the top quark at the Tevatron and at the LHC~\cite{almeida,kuhn,nlotheory}.

The asymmetry gets a positive contribution from interference of the tree-level and box diagrams, as in the upper diagrams in Fig.~\ref{fig:nlo} and a negative contribution from the interference of initial and final state radiation in 
$\ttbar$ + jet ($\ttbar j$) final states, as in the lower diagrams. The total inclusive asymmetry is the sum of these opposing contributions. An intuitive picture of the first effect is that the QCD Coulomb field of an incoming light quark repels the $t$ quark to larger rapidities, while attracting the $\bar{t}$ quark to smaller rapidities, thus creating a positive asymmetry at large $\eta$, as defined by the quark direction~\cite{sterman}. The second effect can be pictured in terms of color-flow: when a forward gluon is radiated by the incoming quark, the large acceleration of the color charge biases the top quark towards the backward direction.  

Predictions for the NLO QCD asymmetry are derived using version 5.7 of {\sc mcfm} with CTEQ6.1(NLO)~\cite{cteq} and $\mtop = 172.5 ~\gevcc$. The forward-backward asymmetry in the $\ttbar$ rest frame is found to be $\ad = 0.058\pm 0.009$. In the laboratory frame the top quark rapidities are broadened by the varying boost of the $\ttbar$ system along the beamline, and the asymmetry is diluted to $\al = 0.038\pm 0.006$. Our {\sc mcfm} predictions are in accord with other recent calculations~\cite{almeida, kuhn, nlotheory}. These predictions are for top quarks as they emerge from the $\qqbar$ collision, before any modifications by detector acceptance and resolution. We will call this the parton-level. Based on our own studies of scale dependence in {\sc mcfm} and also the studies in the references above, we assign a $15\%$ relative uncertainty to all NLO {\sc mcfm} predictions.

An NLO calculation for inclusive $t\bar{t}$ production is an LO  calculation for the production of a $t\bar{t}$ + jet 
final state, and thus an LO calculation for the asymmetry in final states containing an extra jet. A new NLO calculation for 
$t\bar{t}j$ production (and thus for the asymmetry) suggests that the negative asymmetry in this final state is greatly reduced from 
leading-order~\cite{nnlo}. This new result for the $t\bar{t}j$ asymmetry can be incorporated into an analysis of the asymmetry for inclusive $t\bar{t}$ 
production only within the context of a full NNLO calculation of $t\bar{t}$ production. Such calculations are underway but are not 
complete. Threshold resummation calculations indicate that the inclusive asymmetry at NNLO should not differ greatly from that predicted at NLO~\cite{almeida,sterman}. In this paper, we compare to the NLO predictions for $t\bar{t}$ production. We include a $15\%$ scale dependence uncertainty, but note that there is an overall unknown 
systematic uncertainty on the theoretical prediction pending the completion of the NNLO calculation.

In the near-threshold form of the cross section~\cite{almeida} the $\ttbar$ frame asymmetry can be seen to increase with the top quark production angle and velocity ($\beta$), and these are thus key variables for understanding the source of the asymmetry. In this analysis, the proxies for these variables are the top quark rapidities and the mass $\mttb$ of the $\ttbar$ system. Measurements of the rapidity and mass dependence of $\ad$ are described in Sections~\ref{sec:avy} and ~\ref{sec:akin}.

\subsection{NLO QCD Simulation with MC@NLO}\label{sec:mcnlo}
We use the event generator {\sc mc@nlo} to create a simulated sample that includes the QCD asymmetry as predicted by the standard model at NLO. In addition to including the asymmetric processes this generator properly estimates the amount of $gg$, and thus the dilution of the asymmetry from these symmetric processes.

Some naming conventions for the data-to-simulation comparison are given in Table~\ref{tab:levels}. All Monte Carlo (MC) generators will have the same conventions: the truth information is the parton level; the pure top signal after simulation, selection, and reconstruction is the $\ttbar$ level, and the full prediction including backgrounds is $\ttbar$ + bkg level. The reconstructed lepton+jets sample is the data. Subtracting the backgrounds from the data yields the reconstructed $\ttbar$ signal-level. Correcting the data for acceptance and resolution produces a measurement at the parton-level.

\begin{table}[!th]
\begin{center}
\caption{Naming conventions for data and simulation samples.} \label{tab:levels}
\begin{tabular}{c c c c}
\hline
\hline
sample   &    level          &  definition                  & comparable to \\
\hline
data     &    data           &  reco l+jets  &  \\
data     &    signal         &  data minus bkg              &  $\ttbar$ in data\\
data     &    parton         &  corrected signal            &  $\ttbar$ at creation \\
\hline
MC       &    $\ttbar$+bkg   &  reco $\ttbar$ + bkg          & data  \\
MC       &    $\ttbar$       &  reco $\ttbar$ no bkg          & data signal  \\   
MC       &    parton         &  truth level                  & data parton  \\
\hline
\hline                
\end{tabular}
\end{center}
\end{table}

\begin{table*}[!th]
\begin{center}
\caption{NLO QCD asymmetries in two frames. Uncertainties includes MC statistics and scale dependence.} \label{tab:nloframes}
\begin{tabular}{l l c c c }
\hline
\hline
model         & level               &  $\al$            &  $\ad$            & $\al/\ad$     \\
\hline
{\sc mcfm}    & parton              & $ 0.038\pm 0.006$  & $ 0.058\pm 0.009 $ & $0.66\pm 0.10$ \\
{\sc mc@nlo}  &  parton             & $ 0.032\pm 0.005$  & $ 0.052\pm 0.008 $ & $0.62\pm 0.09$\\
{\sc mc@nlo}  &  $\ttbar$           & $ 0.018\pm 0.005 $  & $ 0.024\pm 0.005 $  & $0.75\pm 0.11$\\
{\sc mc@nlo}  & $\ttbar$+bkg       & $ 0.001\pm 0.003 $  & $ 0.017\pm 0.004 $  & $0.06\pm 0.01$\\
\hline
\hline                
\end{tabular}
\end{center}
\end{table*}

\begin{figure*}[!t]
\begin{center}
\mbox{
 \hspace*{-0.3in}  
\includegraphics[height=2.8in, clip]{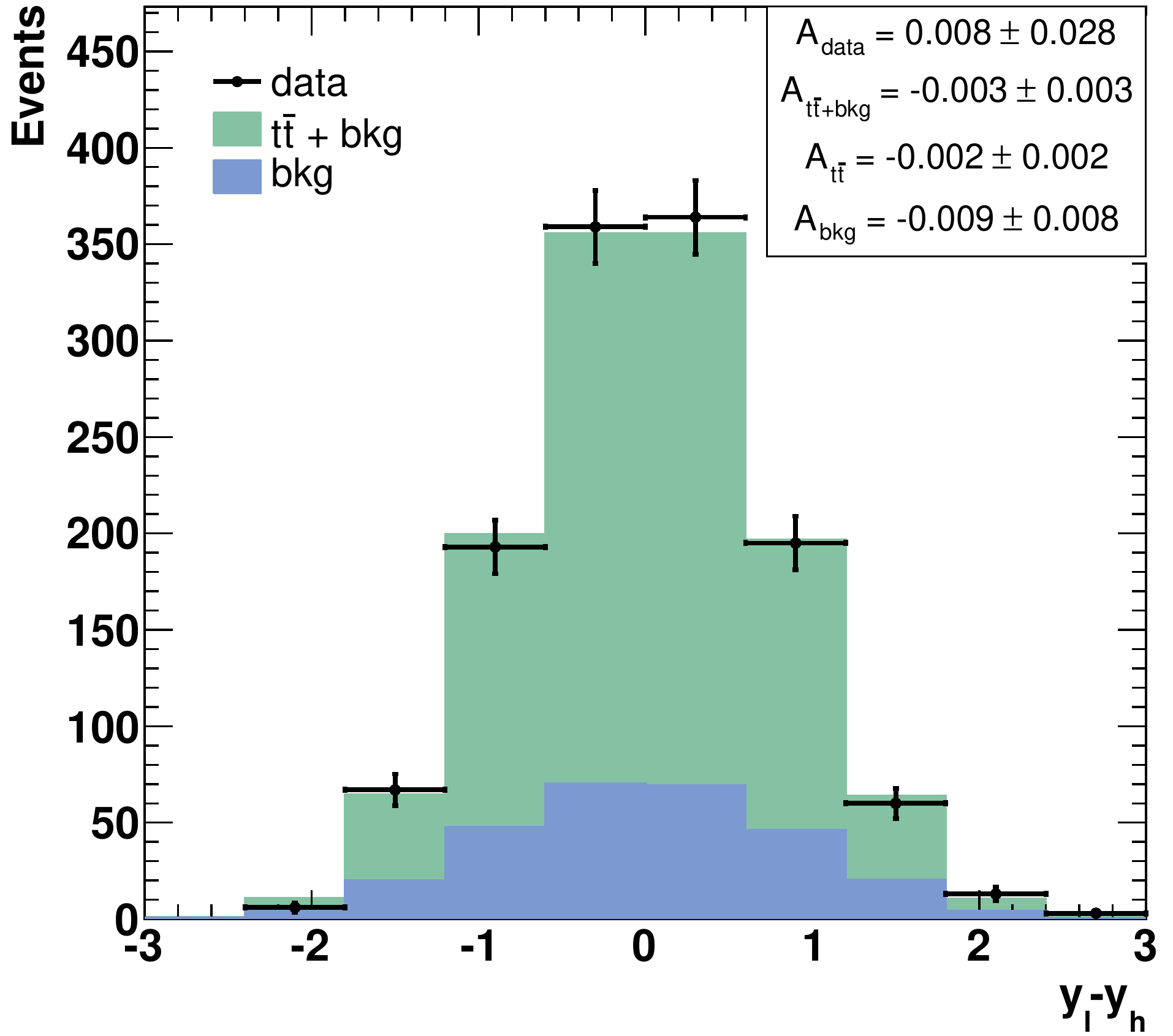}
      \hspace*{-0.1in}\vspace*{0.2in}
  \includegraphics[height=2.8in, clip]{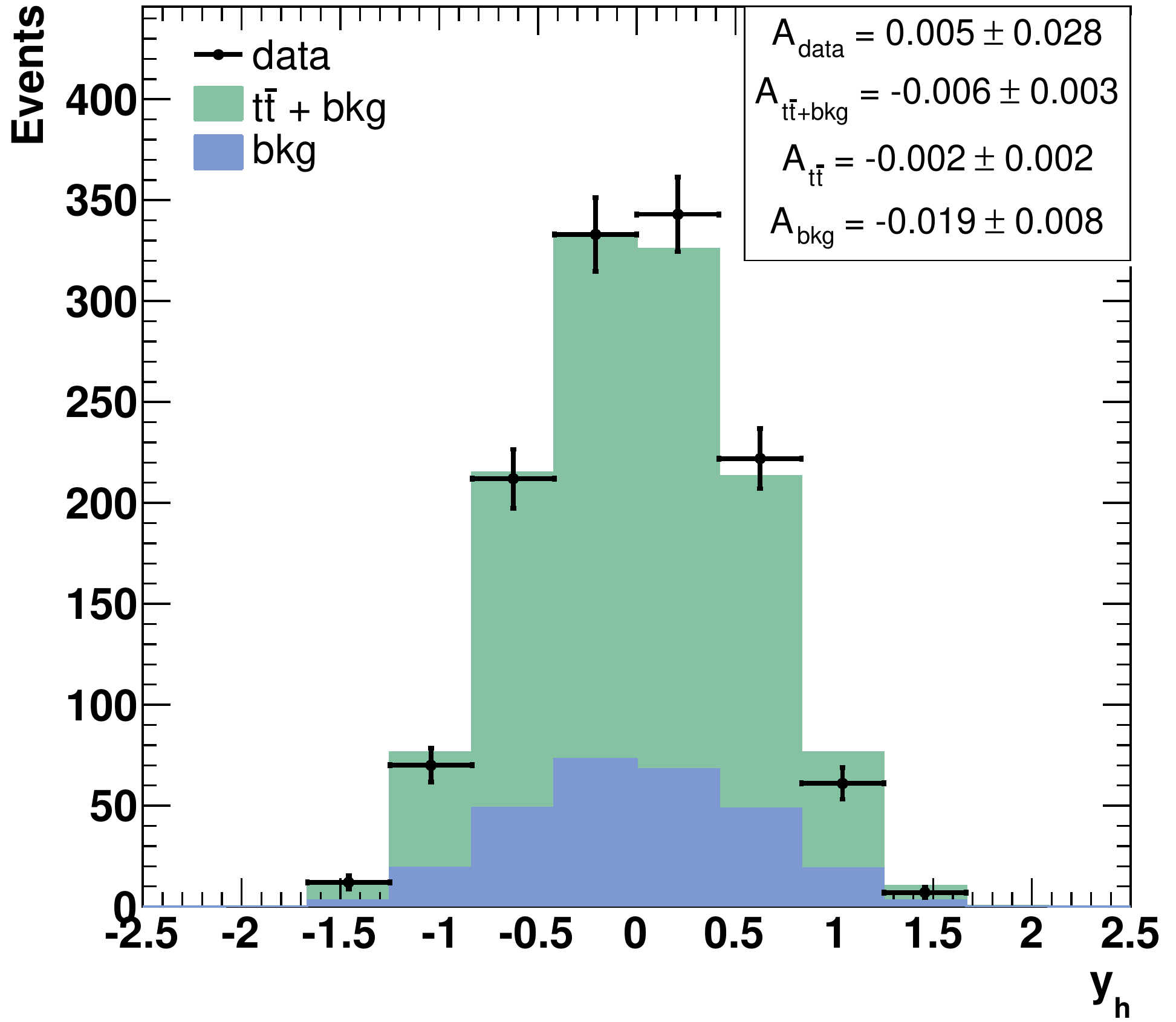}
}
\caption{{\small Rapidity distributions in data compared to predictions. The legend ``$\ttbar$ + bkg'' implies totals in those bins are the sum of the $\ttbar$ and background components. The asymmetries in the data and the predicted $\ttbar$ signal, background, and combination are shown in legends on top right of plots, using the conventions of Table~\ref{tab:levels}.} \label{fig:DYandY}}
\end{center}
\end{figure*}

The {\sc mc@nlo} predictions for the asymmetries at various levels of simulation are shown in Table~\ref{tab:nloframes}. The uncertainties include the Monte Carlo statistics and the NLO theoretical uncertainty. The parton-level {\sc mc@nlo} asymmetries are consistent with {\sc mcfm}, as expected. After CDF detector simulation, event selection, and reconstruction, the asymmetries in the {\sc mc@nlo} $\ttbar$ signal are significantly reduced. In the laboratory frame, the expected asymmetry at the reconstructed $\ttbar$+bkg level is consistent with zero. 

We will see in Sec.~\ref{sec:inclusive} that the statistical error on $\al$ and $\ad$ in the current dataset is $0.028$. Table~\ref{tab:nloframes} shows that, even after background subtraction, the central values of the expected asymmetries are smaller than the experimental resolution. This motivates the continued use of symmetric {\sc pythia} as our default $\ttbar$ model (as discussed in Sec.~\ref{sec:det}), but we will also consider the {\sc mc@nlo} predictions in several specific 
studies.

\subsection{Generic Color-Octet with {\sc madevent}}\label{sec:axig_samples}

It is important that we test our measurement procedures in the regime of the observed asymmetries. We have used {\sc madgraph} and the model of Ref.~\cite{rodrigo} to create asymmetric test samples that can be used as input to our analysis~\cite{tait}. A massive axial color-octet G mixes with the gluon to give a production cross section including pole and interference terms linear in $\cos(\theta^*)$, where $\theta^*$ is the $t$ production angle in the $\ttbar$ rest frame. In these models the asymmetry is an explicit function of the production angle and $\hat{q}^2$, again illustrating the importance of the $\dy$ and $\mttb$ dependence for understanding the source of the asymmetry.

We tuned the octet mass $M_G$ to put the pole out of range and the couplings to give inclusive parton level asymmetries in rough agreement with the data, $\al = 0.110$ and $\ad = 0.157$, while minimizing the effect on the $\ttbar$ cross section and $\mttb$ distribution (see Appendix). After {\sc madgraph} generation, partons are showered with {\sc pythia} and the sample is passed through the complete CDF-II detector simulation. We call this sample OctetA.

\begin{figure*}[!t]
\begin{center}
\mbox{
 \hspace*{-0.3in}  
\includegraphics[height=2.5in,clip]{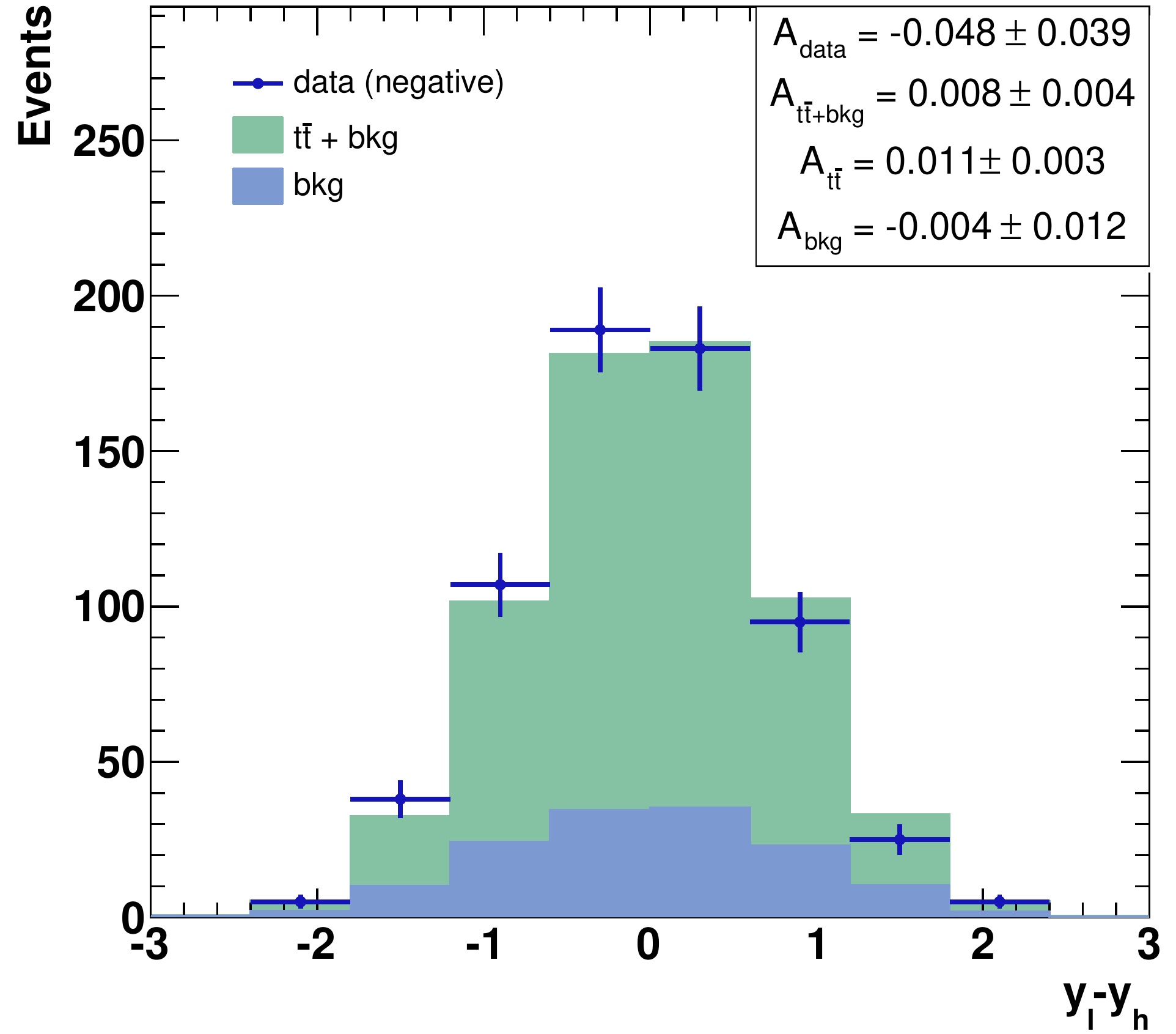}
      \hspace*{-0.1in}\vspace*{0.2in}
  \includegraphics[height=2.5in, clip]{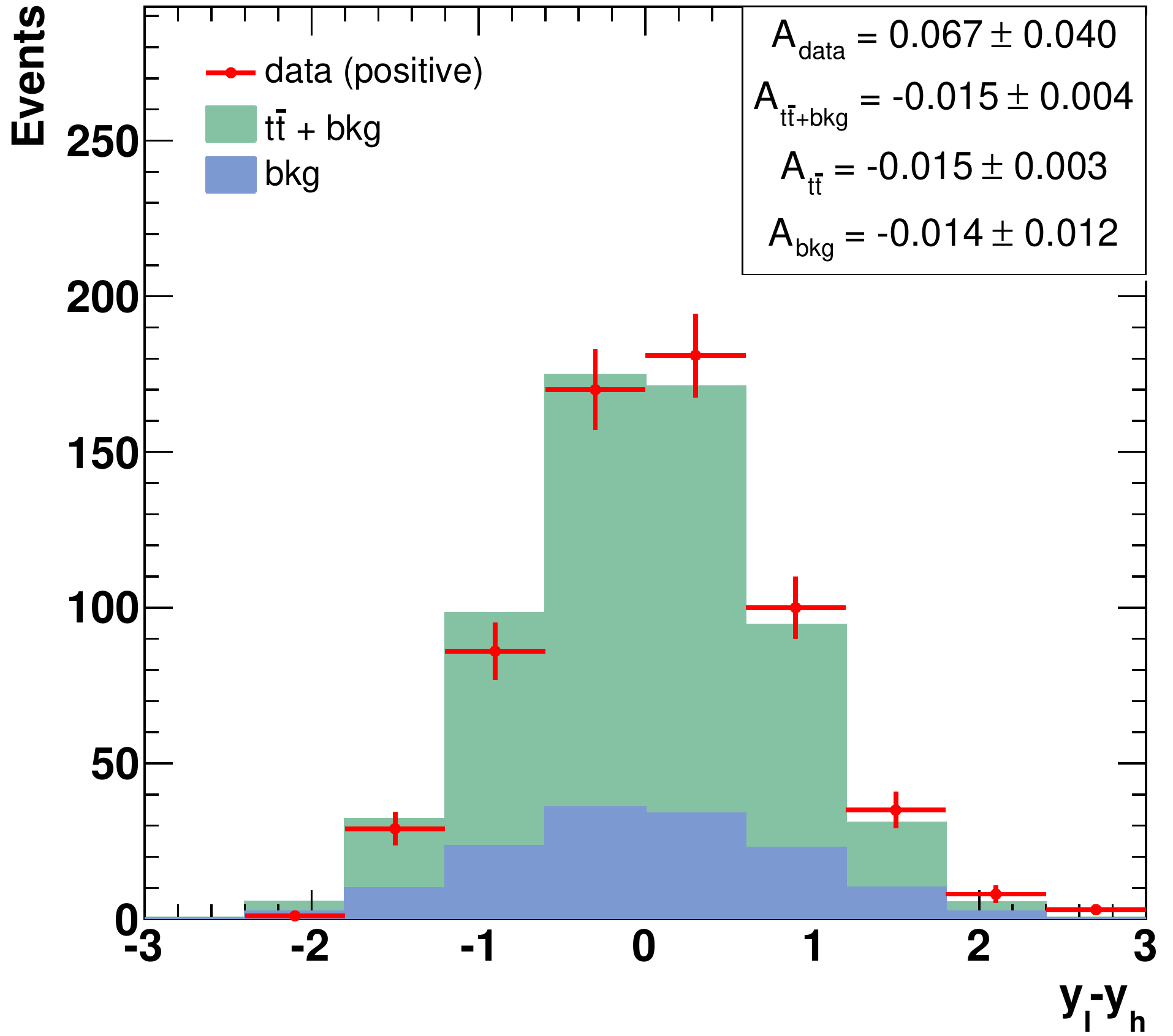}
}
\vspace*{0.15in}
\mbox{
 \hspace*{-0.3in}  
\includegraphics[height=2.5in,clip]{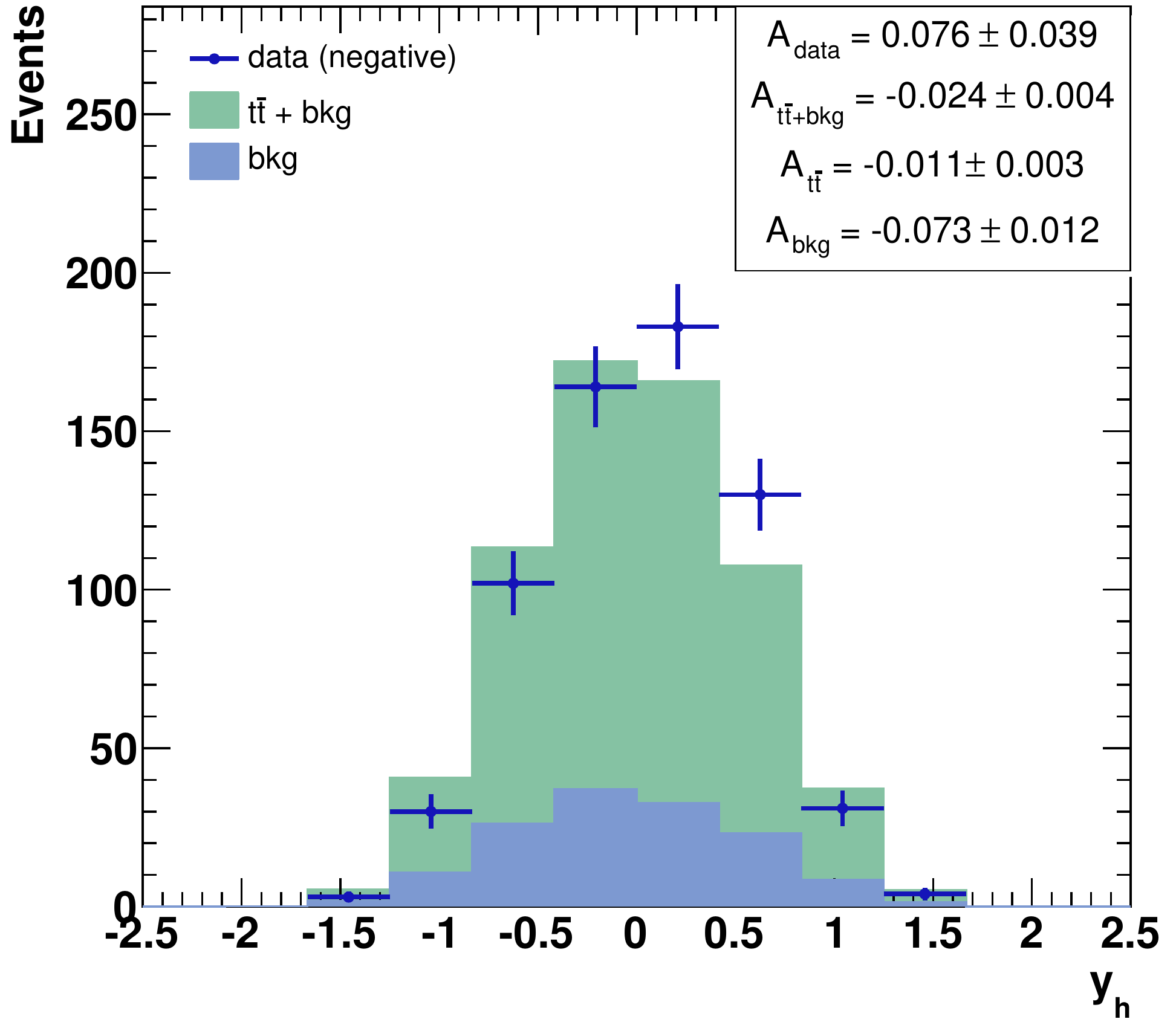}
      \hspace*{-0.1in}\vspace*{0.2in}
  \includegraphics[height=2.5in, clip]{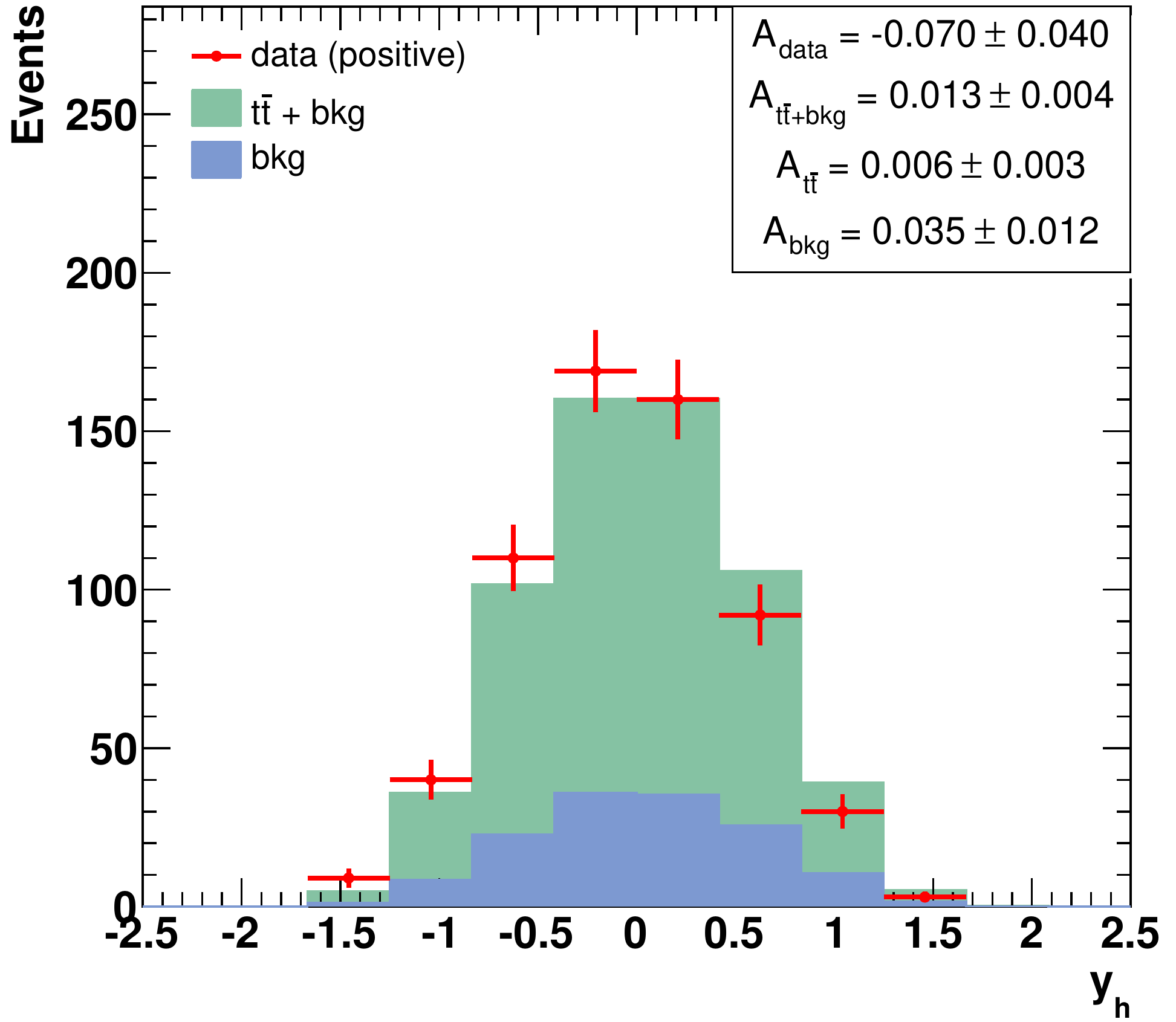}
}

\caption{{\small Distributions of $\dy_{lh}$ (top) and $y_{h}$ (bottom) for events with negative leptons (left) and positive leptons (right).} \label{fig:DYqsep}}
\end{center}

\end{figure*}

A second sample, OctetB, has the same couplings and lower $M_G$, to give inclusive parton level asymmetries $\al = 0.205$ and $\ad = 0.282$, with slight ($\sim 5\%$) increases in the $\ttbar$ cross section and in the high $\mttb$ tail. We consider that OctetA is a better model for understanding the experimental response, but we will appeal to both models in order to span an asymmetry range extending beyond the experimental values. 
 
We emphasize that our use of the Octet models is to study sensitivities and systematic effects in the presence of large asymmetries, and should not be construed as tests of physics hypothesis. More detail on these samples can be found in the Appendix.

\section{Measurement of the Inclusive Asymmetries}\label{sec:inclusive}

We now turn to the rapidity distributions in the data. The inclusive distributions of the $\dy_{lh}$ and $\yh$ variables are shown in Fig.~\ref{fig:DYandY}, compared to the standard {\sc pythia} $\ttbar$ + background prediction. 
These distributions contain the full sample of both lepton signs and should be symmetric. The legend on the top right shows the asymmetries in all components. The data agrees well with $\ttbar$+bkg prediction in both variables, and, in particular, the asymmetries are consistent with zero.

A forward-backward asymmetry becomes apparent when the sample is separated by charge. The top row of Fig.~\ref{fig:DYqsep} shows the $\dy$ distributions for events with negative leptons (left) and positive leptons(right).  We find $A_{lh}^{+}= 0.067\pm 0.040$ and  $A_{lh}^{-} = -0.048\pm 0.039$, where the uncertainties are statistical only. With limited significance, the asymmetries are equal in magnitude and opposite in sign. 

The bottom plots of Fig.~\ref{fig:DYqsep} shows the $\yh$ distributions for events with negative leptons (left) and positive leptons (right). An indication of asymmetry is also observed in this figure: $t$ quarks are dominant in the forward (proton) direction and the $\tbar$ quarks in the backward ($\bar{p}$) direction. The measured asymmetries are $A_{h}^+= -0.070\pm 0.040$ and  $A_{h}^- = 0.076\pm 0.039$, again equal and opposite within uncertainties.

\begin{figure*}[!ht]
\begin{center}
\mbox{
 \hspace*{-0.3in}  
\includegraphics[height=2.65in, clip]{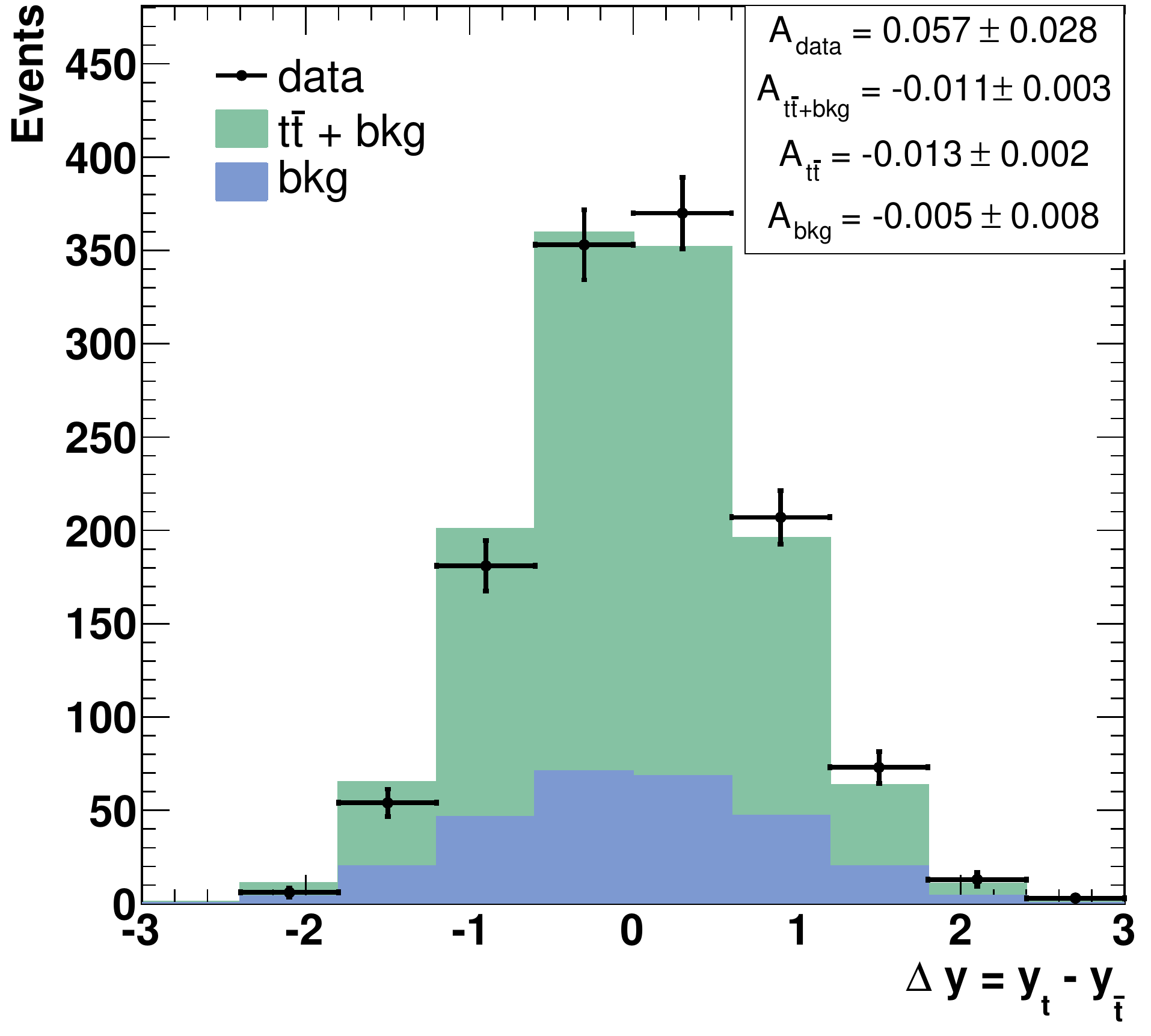}
      \hspace*{-0.1in}\vspace*{0.2in}
  \includegraphics[height=2.65in, clip]{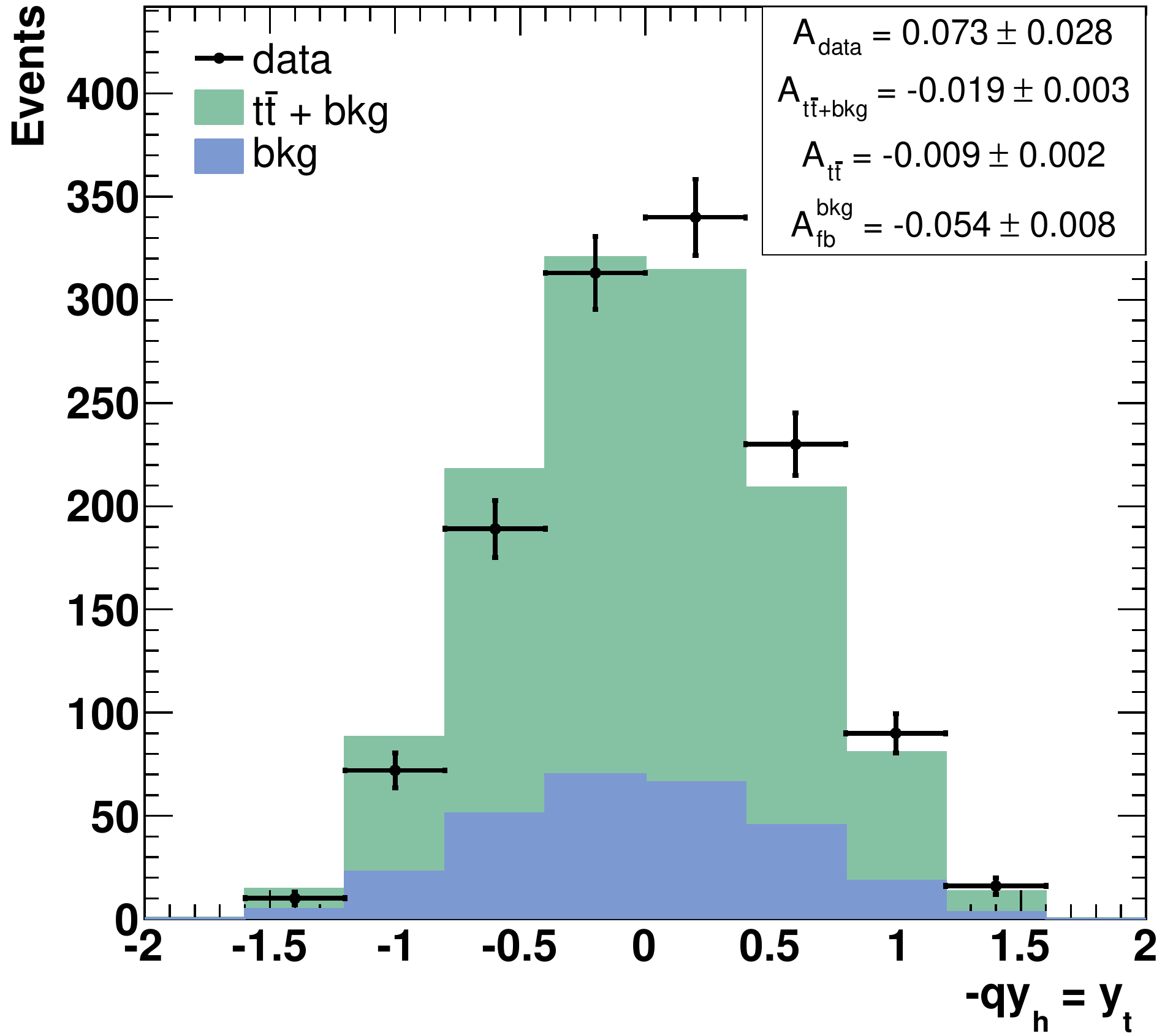}
}
\caption{{\small $\dy$ and $\ytlab$ distributions in data vs prediction.} \label{fig:qDY_validation}}
\end{center}
\end{figure*}

The sign reversal of the asymmetry under interchange of the lepton charge (or, in our formalism, under interchange of $t$ and $\bar{t}$) is consistent with $CP$ conservation. With larger samples and improved precision, the comparison of the charge separated distributions will provide a strict test of $CP$ conservation in $\ttbar$ production. If we assume $CP$ conservation we can calculate the total asymmetry in each frame using Eqs. (1) and (2). The distributions of these variables are shown in Fig.~\ref{fig:qDY_validation}. The asymmetry in the $\ttbar$ frame is $\ad = 0.057 \pm 0.028$, and in the laboratory frame is $\al= 0.073 \pm 0.028$, where both uncertainties are statistical. 

\subsection{The Parton-Level Asymmetry}\label{sec:inc_unfold}

In order to compare our results to theoretical predictions we must correct the data for backgrounds, for incomplete detector acceptance, and for the finite rapidity resolution of the reconstruction. 

We derive the signal level $\ttbar$ distributions by subtracting the expected background from the reconstructed data. This correction is most important in the laboratory frame, where, as shown on the right in Fig.~\ref{fig:qDY_validation}, the backgrounds show a significant negative asymmetry originating in the $W$ production asymmetry in $W$+jets events. 


The reliability of the background model is verified in the subset of the lepton+jets selection that has no $b$-tagged jets. This ``anti-tag'' sample is background enriched, with S:B $\sim 0.3$, and is also fully modeled in our analysis. The predicted $\ttbar$ and lab frame asymmetries in the anti-tag data sample are in excellent agreement with observation, as shown in Table~\ref{tab:inc_anti}.  The absence of asymmetry in this background enriched sample, and the consistency between prediction and observation, suggest that the asymmetry in the $b$-tagged sample is correlated with the $\ttbar$ signal and not the backgrounds. 

\begin{table}[h]
\begin{center}
\caption{Asymmetries in the anti-tag sample of the data and $\ttbar$ + bkg level prediction.} \label{tab:inc_anti}
\begin{tabular}{l c c}
\hline
\hline
selection              &     $\ad$          & $\al$      \\
\hline
anti-tag data          & $0.033\pm 0.018$   & $-0.016\pm 0.018$ \\ 
anti-tag prediction          & $0.010\pm 0.007$   & $-0.023\pm 0.007$ \\
\hline
\hline                
\end{tabular}
\end{center}
\end{table}

\begin{figure*}[!hpt]
\begin{center}

\mbox{
\hspace*{-0.3in}
\includegraphics[height=2.7in, clip]{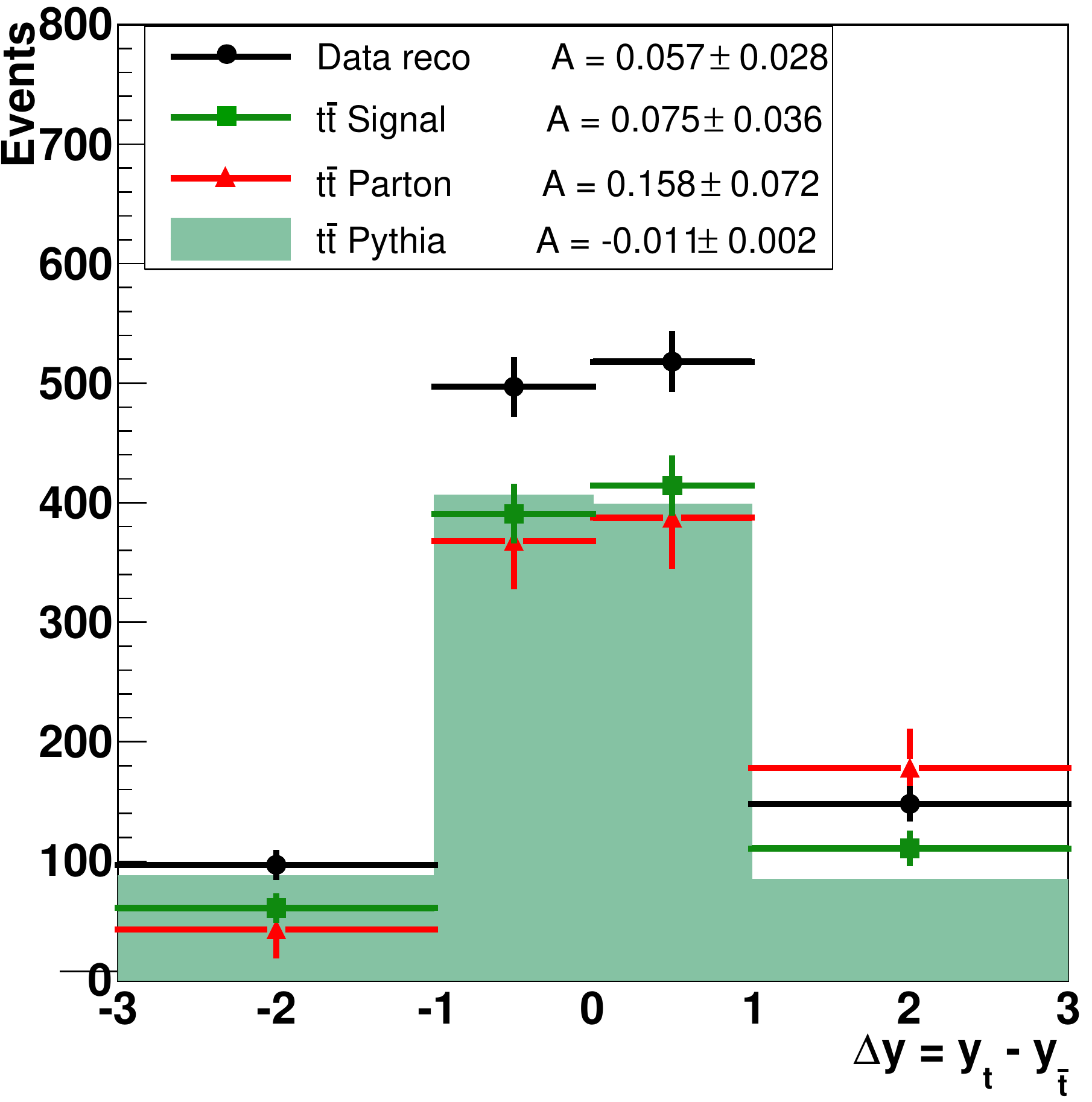}
      \hspace*{-0.1in}\vspace*{0.3in}
  \includegraphics[height=2.7in, clip]{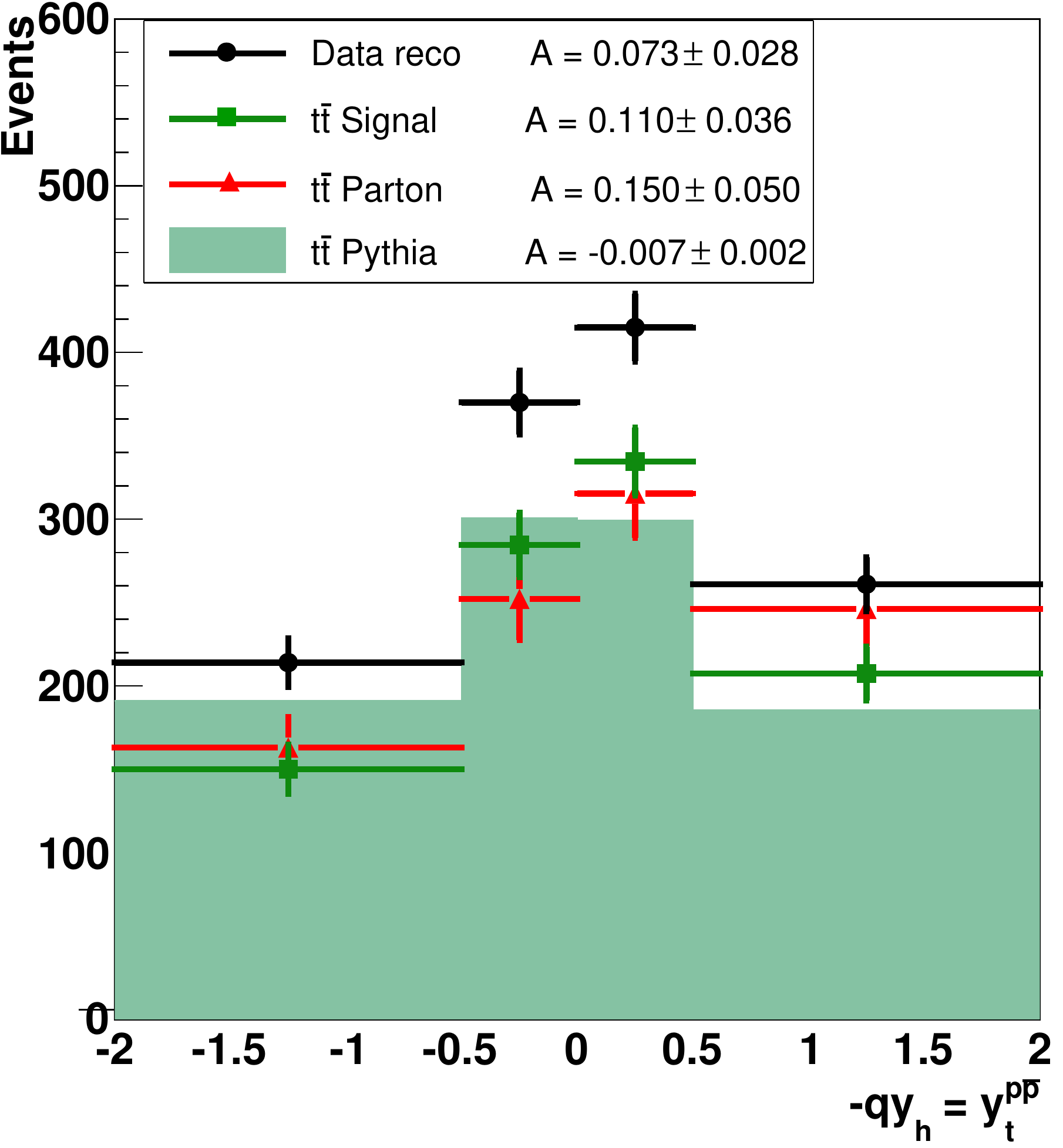}
}
\caption{{\small Four-bin representation of rapidity distributions for all correction levels. Solid histogram is the {\sc pythia} $\ttbar$ model.} \label{fig:unf}}
\end{center}
\end{figure*}

Acceptance and resolution corrections are made with a simple linear unfolding of the $\dy$ and $\ytlab$ distributions using the technique described in Ref.~\cite{afbprlcdf}. Let the binned parton-level rapidity distributions be represented by the vector $\vec{n}$. The $\vec{n}$ distribution is modified by the acceptance and then by the smearing in the reconstruction. These transformations can be expressed as matrices transforming the distribution vector from the parton level to our reconstructed signal: $\vec{n}_{\text{signal}} = \mathbf{S} \mathbf{A} \vec{n}_{\text{parton}}$.

The matrices $\mathbf{A}$ and $\mathbf{S}$ are derived from {\sc pythia} samples by comparing distributions at the Monte Carlo truth level to the same distributions after reconstruction. The acceptance matrix $\mathbf{A}$ is diagonal. The smearing matrix $\mathbf{S}$ measures the bin-to-bin migration arising from the finite resolution of reconstructing the events in the $\ttbar$ hypothesis. To measure the parton-level value, we subtract backgrounds to recover the signal from the data, and then invert the transformation:

\begin{equation}
\vec{n}_{\text{parton}} = \mathbf{A}^{-1} \mathbf{S}^{-1} ( \vec{n}_{\text{data}} - \vec{n}_{\text{bkg}})
\end{equation}

\noindent With the assumption of the $\mathbf{A}$ and $\mathbf{S}$ response as computed with {\sc pythia}, this technique gives a model independent measure of the parton-level asymmetry. The result was found to be robust and the uncertainty minimized when the distributions are separated into four bins with bin edges at 0.0, and $|\dy|=1.0$ or $|\ytlab|=0.5$ ~\cite{afbprlcdf,schwarz,hirsch,glenn}. 


The measurement is affected by uncertainties in our models for the amount and shape of the backgrounds, the amount of initial state and final state radiation (ISR and FSR) in {\sc pythia}, the jet energy scale (JES) of the calorimeter, the parton distribution functions (PDF), and the color reconnection in the final state. These additional systematic uncertainties are studied by repeating the analysis with reasonable variations in the model parameters. We also test the result of substituting the other LO generators {\sc herwig} and {\sc alpgen} for {\sc pythia} in the model for the matrix unfold. The effect of these model variations on the parton-level asymmetry is small, as seen in Table~\ref{tab:inc_sys}. 

\begin{table}[!th]
\begin{small}
\begin{center}
\caption{Systematic uncertainties on parton-level asymmetries in both frames.} \label{tab:inc_sys}
\begin{tabular}{l l l}
\hline
\hline
effect                      &     $\delta\al$  & $\delta\ad$      \\
\hline
background magnitude    & $0.015$         & $0.011$         \\
background shape        & $0.014$         & $0.007$         \\
ISR/FSR                 & $0.010$         & $0.001$          \\
JES                     & $0.003$         & $0.007$ \\
PDF                     & $0.005$         & $0.005$ \\
color reconnection      & $0.001$         & $0.004$ \\
LO MC generator         & $0.005$         & $0.005$ \\
\hline
total                   & $0.024$         & $0.017$ \\
\hline
\hline                
\end{tabular}
\end{center}
\end{small}
\end{table}

It is possible that the corrections in the presence of a large asymmetry would differ from the corrections derived from the symmetric {\sc pythia}. We have studied this possibility by applying the {\sc pythia}- based response corrections to the OctetA model, which has an asymmetry like the data and a resemblance to the data in all other respects. We find that the bias in the corrected inclusive asymmetries is small, roughly $0.02$, and we take this as evidence that the technique is essentially robust against perturbations of this kind. Since we have no reason to prefer the prediction of this or any other model, we do not include a modeling uncertainty. Our inclusive results assume the corrections and uncertainties calculated with the standard {\sc pythia} model. 

Fig.~\ref{fig:unf} shows the $\dy$ and $\ytlab$ distributions at all of the correction levels in the four-bin representation. The effect of the background subtraction is clear. The $\ttbar$ signal (squares) derived from the background subtracted data can be directly compared with the {\sc pythia} signal prediction, and continues to show the asymmetries. The corrected distribution at the parton-level (triangles) can also be compared to the symmetric {\sc pythia} prediction.

\begin{table}[!th]
\begin{center}
\caption{Summary of inclusive asymmetries. Uncertainties include statistical, systematic, and theoretical uncertainties.} \label{tab:asym_summary}
\begin{tabular}{ l c c c}
\hline
\hline
sample       & level          &     $\ad$       & $\al$      \\
\hline
data         & data          & $  0.057\pm 0.028$ & $  0.073\pm 0.028$ \\
{\sc mc@nlo} & $\ttbar$+bkg   & $  0.017\pm 0.004$ & $  0.001\pm 0.003$\\
\hline
data         & signal        & $  0.075\pm 0.037$ & $  0.110\pm 0.039$ \\
{\sc mc@nlo} & $\ttbar$    & $  0.024\pm 0.005 $          & $  0.018\pm 0.005$ \\
\hline
data         & parton        & $0.158\pm 0.074$   & $0.150\pm 0.055$ \\
{\sc mcfm}   & parton                & $0.058\pm 0.009$          & $0.038\pm 0.006$ \\
\hline
\hline                
\end{tabular}
\end{center}
\end{table}

Table~\ref{tab:asym_summary} summarizes the measured asymmetries for the different levels of correction. It is interesting that at the data-level in the laboratory frame we compare to a model prediction that is consistent with zero. When the backgrounds are subtracted from the reconstructed data we can calculate the asymmetry for a pure $\ttbar$ sample at the signal level, and compare directly to {\sc mc@nlo} $\ttbar$. The signal uncertainty here includes the uncertainty on the background correction. Correcting for acceptance and reconstruction resolution yields the $\ttbar$ parton-level asymmetry, where the uncertainty includes the effects listed in Table~\ref{tab:inc_sys}. The parton-level asymmetry may be directly compared with the standard model prediction of {\sc mcfm}. 

The experimentally simple laboratory frame asymmetry exceeds the prediction by more than two standard deviations at all correction levels.   The $\ttbar$ frame asymmetries are similar in magnitude to the laboratory frame, but less significant because of the larger uncertainties. The ratio of the parton-level asymmetries in the two frames is $\al/\ad = 0.95\pm 0.41$, where the error is corrected for the expected correlation across frames in the NLO QCD assumption. This measured ratio is consistent with the expected SM NLO value of 0.6, but the uncertainty is large.   

\subsection{Cross-Checks of the Inclusive Asymmetry}\label{sec:inc_cross}

Table~\ref{tab:inc_cross} shows the asymmetries in the data when the sample is separated according to the lepton flavor and the number of $b$-tagged jets in the event. All of our simulated models predict asymmetries that are independent of the lepton type. Within the large errors, the data are consistent with this expectation.

The $b$-tagged sample contains 281 events with two $b$-tags. This double-tag sample is small, but has minimal backgrounds and robust jet-parton assignment. The double-tag sample is a special category of $\ttbar$ decays where both the $b$ and $\bar{b}$ jet have $\abseta 1.0$, but all of our simulation models predict similar asymmetries in single tags and double-tags. In the data the results are consistent across single and double-tags, albeit with reduced agreement in $\al$. We will discuss the double-tag consistency in the laboratory frame in more detail in Sec.~\ref{sec:frames}.
 
\begin{table}[!ht]
\begin{small}
\begin{center}
\caption{Measured asymmetries at the data-level for different lepton and $b$-tag selections.} \label{tab:inc_cross}\begin{tabular}{ l c c }
\hline
\hline
selection               &     $\ad$          & $\al$      \\
\hline
inclusive              & $0.057\pm 0.028$   & $0.073\pm 0.028$ \\
\hline
electrons              & $0.026\pm 0.037$   & $0.053\pm 0.037$ \\
muons                  & $0.105\pm 0.043$   & $0.099\pm 0.043$ \\
\hline
single $b$-tags            & $0.058\pm 0.031$   & $0.095\pm 0.032$ \\
double $b$-tags            & $0.053\pm 0.059$   & $-0.004\pm 0.060$ \\
\hline
\hline                
\end{tabular}
\end{center}
\end{small}
\end{table}

\section{Rapidity Dependence of the Asymmetry in the $\ttbar$ Rest Frame}\label{sec:avy}

In Sec.~\ref{sec:models} we discussed the importance of measuring the rapidity and $\mttb$ dependence of the asymmetry. The correlated dependence on both variables would be most powerful, but, given the modest statistical precision of our current dataset, we begin with separate measurements of each. In this section we show how a $\dy$-dependence may be calculated from the results of Sec.~\ref{sec:inc_unfold}. The $\mttb$-dependence (as well as the correlation of $\mttb$ and $\dy$) will be discussed in the sections following.

In the standard model at NLO the $\ttbar$ frame asymmetry increases linearly with $\dy$, as seen in Fig.~\ref{fig:mcfmy}. The slope is significant, with the asymmetry reaching values of roughly $20\%$ at large $\dy$. 


The $\dy$ dependence of the asymmetry in our binned data can be calculated in each bin i of positive $\dy$ as
\begin{equation}
\daddy = \frac{N(\dy_i)-N(-\dy_i)}{N(\dy_i)+N(-\dy_i)}\\  
\end{equation}

\begin{figure}
\begin{center}
\includegraphics[height=2.7in, clip]{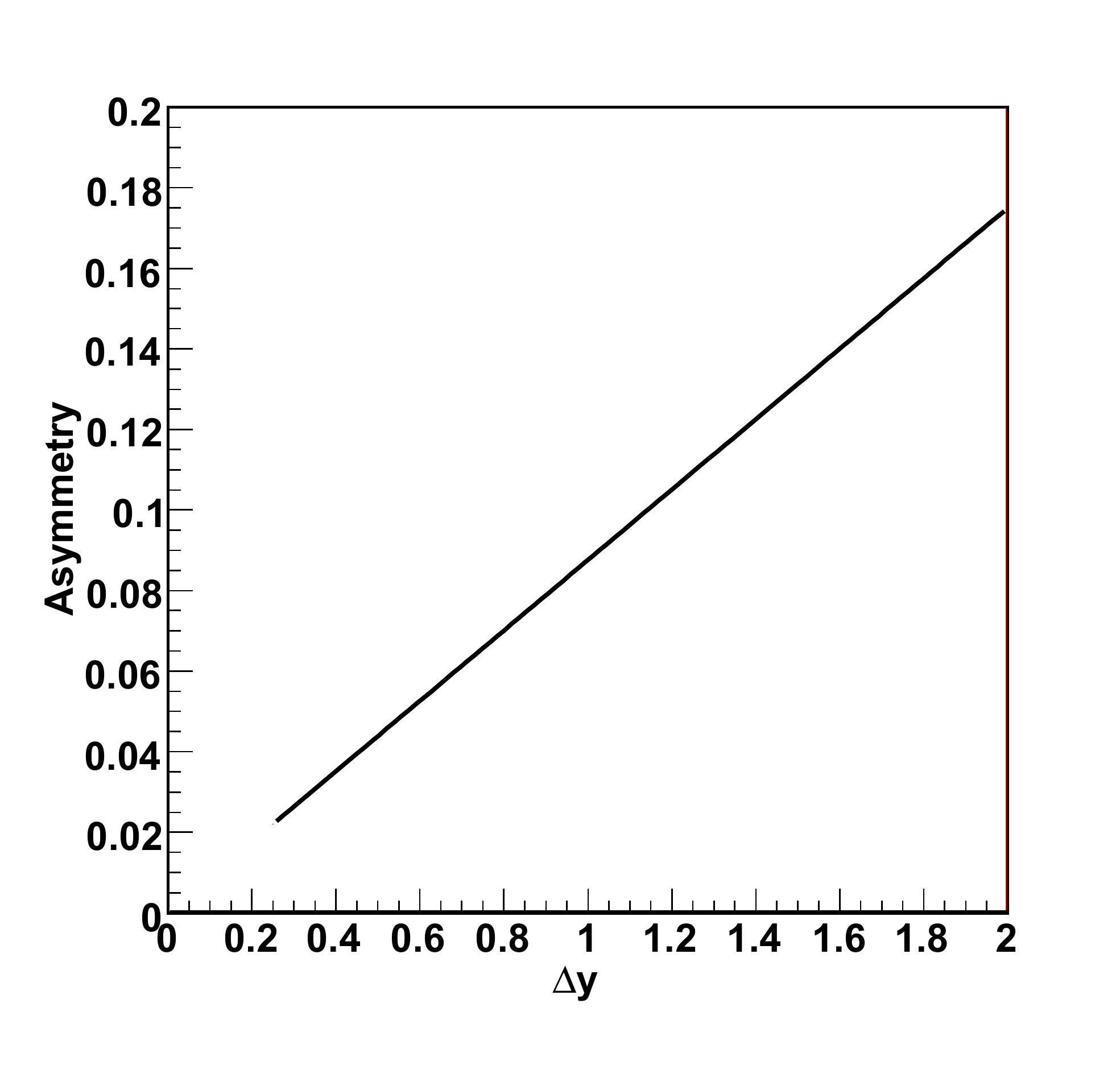}
\caption{{\small $\dy$-dependence of $\ad$ according to {\sc mcfm}.} \label{fig:mcfmy}}
\end{center}
\end{figure}

\begin{figure}[!ht]
\begin{center}
\includegraphics[height=3.0in, clip]{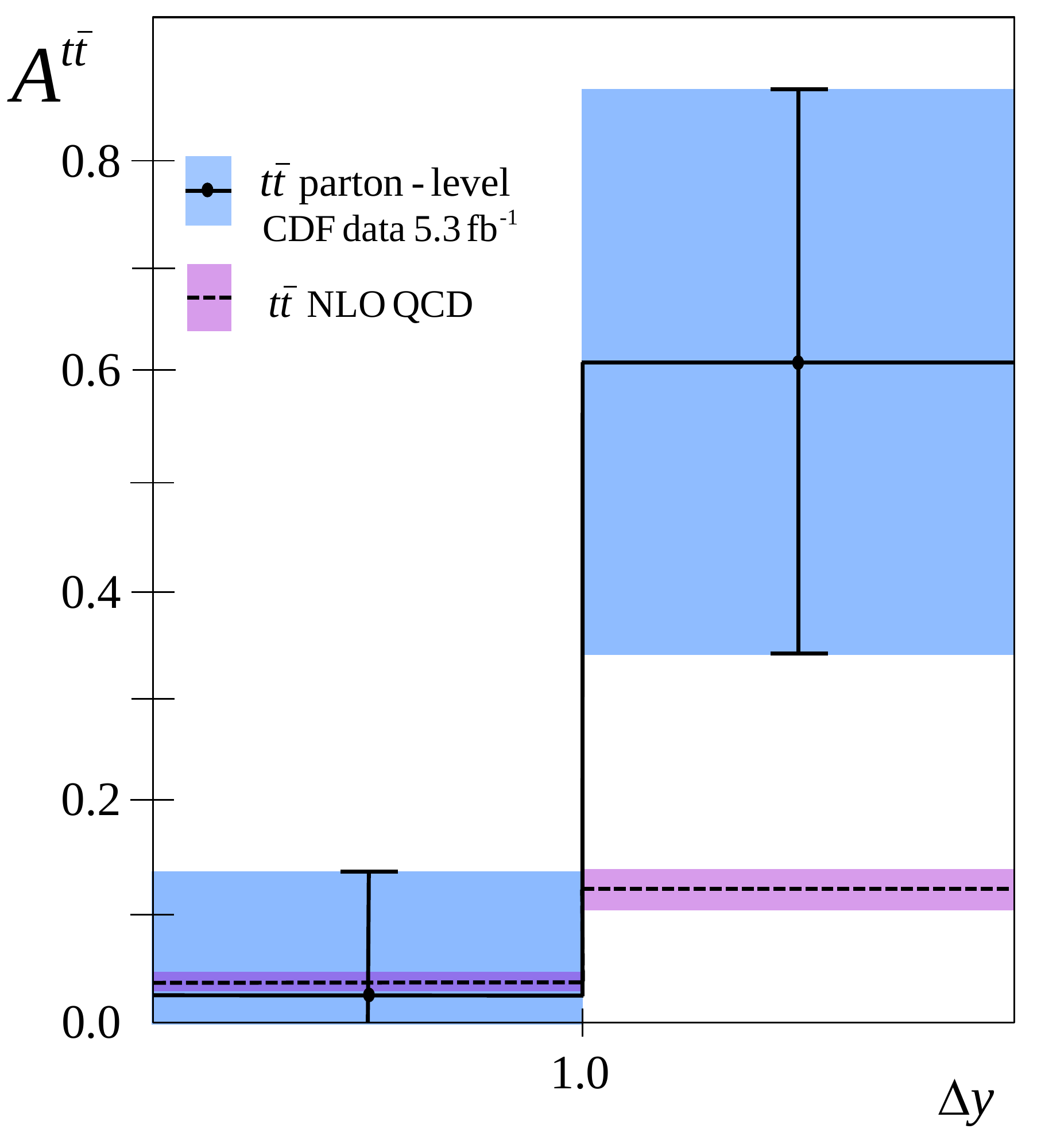}
\caption{{\small Parton level asymmetries at small and large $\dy$ compared to SM prediction of {\sc mcfm}. The shaded bands represent the total uncertainty in each bin. The negative going uncertainty for $\dy < 1.0$ is suppressed.} \label{fig:dy_diff}}
\end{center}
\end{figure}

\begin{table*}[!th]
\begin{center}
\begin{minipage}{3.5in}
\caption{The $\ttbar$ frame asymmetry $\ad$ at small and large rapidity difference, compared to the SM prediction of {\sc mcfm}.} \label{tab:rap_v_Y}
\begin{tabular}{ l l l l }
\hline
\hline
sample     &level     &  $|\dy| < 1.0$            &    $|\dy| \geq 1.0$   \\
\hline
data       &data      & $0.021\pm 0.031$           & $0.208\pm 0.062$   \\
\hline
data       & parton   & $0.026\pm 0.104\pm 0.056$  & $0.611\pm 0.210\pm 0.147$  \\
{\sc mcfm} & parton   & $0.039\pm 0.006$           & $0.123\pm 0.018$  \\
\hline
\hline                
\end{tabular}
\end{minipage}
\end{center}
\end{table*}

A parton-level measurement of $\daddy$ in two bins of high and low $\dy$ is available from the corrected $\dy$ distribution in Fig.~\ref{fig:unf}. We calculate the asymmetry separately for the low rapidity difference inner bin pair $|\dy|<1.0$ and the large rapidity difference outer bin pair $|\dy| \geq 1.0$. The systematic uncertainties in the bin-by-bin comparison are evaluated using the same techniques as in the inclusive measurement. Uncertainty in the background shape and normalization assumptions cause a significant systematic uncertainty in the high $\dy$ bin. 

The $\dy$-dependent asymmetries are shown in Table ~\ref{tab:rap_v_Y}. For the parton-level data, the first uncertainty is statistical and the second is systematic. The uncertainty on the {\sc mcfm} prediction is dominated by the NLO theory uncertainty. For $|\dy|\leq 1.0$, the small data-level asymmetry maps into a small parton-level value with large error. In the large $\dy$ region the parton-level asymmetry is $\ad(|\dy|>1.0)= 0.611\pm 0.270$ (statistical and systematic errors added in quadrature) compared to the {\sc mcfm} prediction of $0.123\pm 0.018$. Fig.~\ref{fig:dy_diff} displays the parton level comparison of asymmetries in data in the two $\dy$ regions. 

\section{ Mass Dependence of the Asymmetry in the $\ttbar$ Rest Frame}\label{sec:akin}

\begin{figure}[!htbp]
\begin{center}
\includegraphics[height=2.7in, clip]{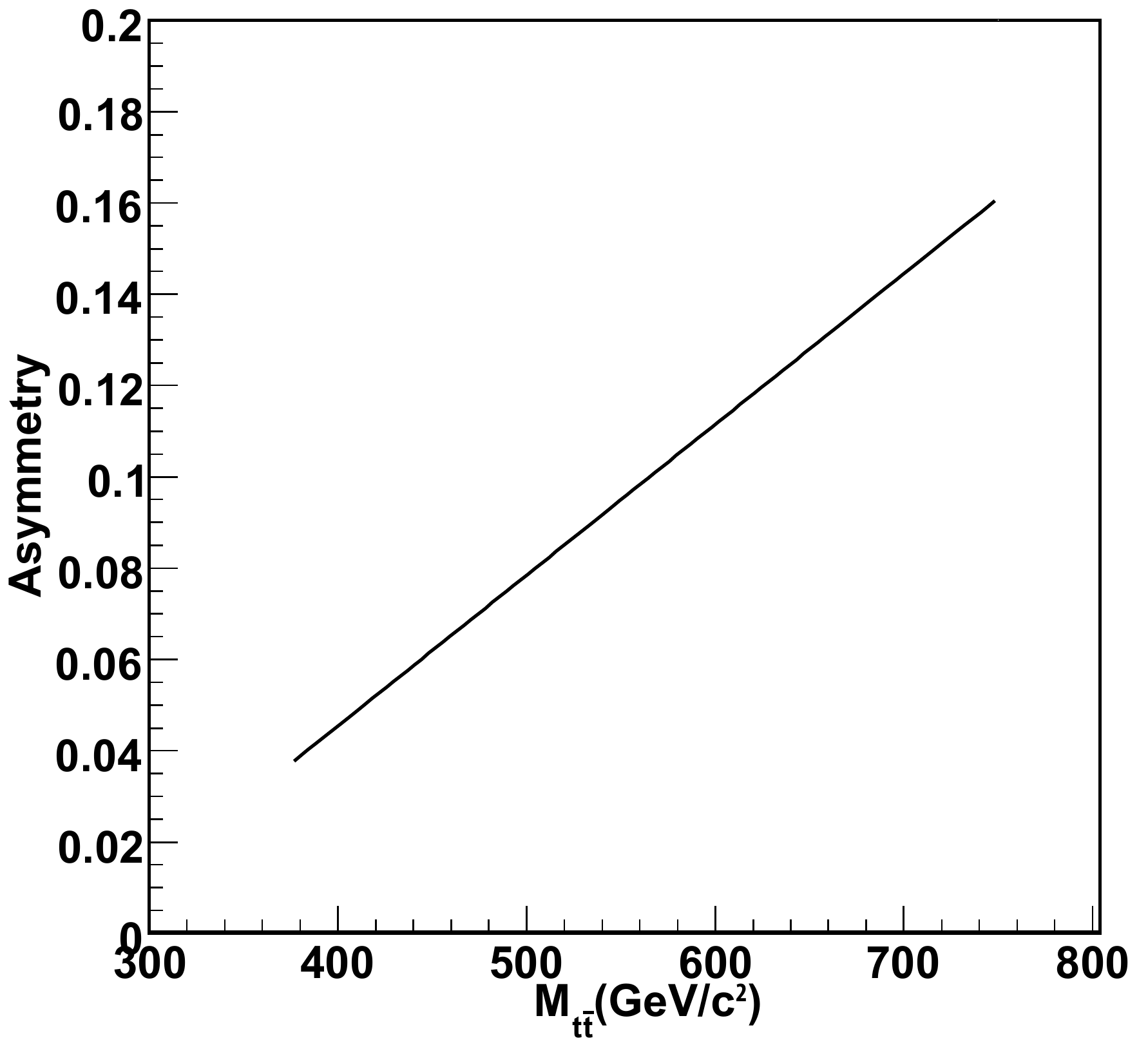}
\caption{{\small $\mttb$-dependence of $\ad$ according to {\sc mcfm}.} \label{fig:mcfm_m}}
\end{center}
\end{figure}

\begin{figure}[!htbp]
\begin{center}
\mbox{
\includegraphics[height=2.0in, clip]{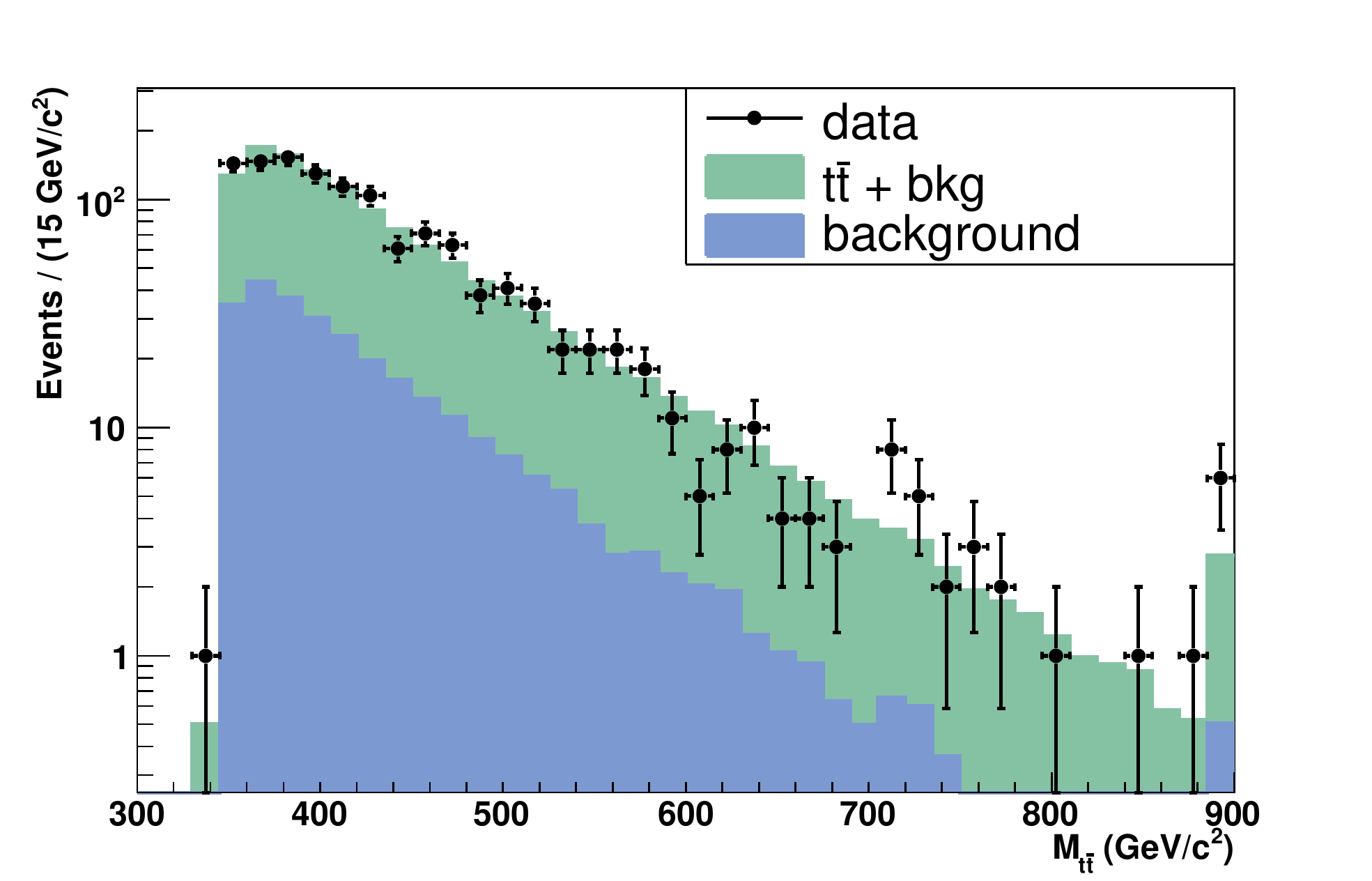}
}
\caption{{\small Event distribution as a function of the total invariant mass $\mttb$.} \label{fig:mttb}}
\end{center}
\end{figure}

\begin{figure*}[!t]
\begin{center}
\mbox{
 \hspace*{-0.3in}  
\includegraphics[height=2.6in, clip]{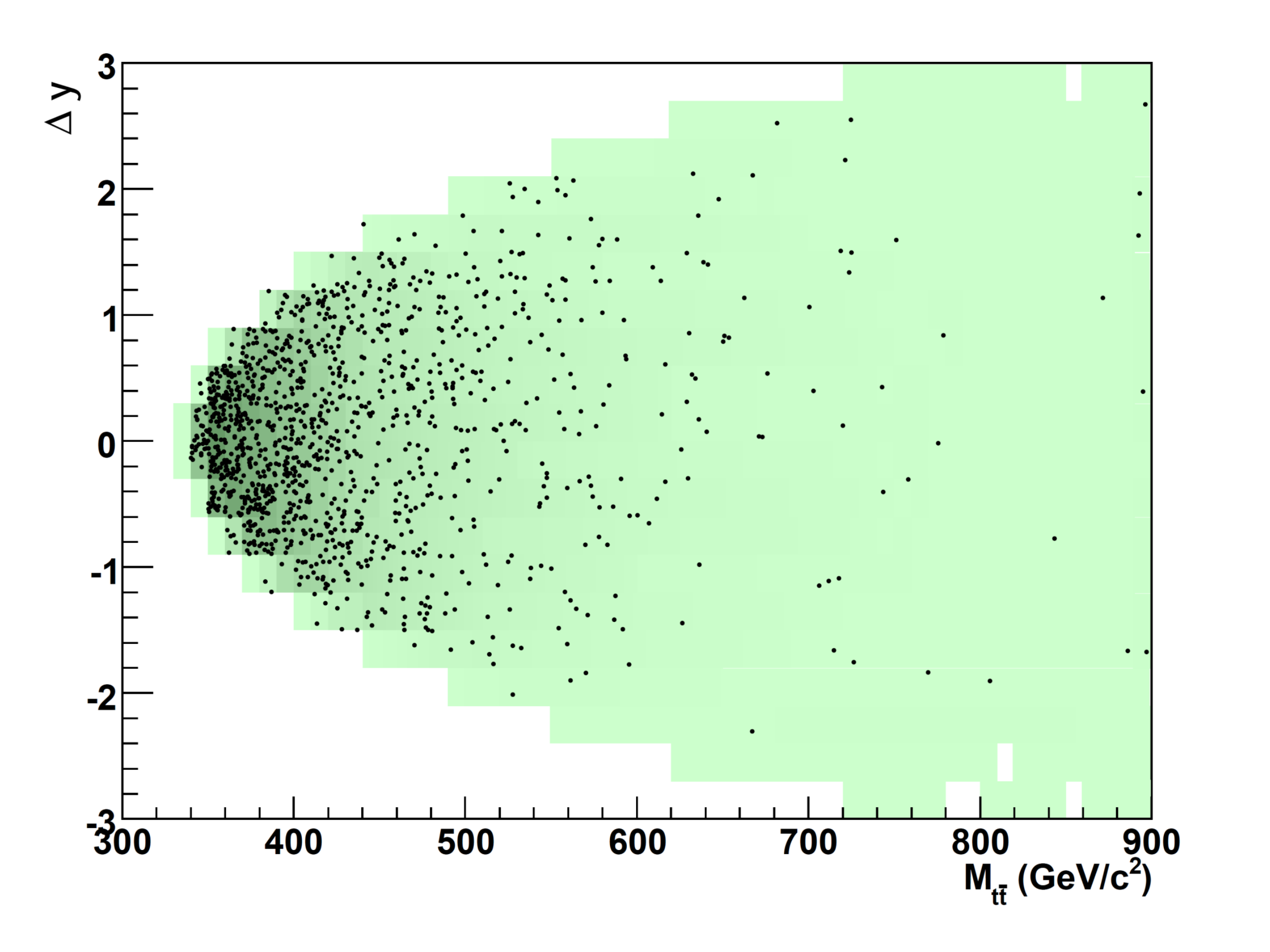}
      \hspace*{-0.1in}\vspace*{0.2in}
  \includegraphics[height=2.6in, clip]{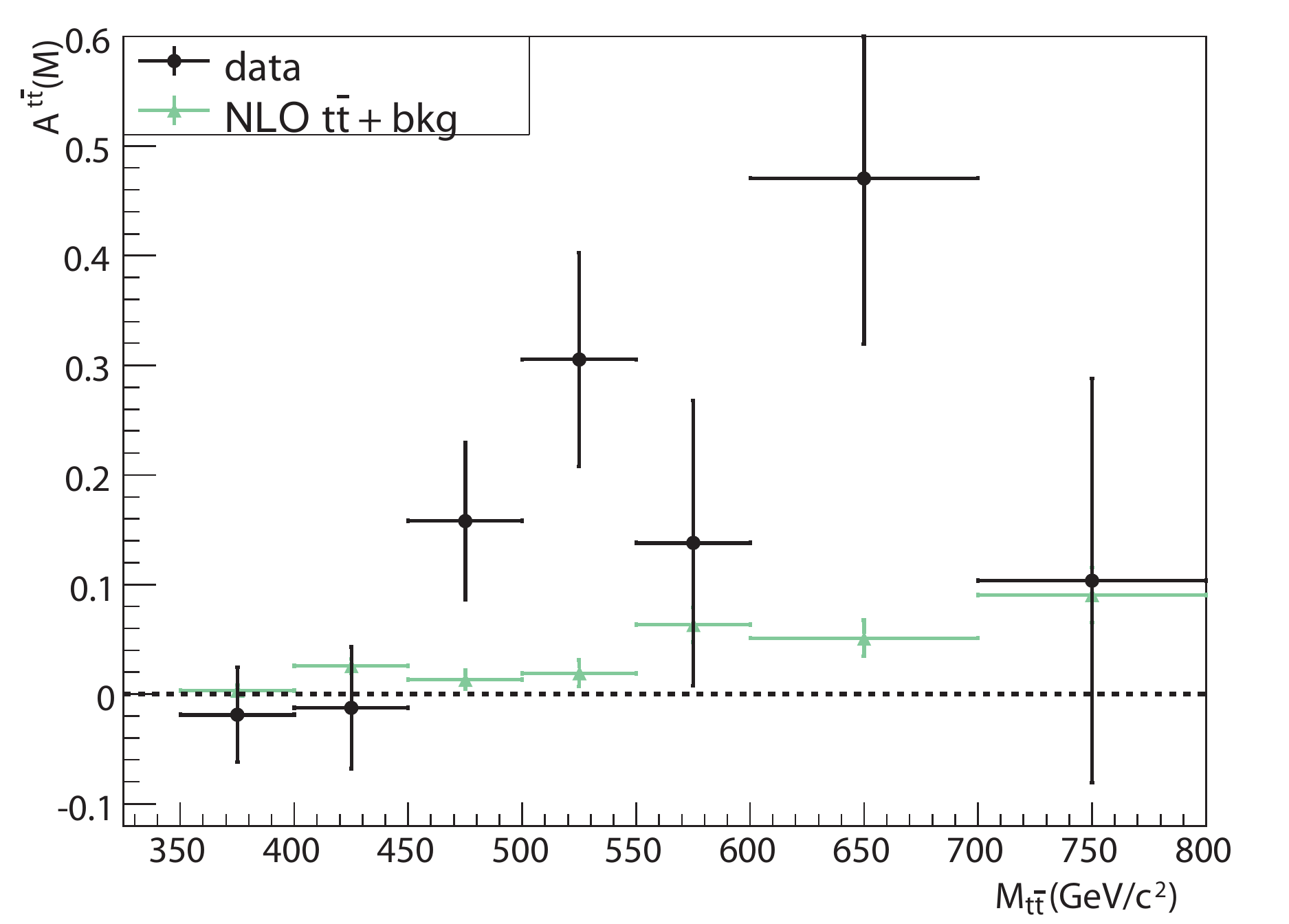}
}
\caption{{\small Left: The $\dy$---$\mttb$ plane. Each dot represents one event, while the intensity of the shading shows approximately the event probability in the standard {\sc pythia} based prediction.Right: The $\ttbar$ frame asymmetry in the data in bins of invariant mass $\mttb$, compared to the prediction of {\sc mc@nlo} $\ttbar$ + backgrounds. The last bin includes all events with 
$\mttb \geq 700 \gevcc$.} \label{fig:AvM2Dandslices}}
\end{center}
\end{figure*}

\begin{figure}[!htbp]
\begin{center}
\includegraphics[width=0.5\textwidth, clip]{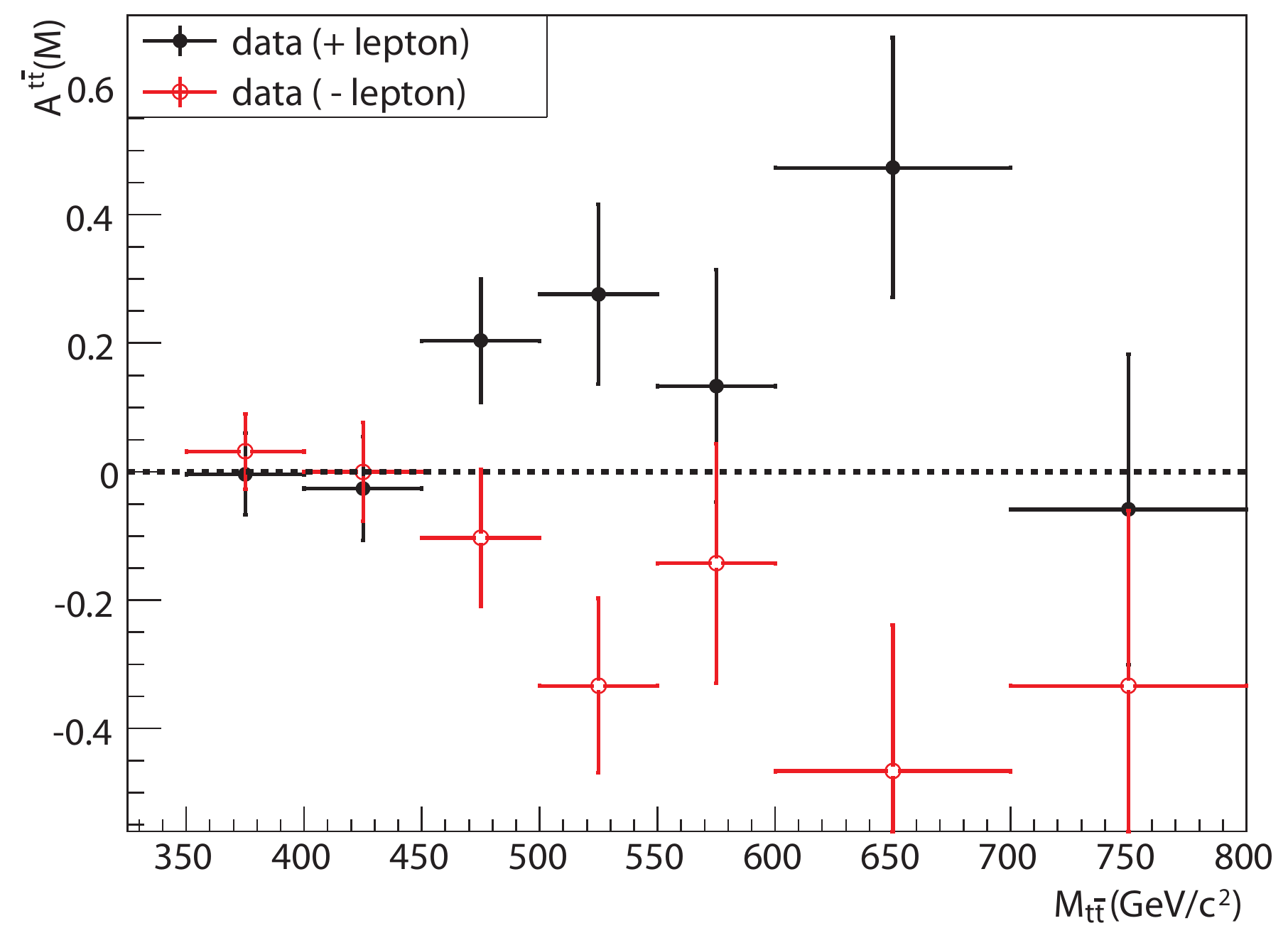}
\caption{{\small  The $\ttbar$ frame asymmetries in bins of invariant mass $\mttb$ when the data is partitioned by lepton charge.} \label{fig:AvMdiffQ}}
\end{center}
\end{figure}

We now turn to the dependence of the asymmetry on the $\ttbar$ invariant mass $\mttb$. The NLO QCD asymmetry also has a strong $\mttb$ dependence, as shown in Fig.~\ref{fig:mcfm_m}. We generally expect the $\mttb$ dependence to contain characteristic information on the fundamental asymmetry mechanism.
 
In this analysis, the value of $\mttb$ is derived from the same reconstruction used to compute the top quark rapidities. The $\mttb$ distribution in our sample, shown in Fig.~\ref{fig:mttb}, is agreement with the standard prediction. Other recent studies of the top pair mass spectrum, including the parton-level differential cross section $d\sigma/d\mttb$, show good agreement with the standard model ~\cite{alice, dlm, D0_tail}. 


Since the mass dependent behavior is usually described in the $\ttbar$ rest frame we focus on the asymmetry in rapidity difference $\dy$ as a function of $\mttb$. The laboratory frame asymmetry derived with $\yh$ is discussed in Sec.~\ref{sec:avm_cross}.

The underlying 2-dimensional distribution of $\dy$ vs.$\mttb$ is shown on the left in Fig.~\ref{fig:AvM2Dandslices}. We expect these variables to obey the simple kinematic relationship $\mttb = 2m_T\cosh(\dy)$, where $m_T$ is the transverse mass of the $\ttbar$ system, and we see this in both the data and the prediction. It is clear that the prior measurement at large $\dy$ captures only part of the region at large $\mttb$. Consequently, the separate measurements of the $\dy$- and $\mttb$-dependence of the asymmetry provide complementary information.

Because $\cosh(\dy)$ is symmetric, this kinematic correlation is independent of the $\mttb$-dependence of any asymmetry in $\dy$. Because of the independence of $m_T$,the measurement at large $|\dy|>1.0$ captures only part of the region at large $\mttb$. The separate measurements  therefore provide complementary information. 

A mass dependent asymmetry $\daddm$ is found by dividing the $\dy$---$\mttb$ plane into bins of mass $\mttbi$ and calculating the asymmetry in each:

\begin{equation}
\daddm= \frac{N(\dy>0,\mttbi - N(\dy<0, \mttbi )}{N(\dy>0,\mttbi)+N(\dy<0, \mttbi)}\\ 
\end{equation}

We use $50 ~\gevcc$ bins of $\mttb$ below $600 ~\gevcc$, and $100 ~\gevcc$ bins above that.  The $\mttb$-dependent asymmetry in $\dy$ is shown on the right in Fig.~\ref{fig:AvM2Dandslices} and Table~\ref{tab:AvMdiff}, compared to the prediction of {\sc mc@nlo} in combination with the standard background. The uncertainties in the plot are the statistical errors only; in the table the {\sc mc@nlo} uncertainty contains both the statistical and theoretical component. In the bulk of the data at low mass the asymmetry is consistent with zero, while at high mass the asymmetry is consistently above the prediction. Fig.~\ref{fig:AvMdiffQ} shows that when the data are separated by lepton charge, the asymmetries in the two independent samples behave in approximately opposite fashion.

\begin{table}
\begin{center}
\caption{The data-level asymmetry $\ad$ in bins of $\mttb$ compared to the prediction of {\sc mc@nlo} + backgrounds.}\label{tab:AvMdiff}
\begin{tabular}{c c c c}
\hline
\hline
bin-center   &         &  \multicolumn{2}{c}{$\ad$ }      \\
($\gevcc$)     &    N events      &        data             &   {\sc mc@nlo}\\
\hline
375          &    532   &   -0.019 $\pm$ 0.043    &  $  0.003\pm 0.006$    \\
425          &    322   &   -0.012 $\pm$ 0.056    &  $  0.026\pm 0.008$   \\
475          &    190   &   0.158 $\pm$ 0.072     & $  0.013\pm 0.010$    \\
525          &     95   &   0.305 $\pm$ 0.097     & $  0.019\pm 0.013$    \\
575          &     58   &   0.138 $\pm$ 0.130     & $  0.063\pm 0.020$    \\
650          &     34   &   0.471 $\pm$ 0.151     & $  0.051\pm 0.020$    \\
750          &     29   &   0.103 $\pm$ 0.185     & $  0.091\pm 0.022$   \\
\hline
\hline                
\end{tabular}
\end{center}
\end{table}

\begin{figure*}[ht]
\begin{center}
\mbox{
\includegraphics[height=2.7in, width=0.49\textwidth, clip]{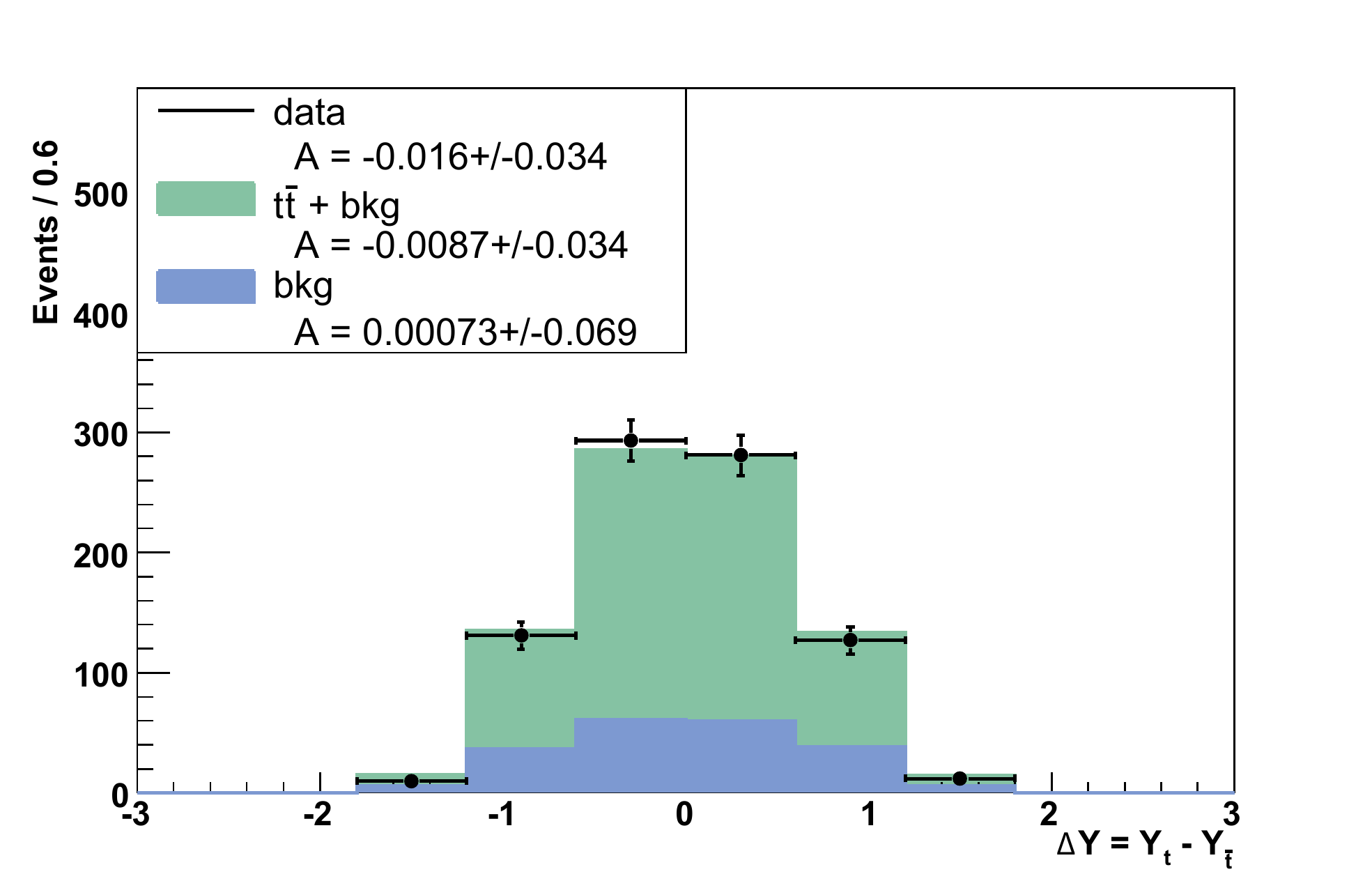}
      \hspace*{-0.1in}\vspace*{0.2in}
  \includegraphics[height=2.7in, width=0.49\textwidth, clip]{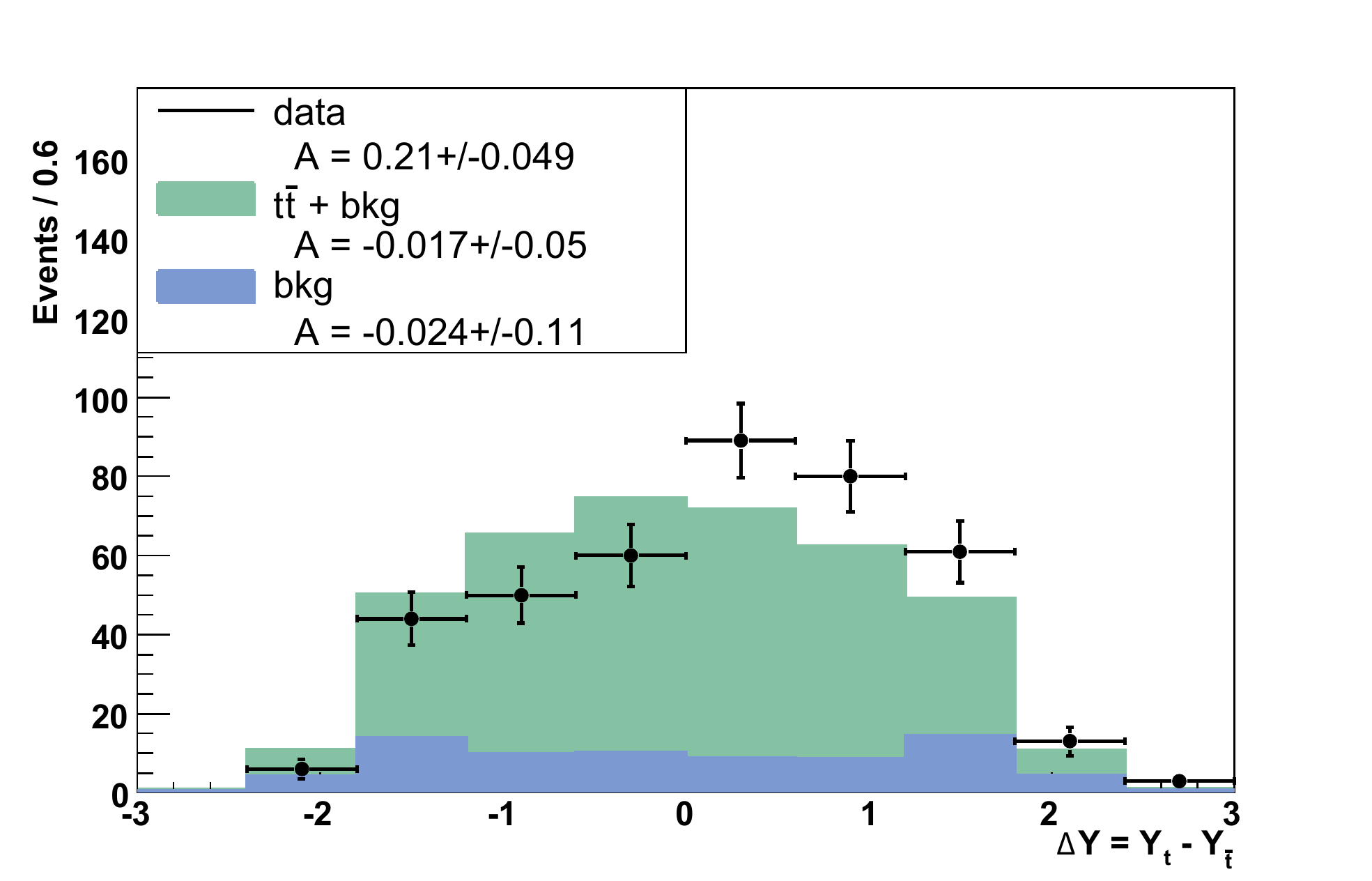}
}
\caption{{\small Top: The distribution of $\dy$ at low mass (left) and high mass (right).} \label{fig:AvM2bins}}
\end{center}
\end{figure*}

\begin{figure*}[ht]
\begin{center}
\mbox{
\includegraphics[height=2.7in, width=0.49\textwidth, clip]{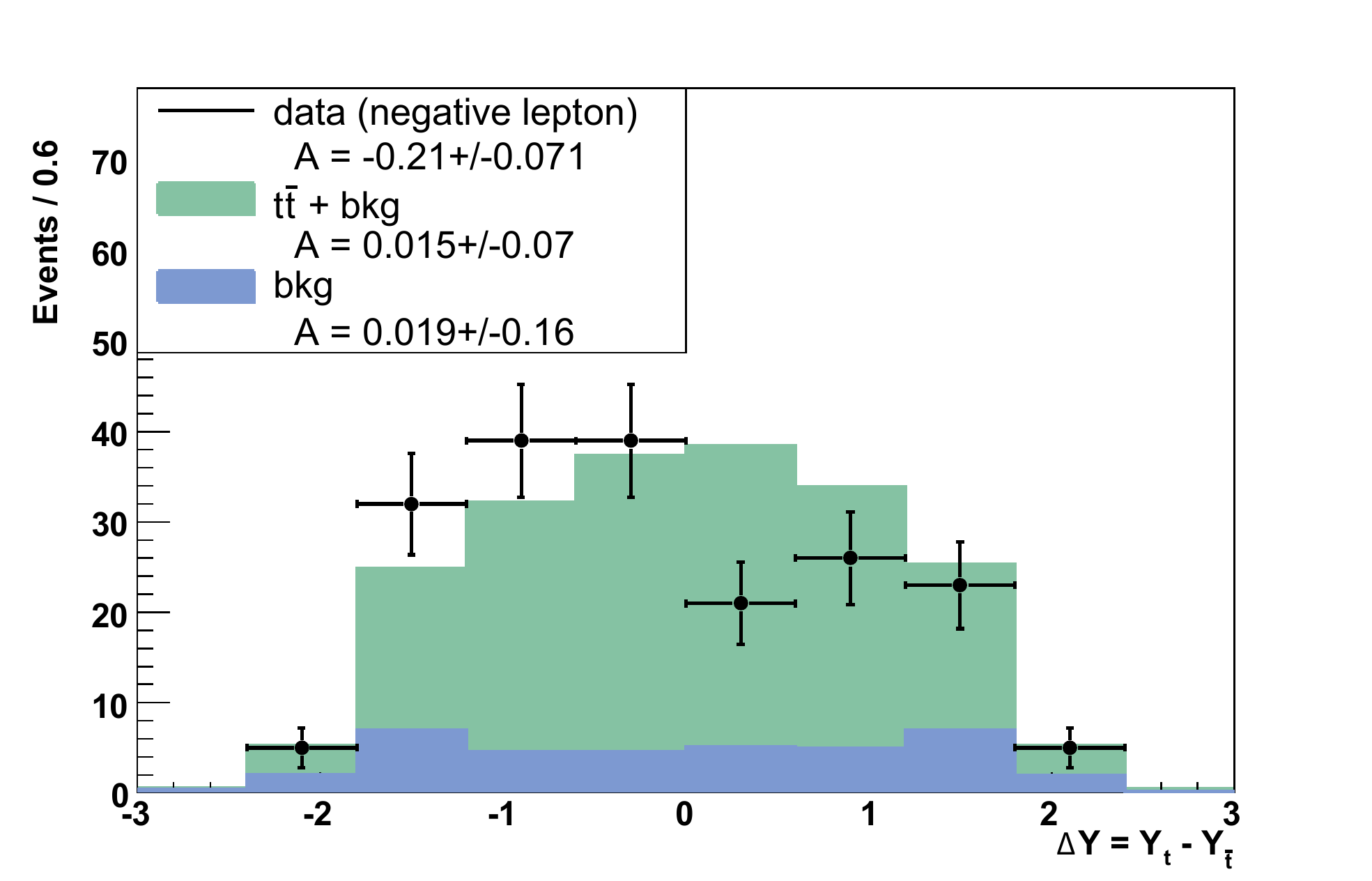}
      \hspace*{-0.1in}\vspace*{0.2in}
  \includegraphics[height=2.7in, width=0.49\textwidth, clip]{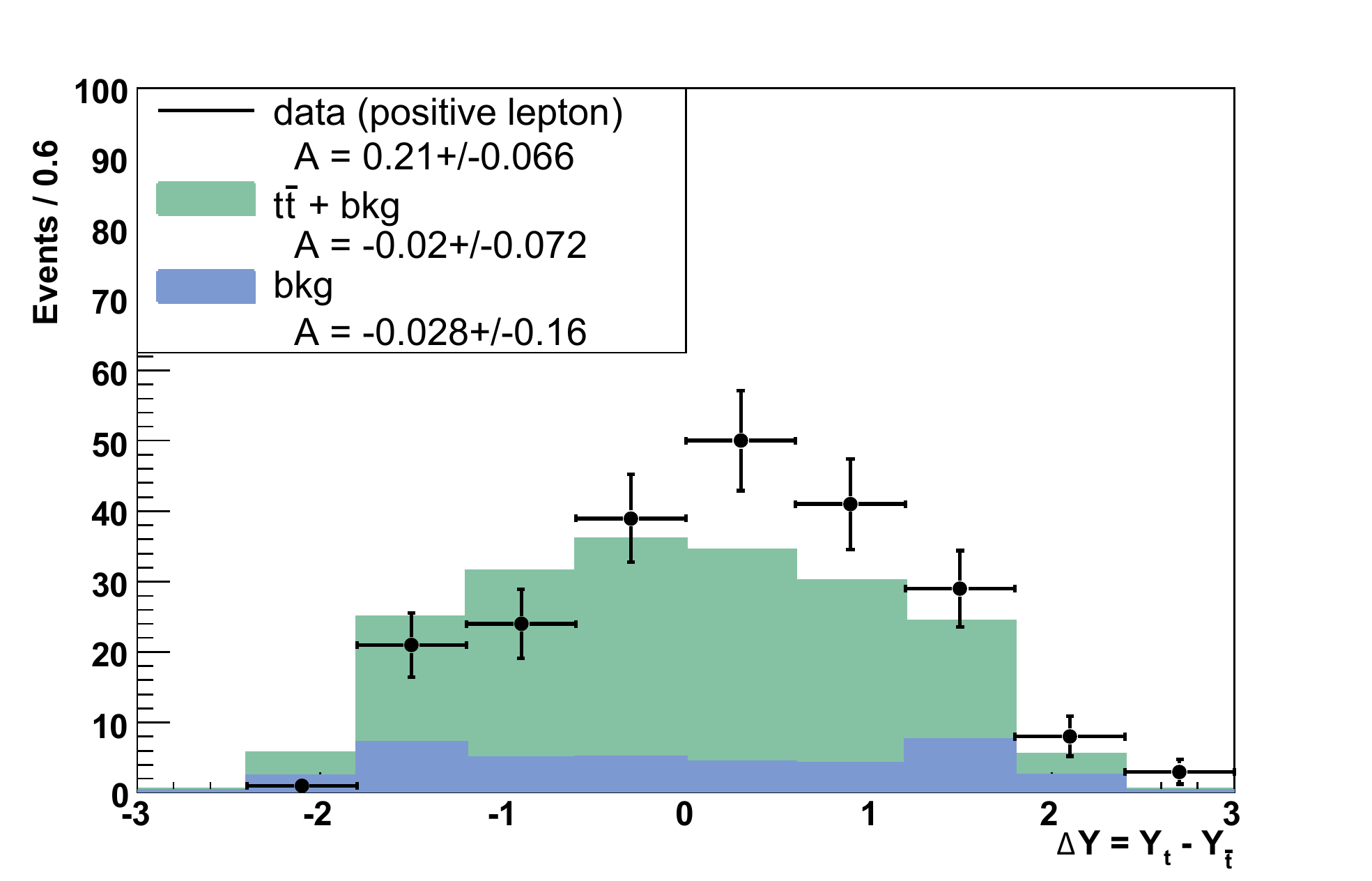}
}
\caption{{\small The distribution of $\dy$ at high mass for events with negative leptons (left) and positive leptons (right).} \label{fig:AvM2binsQ}}
\end{center}
\end{figure*}

\begin{table*}[!htb]
\begin{center}
\caption{Charged and total asymmetries at the data-level, for all, low, and high  $\mttb$.}\label{tab:Ahilowcharges}
\begin{tabular}{c c c c c }
\hline
\hline
selection                       & all $\mttb$       & $\mttb < 450~\gevcc$   & $\mttb\ge 450~\gevcc$  \\
\hline
reco data                        &  0.057$\pm$0.028  & -0.016$\pm$0.034        & 0.210$\pm$0.049 \\
{\sc mc@nlo}                        &  0.017$\pm$0.004  &  0.012$\pm$0.006        & 0.030$\pm$0.007 \\
\hline
$A_{lh}^+$                      &  0.067$\pm$0.040  & -0.013$\pm$0.050        & 0.210$\pm$0.066 \\
$A_{lh}^-$                      & -0.048$\pm$0.039  &  0.020$\pm$0.047        & -0.210$\pm$0.071 \\
\hline
\hline
\end{tabular}
\end{center}
\end{table*}

\subsection{Asymmetries at High and Low Mass}\label{sec:hilowmass}

The large statistical errors in the $\daddm$ distribution of Fig.~\ref{fig:AvM2Dandslices} do not allow any conclusion on the functional dependence. In order to make a quantitative measurement of $A^{\ttbar}(\mttb)$ in a simple, statistically meaningful way, we use a compact representation of $\daddm$ into just two $\mttb$ bins, below and above a given mass boundary. 

The boundary between the low and high mass regions is chosen based on a study of the color-octet samples described in the Appendix. These samples have $\daddm$ distributions that are comparable to the data and reasonable for modeling the sensitivity in that variable. We find that the significance of the asymmetry at high mass is maximized when the bin division is at $\mttb = 450~\gevcc$, and therefore adopt this cut. 

Fig.~\ref{fig:AvM2bins} shows the $\dy$ distributions when the data is divided into two regions, below and above $\mttb = 450~\gevcc$. At low mass the asymmetry is consistent with zero. At high mass, the rapidity difference is 
broader, as expected from the kinematics, and an asymmetry is apparent. The top two lines of Table~\ref{tab:Ahilowcharges} compare the high and low mass asymmetries with the {\sc mc@nlo} prediction. The uncertainty on the prediction combines the statistical and the theoretical uncertainties. At high mass the reconstructed asymmetry $\ad = 0.210 \pm 0.049$ (stat) is more than three standard deviations above the prediction.

\begin{table*}[!th]
\begin{center}
\caption{Tests of the combined mass and rapidity correction procedure. True, reconstructed, and fully corrected asymmetries as found in the two mass regions. Uncertainties on predictions are statistical errors in the MC samples; at truth level these are negligible.} \label{tab:mctest}
\begin{tabular}{l l l l}
\hline
\hline
Sample &  $\ad$ level       & $\mttb < 450~\gevcc$   &$\mttb \ge 450~\gevcc$  \\
\hline
{\sc pythia} & MC truth           & $0.002$          &   $0.001$      \\
       & reconstructed               & $-0.011\pm 0.006$   & $-0.013 \pm 0.008$    \\
       & corrected            & $0.001\pm 0.018$   & $0.006 \pm 0.014$   \\       
\hline
{\sc mc@nlo} & MC truth             &    $0.043$       &  $0.070$               \\
        & reconstructed                 & $0.015\pm 0.006$     &$0.043\pm 0.009$   \\
       & corrected      & $0.066\pm 0.014$       & $0.086\pm 0.011$       \\
\hline
Octet A      & MC truth             & $0.081$             &   $0.276$   \\
             & reconstructed              & $0.024\pm 0.035$     & $0.183\pm0.010$   \\
             & corrected      & $0.054\pm 0.022$   &$0.308\pm0.016$       \\

\hline
Octet B      & MC truth        &  $0.150$           &  $0.466$    \\    
             & reconstructed          & $0.078\pm 0.036$  &  $0.310\pm0.009$    \\
             & corrected    & $0.187\pm 0.024$ &   $0.476\pm0.015$    \\
\hline
\hline                
\end{tabular}
\end{center}
\end{table*}

The high mass $\dy$ distributions for the two separate lepton charges are shown in Fig.~\ref{fig:AvM2binsQ}, and the asymmetries in those distributions are summarized in the bottom part of Table~\ref{tab:Ahilowcharges}. Under the interchange of lepton charge, or, equivalently, under the interchange of  $t$ and $\bar{t}$, the asymmetry at high mass is approximately reversed. This is consistent with $CP$ conservation, and also a strong argument against a false asymmetry arising in event selection or $\ttbar$ reconstruction, as neither the event selection nor reconstruction make reference to the lepton charge.

The results here suggest that the modest inclusive asymmetry in the $\ttbar$ rest frame originates with a large asymmetry in a small population at high $\mttb$. 

\subsection{The Mass Dependent Asymmetry at the Parton-Level}\label{sec:unfold}

In the measurement of the inclusive asymmetry we used a simple matrix technique to correct the rapidity distributions for acceptance and resolution and derive parton-level asymmetries that could be compared with theory. We do this now for the mass dependent asymmetry in the $\ttbar$ frame. We divide the data into two bins in $\dy$, forward and backward, and two bins in mass, above and below 450~$\gevcc$ and re-apply the well tested $4\times4$ unfold machinery of the inclusive analysis. The procedure yields fully corrected, model-independent asymmetries that can be compared with theoretical predictions. 

We represent the four bins of the parton-level distribution of $\dy$ and $\mttb$ by a single vector $\vec{n} = [n_{LF},n_{LB},n_{HF},n_{HB}]$ where, for example, $n_{LF}$ is the number of forward events at low mass. As in the inclusive case, we know that the true $\vec{n}$ distribution is modified by matrices representing the acceptance and then by the smearing in the reconstruction, so that $\vec{n}_{\text{signal}} = \mathbf{S} \mathbf{A} \vec{n}_{\text{parton}}$. To measure the parton-level value, we subtract backgrounds to recover the signal from the data, and then invert the transformation as in Eq. (5). 





As before, the matrices $\mathbf{A}$ and $\mathbf{S}$ are derived from Pythia Monte Carlo samples by comparing truth distributions to the same distributions after reconstruction. The bin-to-bin migration measured in the smearing matrix now includes the cross-terms between high and low mass and forward and backward $\dy$. The most significant migration is caused by mis-reconstructions that underestimate $\mttb$ and smear the shape of the $\mttb$ spectrum towards lower masses. 


\begin{table}[!ht]
\begin{small}
\begin{center}
\caption{Systematic asymmetry uncertainties in the two-mass bin unfold} \label{tab:systematics}
\begin{tabular}{l c c }
\hline
\hline
Source             & $\mttb < 450~\gevcc$   & $\mttb \ge 450~\gevcc$  \\
\hline
background size             & 0.017    &  0.032   \\
background shape            & 0.003    &  0.003  \\
JES                  & 0.005    &  0.012   \\
ISR/FSR              & 0.012    &  0.008    \\
color reconnection     & 0.009    &  0.004    \\
PDF                  & 0.018    &  0.004   \\
physics model        & 0.035    &  0.035   \\
\hline
total                & 0.047    &  0.049   \\
\hline
\hline
\end{tabular}
\end{center}
\end{small}
\end{table}

The accuracy of the procedure is first tested against simulated control samples using {\sc pythia} and {\sc mc@nlo}. 
The {\sc pythia} test uses a $\ttbar$ sample that is independent of the one used to create the response matrices. The top part 
of Table~\ref{tab:mctest} shows that the correction procedure is unbiased when operating on the symmetric {\sc pythia} input. 
The {\sc mc@nlo} sample allows us to study the accuracy of the correction in measuring the NLO QCD effect. 
A small possible bias of $\sim 0.02$ is insignificant compared to the statistical uncertainty in the present data set.  

Next, we use the color-octet samples to test how well the correction derived from symmetric {\sc pythia} can recover large parton-level
asymmetries. The bottom half 
of Table ~\ref{tab:mctest} shows that the correction procedure recovers both the high and low mass 
asymmetries to within a few percent of the true values.  
The corrections in the Octet sample show a possible $\sim 0.02-0.03$ bias that is marginally significant 
compared to the statistical precision of the test. Because the Octet samples match the data well in the two key distributions 
$\dy$ and $\mttb$ (see Appendix) we expect that this is a representative measure of possible model dependence in the correction, and we
assign a systematic uncertainty of $0.035$ for this effect.

Additional systematic uncertainties are evaluated in a manner similar to the inclusive case. These uncertainties are estimated by repeating the analysis while varying the model assumptions within their known uncertainties for background normalization and shape, the amount of initial- and final-state radiation (ISR/FSR) in {\sc pythia}, the calorimeter jet energy scale (JES), the model of final state color connection, and parton distribution functions (PDF). Table~\ref{tab:systematics} shows the expected size of all systematic uncertainties. The physics model dependence dominates. 
 
\begin{table}[!ht]
\begin{small}
\begin{center}
\caption{Asymmetry $\ad$ at high and low mass compared to prediction.} \label{tab:summaryunfold}
\begin{tabular}{ l l l l}
\hline
\hline
 selection                      & $\mttb < 450~\gevcc$ &   & $\mttb \ge 450~\gevcc$ \\
\hline
data               &  $-0.016\pm0.034$  &   & $0.210\pm 0.049$ \\
 $\ttbar$+bkg       &  $+0.012\pm 0.006$          &   & $0.030\pm 0.007$  \\
({\sc mc@nlo})     &                              &   &          \\
\hline
data signal    &  $-0.022\pm0.039\pm 0.017$   &   & $0.266\pm 0.053\pm 0.032$ \\
$\ttbar$       &  $+0.015\pm 0.006$          &   & $0.043\pm 0.009$  \\
({\sc mc@nlo})   &                               & &                   \\
\hline
data parton  &  $-0.116\pm0.146\pm 0.047$  &   & $0.475\pm0.101\pm 0.049$ \\
{\sc mcfm}  &       $+0.040\pm 0.006$           &   & $0.088\pm0.013$   \\
\hline
\hline
\end{tabular}
\end{center}
\end{small}
\end{table}

Table ~\ref{tab:summaryunfold} compares the low and high mass asymmetry to predictions for the data level, the background subtracted signal-level, and the fully corrected parton-level. The MC predictions include the $15\%$ theoretical uncertainty.  At low mass, within uncertainties, the asymmetry at all correction levels agrees with predictions consistent with zero. At high mass, combining statistical and systematic uncertainties in quadrature, the asymmetries at all levels exceed the predictions by more than three standard deviations. The parton-level comparison is summarized in Fig.~\ref{fig:unfold_result}. For $\mttb \geq 450~\gevcc$, the parton-level asymmetry at in the $\ttbar$ rest frame is $\ad = 0.475\pm 0.114$ (stat+sys), compared with the {\sc MCFM} prediction of 
$\ad=0.088\pm 0.013$. 

\begin{figure}
\begin{center}
\includegraphics[height=3.0in, clip]{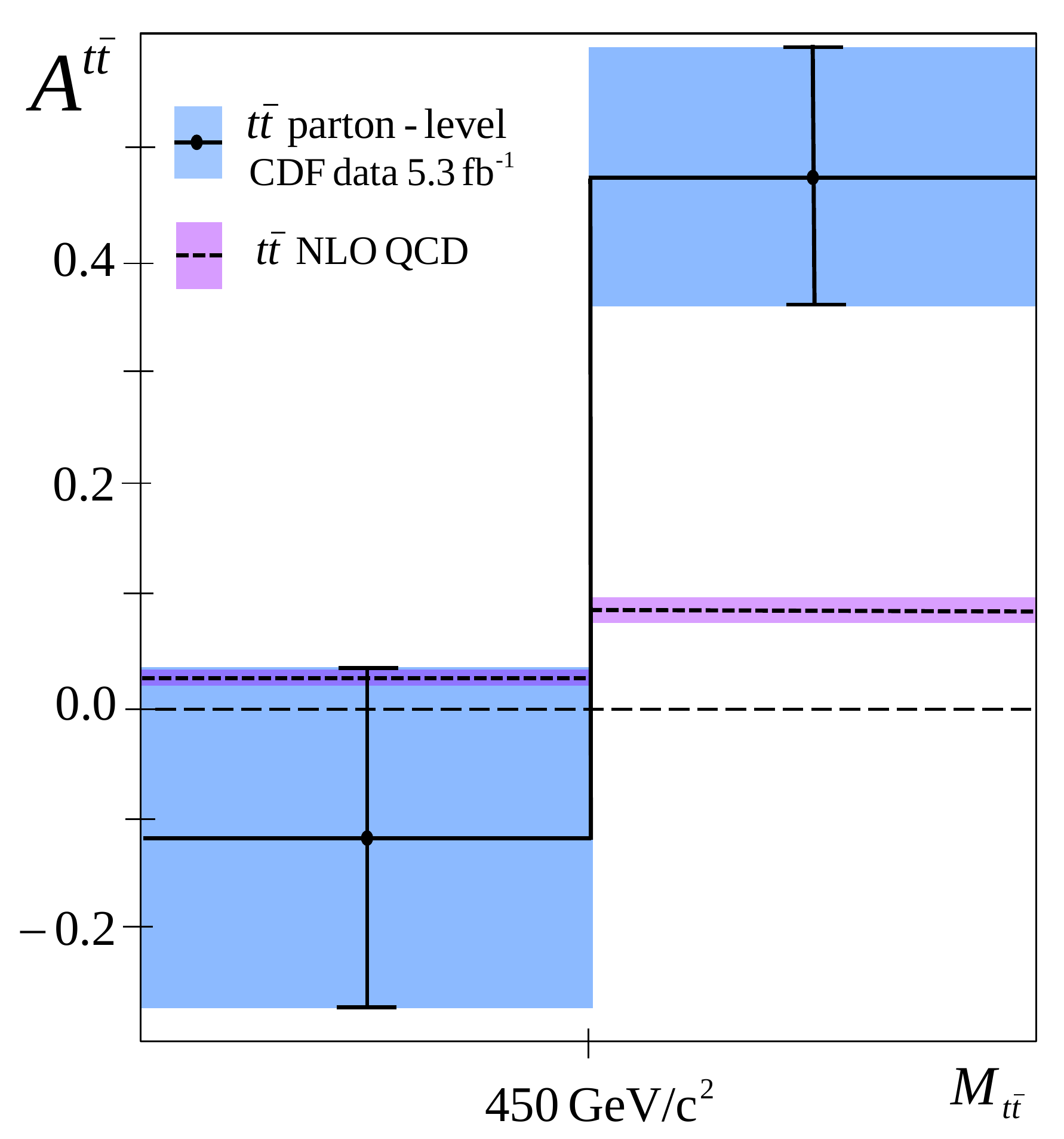}
\caption{{\small Parton-level asymmetry in $\dy$ at high and low mass compared to {\sc mcfm} prediction. The shaded region represents the total uncertainty in each bin.}\label{fig:unfold_result}}
\end{center}
\end{figure}

\section{Cross-Checks of the Mass Dependent Asymmetry}\label{sec:avm_cross}

The large and unexpected asymmetry at high mass demands a broader study of related effects in the $\ttbar$ data. We look for anomalies that could be evidence of a false positive, 
along with correlations that could reveal more about a true positive. In order to avoid any assumptions related to the background subtraction, we make comparisons at the data level, appealing when necessary to the full $\ttbar$ + bkg simulation models. 


\subsection{Lepton Type}
All of our simulated models predict asymmetries that are independent of the lepton type: {\sc pythia} predicts asymmetries that are consistent with zero, and the Octet models predict asymmetries that are consistent with each other. The data are shown in Table~\ref{tab:avm_cross}. At high mass, both lepton types show positive asymmetries consistent within errors. 

\begin{figure}
\begin{center}
\includegraphics[height=2.7in, clip]{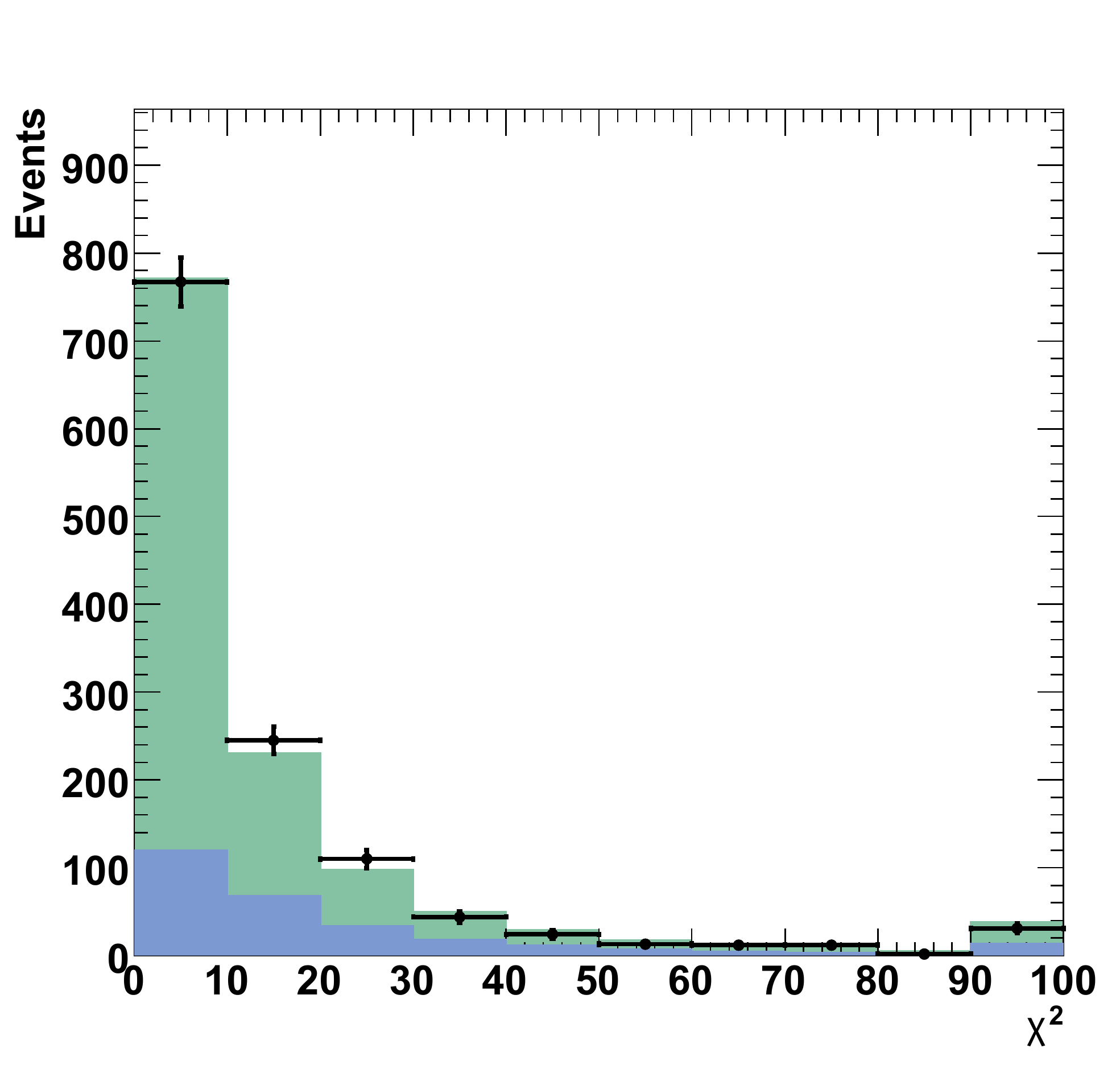}
\caption{{\small Distribution of $\ttbar$ reconstruction $\chi^2$. Black crosses are data, histogram is sig+bkg prediction.The last bin on the right contains all events with $\chi^2 > 100$.} \label{fig:chisq}}
\end{center}
\end{figure}

\begin{table*}[!ht]
\caption{Data level asymmetries $\ad$ for different event selections. In the case of no-b-fit, the $\ttbar$ reconstruction has been run without the constraint that $b$-tagged jets be associated with b-partons.}\label{tab:avm_cross}\begin{center}
\begin{tabular}{c c c c c }
\hline
\hline
selection          & N events    &     all $\mttb$     & $\mttb < 450~\gevcc$   & $\mttb \geq 450~\gevcc$  \\
\hline
standard           & 1260          & 0.057$\pm$0.028 & -0.016$\pm$0.034 & 0.212$\pm$0.049 \\
\hline
electrons         & 735           & 0.026$\pm$0.037   & -0.020$\pm$0.045        & 0.120$\pm$0.063 \\
muons             & 525           & 0.105$\pm$0.043   & -0.012$\pm$0.054        & 0.348$\pm$0.080 \\
\hline
data $\chi^2 < 3.0$ &  338         & 0.030$\pm$0.054   & $-0.033\pm0.065$  & $0.180\pm0.099$\\
data no-b-fit          & 1260         & 0.062$\pm$0.028   & $ 0.006\pm0.034$  & $0.190\pm0.050$\\
\hline
data single b-tag   & 979           & 0.058$\pm$0.031 & -0.015$\pm$0.038 & 0.224$\pm$0.056 \\
data double b-tag   & 281           & 0.053$\pm$0.059 & -0.023$\pm$0.076 & 0.178$\pm$0.095 \\
\hline
data anti-tag     & 3019          & 0.033$\pm$0.018 & 0.029$\pm$0.021 & 0.044$\pm$0.035 \\
pred anti-tag     &  -            & 0.010$\pm$0.007 & 0.013$\pm$0.008 & 0.001$\pm$0.014 \\
\hline
\hline
pre-tag & 4279 & 0.040$\pm$0.015 & 0.017$\pm$0.018 & 0.100$\pm$0.029 \\
pre-tag no-b-fit &  4279 & 0.042$\pm$0.015 & 0.023$\pm$0.018 & 0.092$\pm$0.029 \\
\hline
\end{tabular}
\end{center}
\end{table*}

\subsection{Reconstruction}

It is conceivable that a reconstruction error could produce an asymmetry from symmetric inputs. The quality of the reconstruction is summarized by a $\chi^2$ that measures the consistency of the solution with the $\ttbar$ hypothesis. The distribution of $\chi^2$ in our sample, shown in Fig.~\ref{fig:chisq}, is in very good agreement with the prediction, including a good match on the long tail. When the sample is restricted to high quality fits with $\chi^2 \leq 3.0$, we find 338 events in which $\ad = -0.033\pm0.065$ at low mass and $\ad = 0.180\pm0.099$ at high mass. Although the statistical precision is diminished in this small sample, it suggests that the high mass asymmetry is present in the best reconstructed events. Since the $\chi^2$ requirement rejects a significant fraction of the background, it also suggests that the high mass asymmetry is not a background related effect.

To test for possible reconstruction biases related to $b$-tagging, we re-run the reconstruction algorithm removing the constraint that $b$-tag jets be matched to $b$ partons. We find $\ad = 0.006\pm0.034$ at low mass and $\ad = 0.190\pm0.050$ at high mass. When we further separate the events by lepton charge, the $\dy_{lh}$ asymmetries are $A_{lh}^- = -0.190\pm 0.074$ and $A_{lh}^+ = 0.190\pm 0.069$. The large forward-backward charge asymmetry at high mass is seen to be independent of the use of b-jet identification in the reconstruction. 

\begin{table*}[!th]
\begin{center}
\begin{minipage}{4.0in}
\caption{{\sc mc@nlo} predictions for $\ad$ in reconstructed $\ttbar$ signal (no backgrounds) as a function of $\mttb$ and jet multiplicity. The uncertainties reflect MC statistics only.} \label{tab:mcnlo_njets}
\begin{tabular}{ l c c c }
\hline
\hline
selection            & all $\mttb$      & $\mttb < 450~\gevcc$   & $\mttb\ge 450~\gevcc$  \\ 
\hline
inclusive            & $0.024\pm 0.004$           & $0.015\pm 0.005$      &  $0.043\pm0.007$   \\
4-jet                & $0.048\pm 0.005$           & $0.033\pm 0.006$      &  $0.078\pm 0.009$   \\ 
5-jet                & $-0.035\pm 0.007$          & $-0.032\pm 0.009$     & $ -0.040\pm 0.012$    \\ 
\hline
\hline
\end{tabular}
\end{minipage}
\end{center}
\end{table*}

\begin{table*}[!th]
\begin{center}
\caption{Asymmetries $\ad$ in the data as a function of jet multiplicity.}\label{tab:avm_njets}
\begin{tabular}{ c c c c c }
\hline
\hline
selection          & N events    &     all $\mttb$     & $\mttb < 450~\gevcc$   & $\mttb \geq 450~\gevcc$  \\
\hline
inclusive           &  1260     &  0.057$\pm$0.028  & -0.016$\pm$0.034        & 0.212$\pm$0.049 \\
4-jet          &  939      & 0.065$\pm$0.033   & -0.023$\pm$0.039 & 0.26$\pm$0.057 \\
5-jet          &  321       & 0.034$\pm$0.056  & 0.0049$\pm$0.07 & 0.086$\pm$0.093 \\  
\hline
\hline
\end{tabular}
\end{center}
\end{table*}

\begin{table*}[!th]
\begin{center}
\caption{Reconstruction level asymmetries $\al$ in the laboratory frame.}\label{tab:Alhilowcharges}
\begin{tabular}{ c c c c c}
\hline
\hline
selection                    & all $\mttb$       & $\mttb < 450~\gevcc$   & $\mttb\ge 450~\gevcc$  \\
\hline
data reco                     &  0.073$\pm$0.028  &  0.059$\pm$0.034        & 0.103$\pm$0.049 \\
{\sc mc@nlo}                     &  0.001$\pm$0.003  &  -0.008$\pm$0.005        & 0.022$\pm$0.007 \\
\hline
$A_h^+$                    &  -0.070$\pm$0.040  &  -0.028$\pm$0.050        & -0.148$\pm$0.066 \\
$A_h^-$                    &  0.076$\pm$0.039  &  0.085$\pm$0.047       &  0.053$\pm$0.072 \\
\hline
single $b$-tags              &  0.095$\pm$0.032  &  0.079$\pm$0.034        & 0.130$\pm$0.057 \\
double $b$-tags              &  -0.004$\pm$0.060  &  -0.023$\pm$0.076        & 0.028$\pm$0.097 \\
\hline
\hline
\end{tabular}
\end{center}
\end{table*}

\subsection{B-Jet Identification}

All of our simulated models predict asymmetries that are independent of whether one or two jets are $b$-tagged. In single and double $b$-tagged samples {\sc pythia} predicts asymmetries that are consistent with zero, and the Octet models predict asymmetries that are consistent with each other. In the data, the two cases are consistent with each other, although the statistical precision on the double tagged sample is marginal.

In the background dominated anti-tags, the inclusive and low mass samples have small asymmetries that agree with the prediction. In the high mass anti-tag sample we find $\ad = 0.044\pm0.035$, consistent with either the model prediction of zero or a slight excess due to the $\ttbar$ component there. Mixing backgrounds and $\ttbar$ in the expected ratio and assuming the $\ttbar$ component has an asymmetry of $0.266$ (as in Table ~\ref{tab:summaryunfold}), we find a total expected asymmetry 
in the anti-tag sample of $\ad = 0.079\pm 0.034$ in agreement with the data.

The lepton+jets sample with no $b$-tagging is the ``pre-tag'' sample. Our standard {\sc pythia} + background model predicts pre-tag asymmetries consistent with zero for all mass categories. The asymmetries in the data are shown in Table ~\ref{tab:avm_cross}. At low mass the asymmetry in the pre-tags is consistent with zero. At high mass, the pre-tag sample has a significant asymmetry $0.100\pm0.029$. If we assume that $\ttbar$ signal at high mass has $\ad = 0.266$ as in Table~\ref{tab:summaryunfold} and combine $\ttbar$ with our standard backgrounds in the expected pre-tag ratio, we predict a pre-tag asymmetry of $\ad = 0.111\pm 0.028$, in good agreement with the data.  

As a final check in the pre-tag sample, we repeat the exercise of running the reconstruction without the constraint that $b$-tagged jets are used as b-partons. The results are shown in the bottom row of Table~\ref{tab:avm_cross}. The asymmetry at high mass is $0.092\pm 0.029$, a significant effect in a sample that makes absolutely no reference to $b$-tagging. 

\subsection{Jet Multiplicity}\label{sec:njets}

In Sec.~\ref{sec:mcfm} we discussed the two components of the NLO QCD asymmetry: (1) radiative corrections to
quark-antiquark production and (2) interference between different amplitudes contributing to the $\ttbar j$ final state. The two contributions have opposite signs. At NLO, the first is positive and dominant for the inclusive measurement, while the second is negative and subdominant. Since only the second term produces $\ttbar j$ events, we expect that the QCD asymmetry will be a function of the jet multiplicity. 

We have studied the jet multiplicity dependence of $\ad$ in {\sc mc@nlo}. We define 4-jet events as those with four jets with $\etran \ge 20$ GeV and $\abseta 2.0$ and no other such jets. We define 5-jet events as those with at least five jets with $\etran>20$ GeV and $\abseta 2.0$. The {\sc mc@nlo} prediction for the pure $\ttbar$ signal after reconstruction is shown in Table~\ref{tab:mcnlo_njets}. The 5-jet asymmetries are negative, as expected. The exclusive 4-jet sample shows asymmetries that are roughly double those in the inclusive sample. 

As we discussed in Sec.~\ref{sec:mcfm}, the reliability of the NLO picture has recently been called into question by NNLO calculations of the $\ttbar j$ component \cite{nnlo}, which reduce the negative asymmetry there to close to zero. However, since no NNLO calculation exists for the exclusive 4-jet, inclusive, or mass dependent asymmetries, the {\sc mc@nlo} prediction in Table~\ref{tab:mcnlo_njets} remains our comparison point.\\

The jet multiplicity dependence of the asymmetries in the data is shown in Table~\ref{tab:avm_njets}. Vetoing events with extra jets does not produce a significant increase in the asymmetry. However, in the 5-jet sample, the high mass asymmetry is consistent with zero. With a larger sample and better precision it might be possible to use the jet multiplicity to test whether the observed asymmetry is an amplified version of the QCD charge asymmetry or a different effect altogether.  
  
\subsection{Frame Dependence}\label{sec:frames}

As in the inclusive analysis, it is interesting to compare $\ad$ to $\al$. In the NLO QCD effect, the frame dependence of the asymmetry (see Sec.~\ref{sec:mcfm}) persists at high mass. For $\mttb \geq 450~\gevcc$ our {\sc mc@nlo} model predicts the ratio of reconstructed asymmetries in the two frames $\al/\ad \sim 0.74$. The OctetA model predicts less mass dependence, with a ratio of $0.90$.

The lab frame data asymmetries above and below $\mttb = 450~\gevcc$ are shown in Table ~\ref{tab:Alhilowcharges}. 
The variation of the asymmetry across the $450~\gevcc$ mass edge is not as distinct as in the $\ttbar$ frame, 
and the deviation from the {\sc mc@nlo} prediction is not as significant. Within the large errors, the asymmetries in the two lepton charge samples are consistent with $CP$ invariance. 

Comparing Tables \ref{tab:Alhilowcharges} and~\ref{tab:Ahilowcharges}, the ratio of $\al$ to $\ad$ at high mass is $0.49\pm 0.21$, lower than both the {\sc mc@nlo} and Octet models. We have used pseudo-experiment techniques to evaluate the statistical consistency of this ratio with the models, using a large number of simulated experiments that differ by Poisson fluctuations in the $\dy$ and $-q\yh$ distributions. A $\al/\ad$ ratio of $0.49$ or less occurs in $14\%$ of pseudo-experiments with {\sc mc@nlo}, but in $< 1\%$ of experiments with OctetA. 

Finally, we look at $\al$ as a function of the $b$-tag multiplicity. We observed in Sec.~\ref{tab:inc_cross} that the inclusive $\al$ is zero in the double $b$-tagged events. In Table~\ref{tab:Alhilowcharges}, we see that this pattern persists at high mass, although the statistical precision is poor. Appealing again to pseudo-experiments with Poisson fluctuations, we find that a ratio of double to single tag $\al$ as small as that in the data occurs in $6\%$ of all pseudo-experiments with {\sc mc@nlo}. We conclude that the low value of $\al$ in the double $b$-tagged sample is consistent with a statistical fluctuation.

\section{Conclusions}\label{conclusions}

We have studied the forward-backward asymmetry of top quark pairs produced in 1.96 TeV $\ppbar$ collisions at the Fermilab Tevatron. In a sample of 1260 events in the lepton+jet decay topology, we measure the parton-level inclusive asymmetry in both the laboratory and $\ttbar$ rest frame, and rapidity-dependent, and $\mttb$-dependent asymmetries in the $\ttbar$ rest frame. We compare to NLO predictions for the small charge asymmetry of QCD. 

The laboratory frame measurement uses the rapidity of the hadronically decaying top system and combines the two lepton charge samples under the assumption of CP conservation. This distribution shows a parton-level forward backward asymmetry in the laboratory frame of 
$\al = 0.150\pm 0.055$~(stat+sys). This has less than $1\%$ probability of representing a fluctuation from zero, and is two standard deviations above the predicted asymmetry from NLO QCD. We also study the frame-invariant difference of the rapidities, $\dy = \yt - \ytbar$, which is proportional to the top quark rapidity in the $\ttbar$ rest frame. Asymmetries 
in $\dy$ are identical to those in the $t$ production angle in the $\ttbar$ rest frame. We find a parton-level asymmetry of $A^{\rm{\ttbar}} = 0.158\pm 0.075$~(stat+sys), which is somewhat higher than, but not inconsistent with, the NLO QCD expectation of $0.058\pm 0.009$. 

In the $\ttbar$ rest frame we measure fully corrected asymmetries at small and large $\dy$

\begin{center}
   $\ad(|\dy|<1.0) = 0.026\pm 0.118$ \\
   $\ad(|\dy|\geq 1.0) = 0.611\pm 0.256$ \\
\end{center}
to be compared with {\sc mcfm} predictions of $0.039\pm 0.006$ and $0.123\pm 0.008$ for these $\dy$ regions respectively.

In the $\ttbar$ rest frame the asymmetry is a rising function of the $\ttbar$ invariant mass $\mttb$, with parton level asymmetries 
\begin{center}
   $\ad(\mttb < 450 ~\gevcc) = -0.116\pm 0.153$\\
\hspace*{-0.15in}
   $\ad(\mttb \geq 450 ~\gevcc)=  0.475\pm 0.114$\\
\end{center}
to be compared with {\sc mcfm} predictions of $0.040\pm 0.006$ and $0.088\pm 0.013$ for these $\mttb$ regions respectively. The asymmetry at high mass is $3.4$ standard deviations above the NLO prediction for the charge asymmetry of QCD, however we are aware that the accuracy of these theoretical predictions are under study. The separate results at high mass and large $\dy$ contain partially independent information on the asymmetry mechanism. 

The asymmetries reverse sign under interchange of lepton charge in a manner consistent with CP conservation. The $\ttbar$ frame asymmetry for $\mttb \geq 450~\gevcc$ is found to be robust against variations in $\ttbar$ reconstruction quality and secondary vertex b-tagging. When the high-mass data is divided by the lepton flavor, the asymmetries are larger in muonic events, but statistically compatible across species. Simple studies of the jet multiplicity and frame dependence of the asymmetry at high mass may offer the possibility of discriminating between the NLO QCD effect and other models for the asymmetry, but the statistical power of these comparisons is currently insufficient for any conclusion. 

The measurements presented here suggest that the modest inclusive $\ttbar$ production asymmetry originates from a significant effect at large rapidity difference $\dy$ and total invariant mass $\mttb$. The predominantly $\qqbar$ collisions of the Fermilab Tevatron are an ideal environment for further examination of this effect, and additional studies are in progress.  

We thank T. Tait, G. Sterman, W. Vogelsang, and A. Mitov for valuable conversations and assistance. We thank the Fermilab staff and the technical staffs of the participating institutions for their vital contributions. This work was supported by the U.S. Department of Energy and National Science Foundation; the Italian Istituto Nazionale di Fisica Nucleare; the Ministry of Education, Culture, Sports, Science and Technology of Japan; the Natural Sciences and Engineering Research Council of Canada; the National Science Council of the Republic of China; the Swiss National Science Foundation; the A.P. Sloan Foundation; the Bundesministerium f\"ur Bildung und Forschung, Germany; the World Class University Program, the National Research Foundation of Korea; the Science and Technology Facilities Council and the Royal Society, UK; the Institut National de Physique Nucleaire et Physique des Particules/CNRS; the Russian Foundation for Basic Research; the Ministerio de Ciencia e Innovaci\'{o}n, and Programa Consolider-Ingenio 2010, Spain; the Slovak R\&D Agency; and the Academy of Finland. 

\begin{figure*}[!t]
\begin{center}
\mbox{
\includegraphics[height=2.0in,  clip]{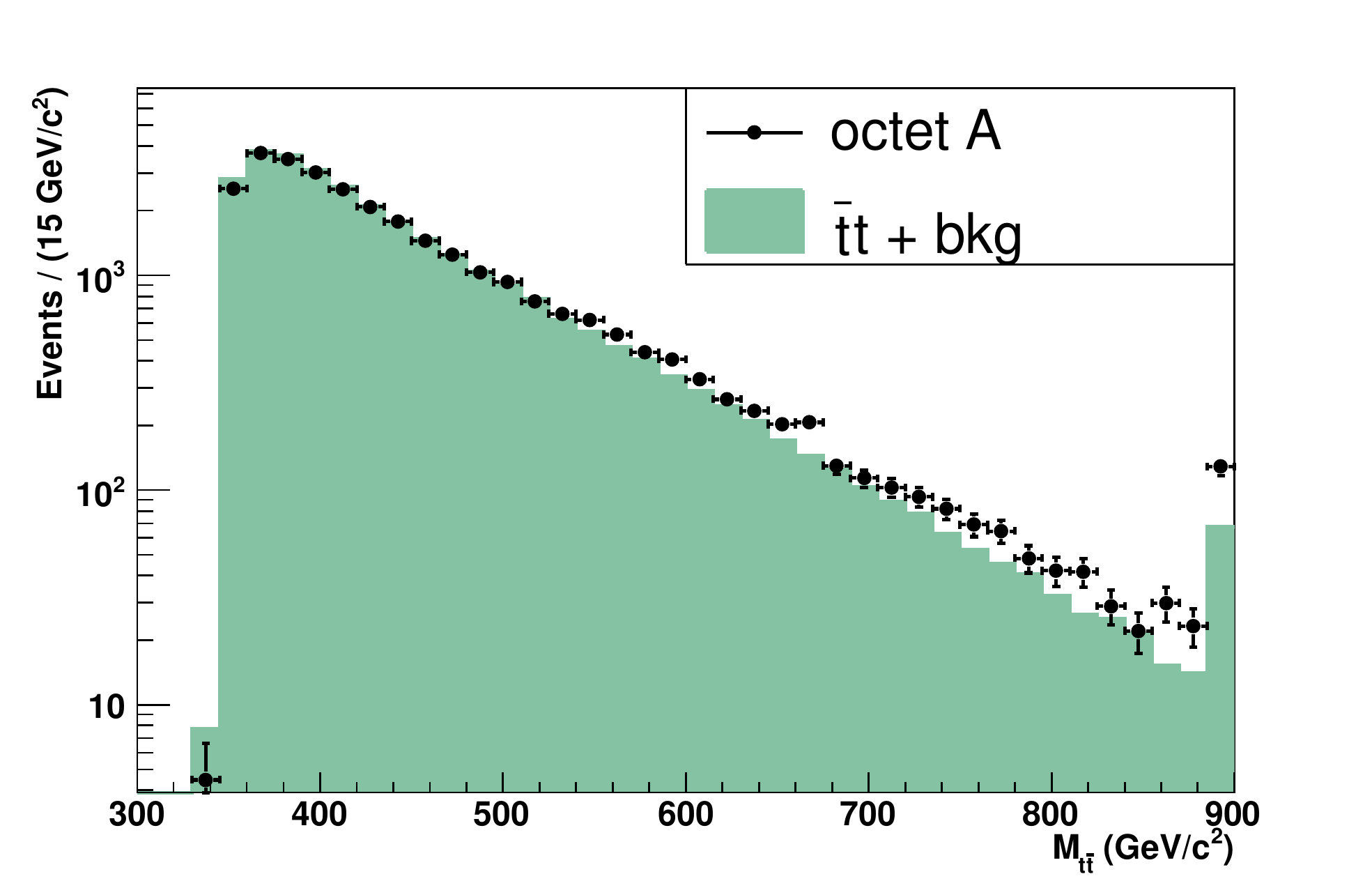}
    \hspace*{-0.1in}\vspace*{0.2in}
\includegraphics[height=2.0in, clip]{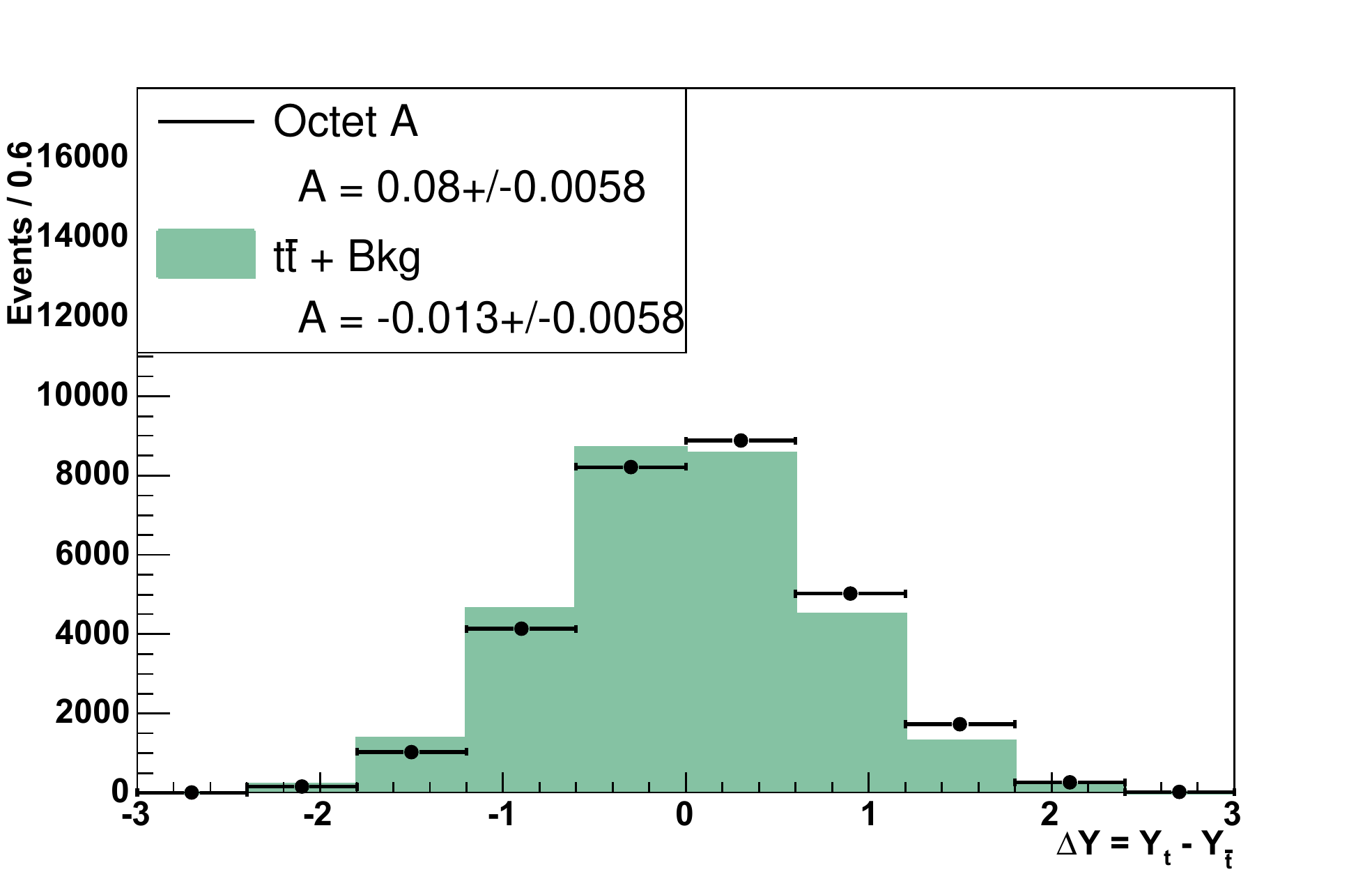}
}
\caption{{\small  The distributions of $\mttb$ (left) and $\dy$ (right) in the OctetA sample compared to {\sc pythia}.} \label{fig:octeta_v_pythia}}
\mbox{
\includegraphics[height=2.2in, clip]{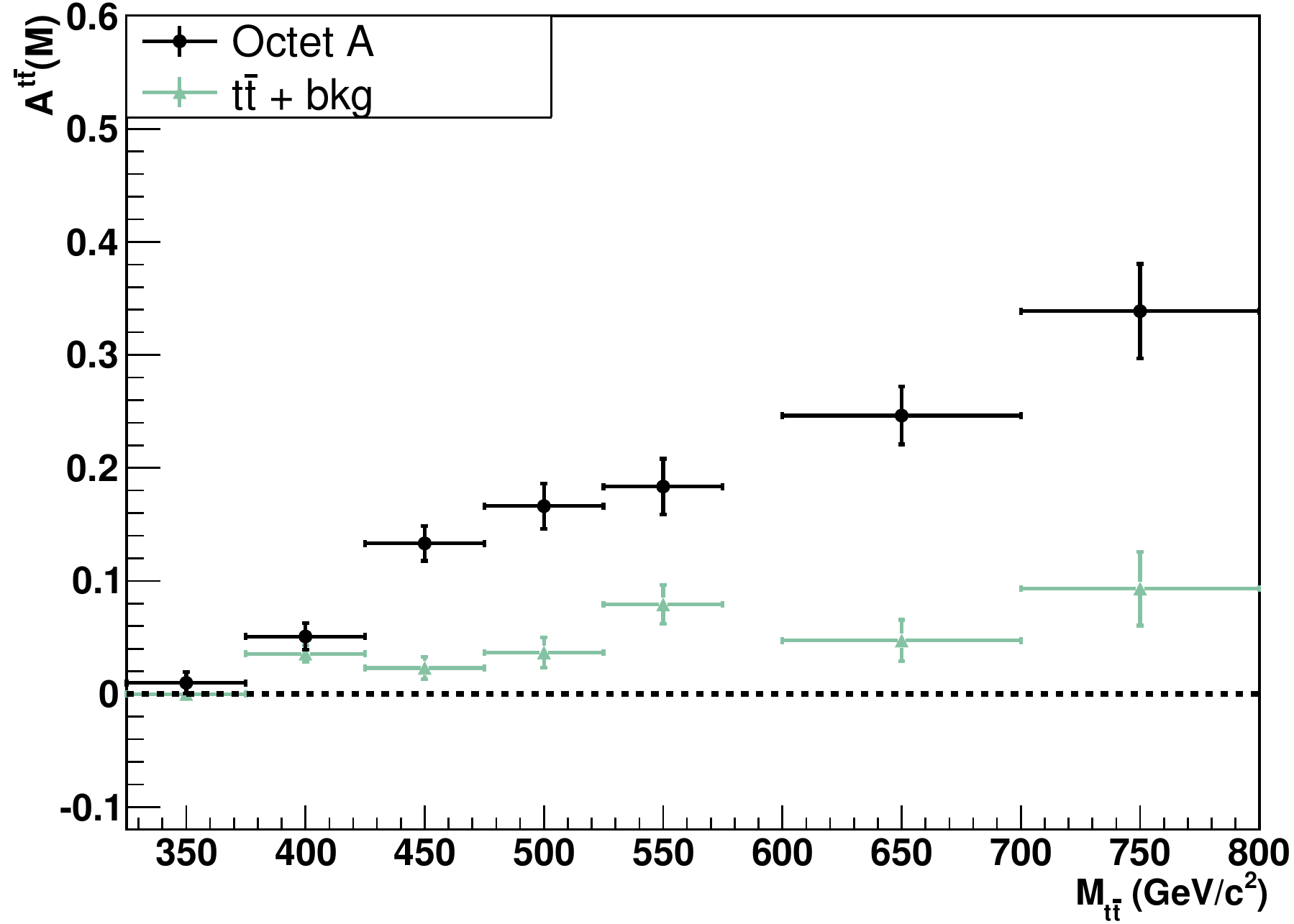}
      \hspace*{-0.1in}\vspace*{0.2in}
  \includegraphics[height=2.2in, clip]{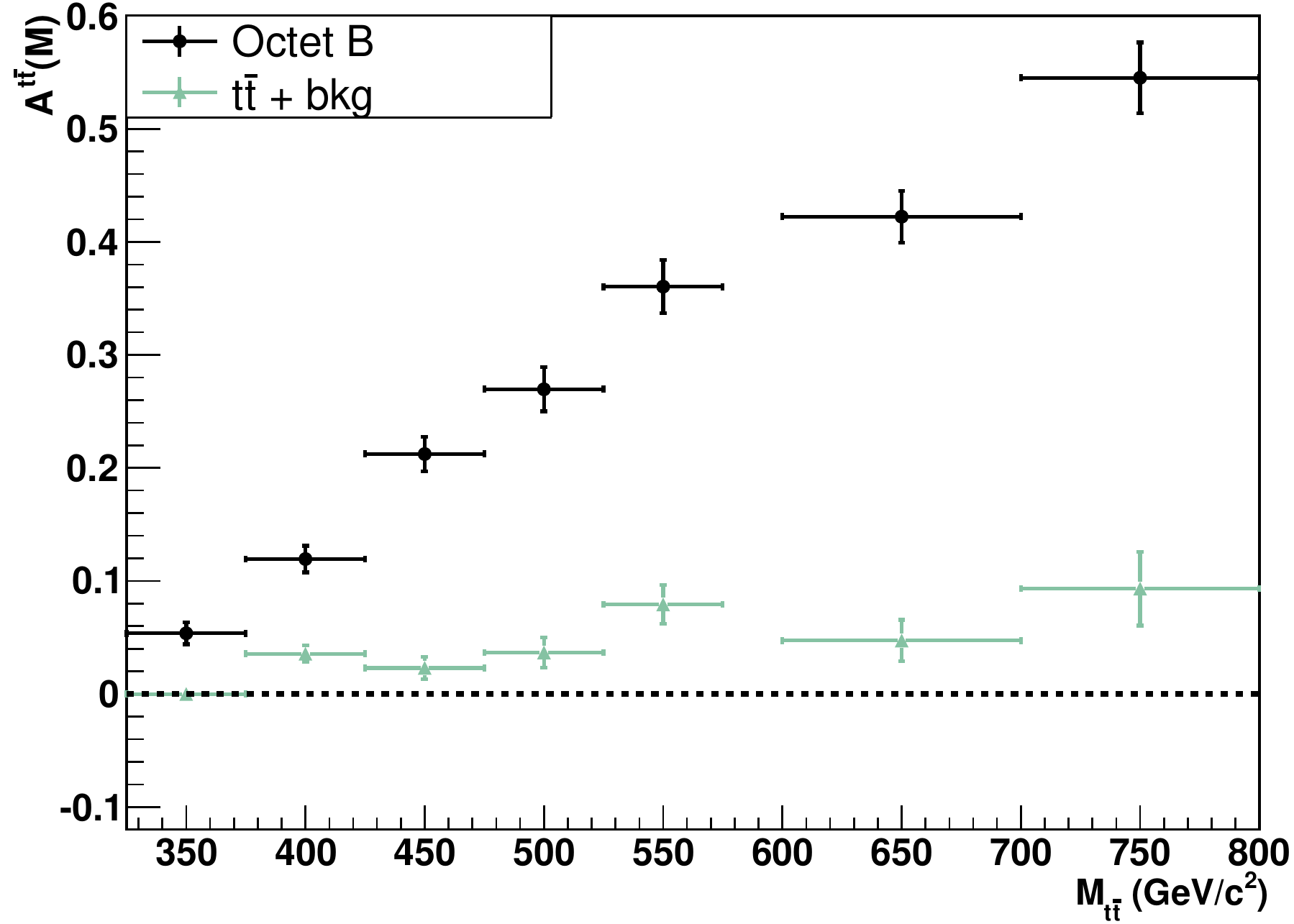}
}
\caption{{\small The reconstructed $M$ dependence of $\ad$ in the color-octet models. Left: OctetA. 
Right: OctetB.}} \label{fig:dadm_axig}
\end{center}
\end{figure*}

\begin{table*}[!th]
\begin{center}
\caption{$\ad$ and significance (in units of standard deviation) in Octet samples for events with $\mttb$ above the bin-edge.}\label{tab:AvM_high_axi}
\begin{tabular}{ l c c c c c }
\hline
\hline
 &\multicolumn{2}{c}{OctetA}   & \multicolumn{2}{c}{OctetB}    \\
\hline
bin-edge        & $\ad$                   & significance  & $\ad$   & significance \\
($\gevcc$)      &                                   &               &                   &      \\
\hline
345            & $0.082 \pm 0.028$    &   $2.90$      & $0.168 \pm 0.028$    &   $5.99$ \\
400            & $0.128 \pm 0.036$    &   $3.55$      & $0.235 \pm 0.035$    &   $6.74$ \\
450            & $0.183 \pm 0.047$    &   $3.91$      & $0.310 \pm 0.044$    &   $7.08$ \\
500            & $0.215 \pm 0.060$    &   $3.60$      & $0.369 \pm 0.054$    &   $6.81$ \\
550            & $0.246 \pm 0.076$    &   $3.25$      & $0.425 \pm 0.066$    &   $6.43$ \\
600            & $0.290 \pm 0.097$    &   $2.97$      & $0.460 \pm 0.081$    &   $5.70$ \\
\hline
\hline                
\end{tabular}
\end{center}
\end{table*}

\section{Appendix: The Color-Octet Models}

In the generic color-octet model of Ref.~\cite{rodrigo}, the gluon-octet interference produces an asymmetric $\cos(\theta^*)$ term in the production cross 
section. The couplings of the top and the light quarks to the massive gluon have opposite sign, giving a positive asymmetry as seen in the data. This was implemented in the {\sc madgraph} framework, and the couplings and $M_G$ were tuned to reasonably reproduce the asymmetries and $\mttb$ distribution of the data~\cite{tait}. The sample called OctetA, with couplings $g_V = 0, ~g_A(q) = 3/2, ~g_A(t)=-3/2$, and mass $M_G = 2.0$ TeV/$c^2$, has parton level asymmetries of $\al = 0.110$ and $\ad = 0.157$.  The LO cross section for this sample is 6.1 pb, in good agreement with the LO {\sc madgraph} cross section for standard model $\ttbar$ production at 6.0 pb. 

\begin{figure*}[!th]
\begin{center}
\mbox{
\includegraphics[height=2.2in, clip]{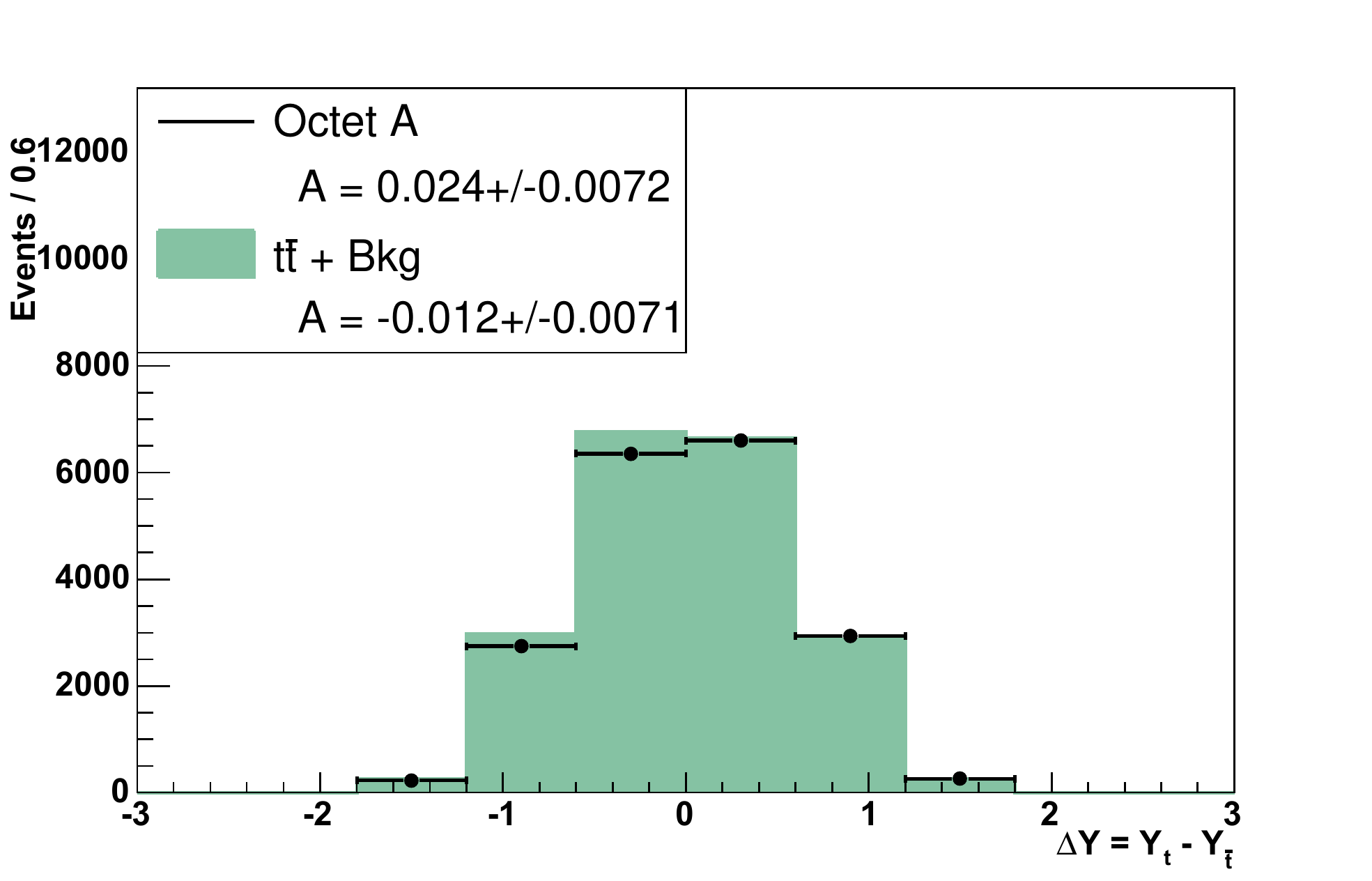}
      \hspace*{-0.1in}\vspace*{0.2in}
  \includegraphics[height=2.2in, clip]{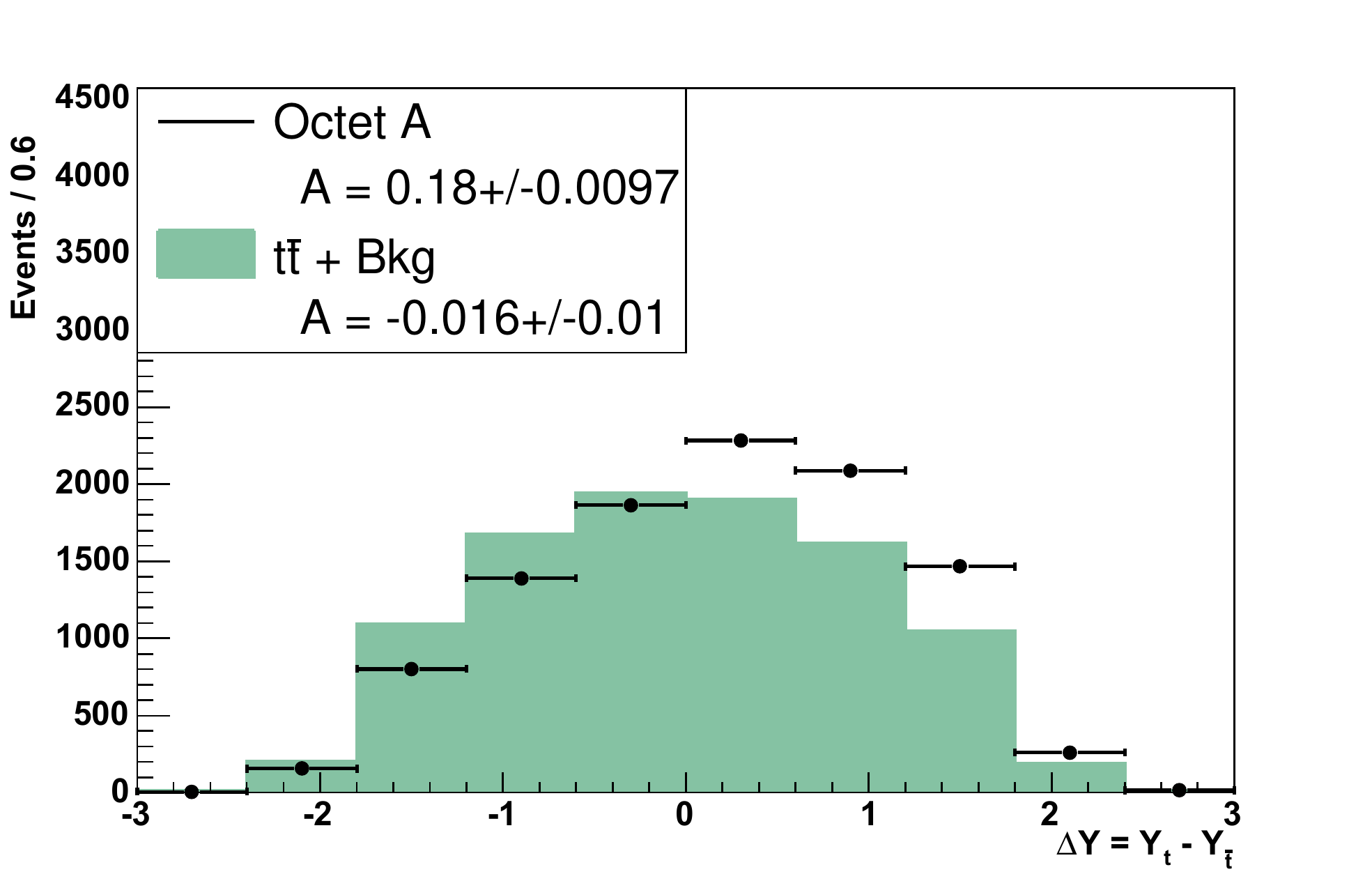}
}
\caption{{\small Top: The distribution of $\dy$ at low mass (left) and high mass (right). The {\sc pythia} baseline model is the filled histogram and the Octet samples are the points.} \label{fig:AvM2bins_octet}}
\end{center}
\end{figure*}

The sample is showered with {\sc pythia}, run through the CDF-II detector simulation, and then subjected to our selection and reconstruction. The $\mttb$ and $\dy$ distributions in OctetA are compared to the {\sc pythia} versions in Fig.~\ref{fig:octeta_v_pythia}. The $\mttb$ distribution is a good match to {\sc pythia}, and we have checked that related transverse variables are also well-modeled. The rapidity distributions after selection and reconstruction have asymmetries $\al = 0.073\pm 0.006$, $\ad = 0.079\pm 0.006$, which are reasonable matches to the data. 

The complementary OctetB sample has the same couplings but $M_G = 1.8$ TeV/$c^2$, giving parton-level asymmetries $\al = 0.205$ and $\ad = 0.282$. The $\ttbar$ cross section increases by $5\%$ and the reconstructed mass distribution has a slight excess at the high mass relative to {\sc pythia}. The data level asymmetry $\ad = 0.16\pm 0.006$, significantly higher than the data. 

\begin{table*}[!th]
\begin{center}
\caption{Reconstructed asymmetry $\ad$ below and above $\mttb = 450~\gevcc$ in Octet models.}\label{tab:Ahilowcharges_axig}
\begin{tabular}{ c c c c }
\hline
\hline
 \ sample                & all $\mttb$         & $\mttb < 450~\gevcc$   & $\mttb\ge~450 ~\gevcc$  \\
\hline
  OctetA                 &  $0.080\pm 0.006$   &  $0.024\pm 0.007$        & $0.180\pm 0.010$      \\
  OctetB                 &  $0.160 \pm 0.006$  &  $0.078\pm 0.007$        & $0.310\pm 0.009$      \\
\hline
\hline
\end{tabular}
\end{center}
\end{table*}

These models both show the same approximate factor of $2$ dilution between data-level and parton-level asymmetries that is seen in the data and in {\sc mc@nlo}.

Since these models have relatively low-lying octet masses near 2 TeV/$c^2$ we expect a significant $\mttb$-dependent asymmetry over our experimental range. The $\daddm$ behavior for the two color-octet samples is shown in Fig.~\ref{fig:dadm_axig}. Both show a smooth and significant rise of the asymmetry with increasing mass. 

In Sec.~\ref{sec:hilowmass} we discussed a simple representation of $\ad(\mttb)$ with two regions of low and high $\mttb$. The question for that representation is how to choose the boundary mass between high and low. Table ~\ref{tab:AvM_high_axi} shows the asymmetry, uncertainty, and significance $\ad/\sigma_{\ad}$ at high mass as a function of the mass threshold for both octet models. The uncertainties are calculated assuming the data sample size of $5.3 ~\ifb$. For both samples, the significance of the asymmetry at high mass is maximum at reconstructed $\mttb = 450~\gevcc$. Looking at Fig.~\ref{fig:mttb}, we see that this is reasonable: $450~\gevcc$ cuts off the bulk of the low mass peak while retaining good statistics on the tail.  

Fig.~\ref{fig:AvM2bins_octet} compares the $\dy$ distributions in the OctetA sample and {\sc pythia} when the events are divided into samples below and above $\mttb = 450~\gevcc$. The $\dy$ distribution is much broader at high mass, as expected, and shows a marked asymmetry.

The reconstructed asymmetries at high and low mass in the color octet samples are given in Table ~\ref{tab:Ahilowcharges_axig}. 
The uncertainties here reflect the Monte Carlo statistics.
At high mass the color octet samples have large asymmetries as seen in Fig.~\ref{fig:AvM2bins_octet}. At low mass, 
the models have small but significant asymmetries, especially in OctetB. We have checked that these asymmetries are charge asymmetries, 
reversing sign under interchange of lepton charge.


\begin{thebibliography}{99}
\bibitem{coords} The polar angle is $\theta$ and the azimuthal angle is $\phi$. With total energy $E$ and momentum $p$, the transverse energy is defined as $\etran = E\sin\theta$ and the transverse momentum is $\ptran = p\sin\theta$. The missing transverse energy ($\met$) is defined by $\vec{\met} = -\Sigma_{i}\etran^{i}\hat{n}_i$ where $\hat{n}_i$ is a unit vector perpendicular to the beam axis and pointing to the $i^{th}$ calorimeter tower. The pseudorapidity is $\eta = -ln(\tan(\theta/2))$.
\bibitem{cc} Unless otherwise stated, the inclusion of the charge-conjugate mode is implied.
\bibitem{towers} The rapidity resolution is independent of polar angle as a consequence of dividing the calorimeter into towers of constant y-width.
\bibitem{tait} We are indebted to T. Tait for the {\sc madgraph} implementation of the color-octet models.
\end{thebibliography}
\end{document}